\documentclass[longauth]{aa}  
\usepackage{graphicx}
\usepackage{txfonts}
\usepackage[dvipsnames]{xcolor}
\usepackage{multirow}
\usepackage{hyperref}
\usepackage{lastpage}
\begin{document}

   \title{The GAPS programme at TNG \thanks{Based on observations made with the Italian \textit{Telescopio Nazionale Galileo} (TNG) operated by the \textit{Fundación Galileo Galilei} (FGG) of the \textit{Istituto Nazionale di Astrofisica} (INAF) at the \textit{Observatorio del Roque de los Muchachos} (La Palma, Canary Islands, Spain).}}

   \subtitle{XLIX. TOI-5398, the youngest compact multi-planet system composed of an inner sub-Neptune and an outer warm Saturn}

   \author{G. Mantovan
          \inst{\ref{inst1},\ref{inst2}}$^{\href{https://orcid.org/0000-0002-6871-6131}{\includegraphics[scale=0.5]{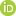}}}$ \and
          L. Malavolta\inst{\ref{inst1},\ref{inst2}}$^{\href{https://orcid.org/0000-0002-6492-2085}{\includegraphics[scale=0.5]{orcid.jpg}}}$ \and S. Desidera\inst{\ref{inst2}}$^{\href{https://orcid.org/0000-0001-8613-2589}{\includegraphics[scale=0.5]{orcid.jpg}}}$\and T. Zingales\inst{\ref{inst1},\ref{inst2}}$^{\href{https://orcid.org/0000-0001-6880-5356}{\includegraphics[scale=0.5]{orcid.jpg}}}$\and L. Borsato\inst{\ref{inst2}}$^{\href{https://orcid.org/0000-0003-0066-9268}{\includegraphics[scale=0.5]{orcid.jpg}}}$\and G. Piotto\inst{\ref{inst1}}$^{\href{https://orcid.org/0000-0002-9937-6387}{\includegraphics[scale=0.5]{orcid.jpg}}}$\and A. Maggio\inst{\ref{inst3}}$^{\href{https://orcid.org/0000-0001-5154-6108}{\includegraphics[scale=0.5]{orcid.jpg}}}$\and D. Locci\inst{\ref{inst3}}$^{\href{https://orcid.org/0000-0002-9824-2336}{\includegraphics[scale=0.5]{orcid.jpg}}}$\and D. Polychroni\inst{\ref{inst4}}$^{\href{https://orcid.org/0000-0002-7657-7418}{\includegraphics[scale=0.5]{orcid.jpg}}}$\and D. Turrini\inst{\ref{inst5}}$^{\href{https://orcid.org/0000-0002-1923-7740}{\includegraphics[scale=0.5]{orcid.jpg}}}$\and M. Baratella\inst{\ref{inst6}}$^{\href{https://orcid.org/0000-0002-1027-5003}{\includegraphics[scale=0.5]{orcid.jpg}}}$\and K. Biazzo\inst{\ref{inst7}}$^{\href{https://orcid.org/0000-0002-1892-2180}{\includegraphics[scale=0.5]{orcid.jpg}}}$\and D. Nardiello\inst{\ref{inst2}}$^{\href{https://orcid.org/0000-0003-1149-3659}{\includegraphics[scale=0.5]{orcid.jpg}}}$\and K. Stassun\inst{\ref{inst8},\ref{inst9}}$^{\href{https://orcid.org/0000-0002-3481-9052}{\includegraphics[scale=0.5]{orcid.jpg}}}$\and V. Nascimbeni\inst{\ref{inst2}}$^{\href{https://orcid.org/0000-0001-9770-1214}{\includegraphics[scale=0.5]{orcid.jpg}}}$\and S. Benatti\inst{\ref{inst3}}$^{\href{https://orcid.org/0000-0002-4638-3495}{\includegraphics[scale=0.5]{orcid.jpg}}}$\and A. Anna John\inst{\ref{inst10}}$^{\href{https://orcid.org/0000-0002-1715-6939}{\includegraphics[scale=0.5]{orcid.jpg}}}$\and C. Watkins\inst{\ref{inst11}}$^{\href{https://orcid.org/0000-0001-8621-6731}{\includegraphics[scale=0.5]{orcid.jpg}}}$\and A. Bieryla\inst{\ref{inst11}}$^{\href{https://orcid.org/0000-0001-6637-5401}{\includegraphics[scale=0.5]{orcid.jpg}}}$\and J. J. Lissauer\inst{\ref{inst12}}$^{\href{https://orcid.org/0000-0001-6513-1659}{\includegraphics[scale=0.5]{orcid.jpg}}}$\and J. D. Twicken\inst{\ref{inst13},\ref{inst14}}$^{\href{https://orcid.org/0000-0002-6778-7552}{\includegraphics[scale=0.5]{orcid.jpg}}}$\and A. F. Lanza\inst{\ref{inst15}}$^{\href{https://orcid.org/0000-0001-5928-7251}{\includegraphics[scale=0.5]{orcid.jpg}}}$\and J. N. Winn\inst{\ref{inst16}}$^{\href{https://orcid.org/0000-0002-4265-047X}{\includegraphics[scale=0.5]{orcid.jpg}}}$\and S. Messina\inst{\ref{inst15}}$^{\href{https://orcid.org/0000-0002-2851-2468}{\includegraphics[scale=0.5]{orcid.jpg}}}$\and M. Montalto\inst{\ref{inst15}}$^{\href{https://orcid.org/0000-0002-7618-8308}{\includegraphics[scale=0.5]{orcid.jpg}}}$\and A. Sozzetti\inst{\ref{inst5}}$^{\href{https://orcid.org/0000-0002-7504-365X}{\includegraphics[scale=0.5]{orcid.jpg}}}$\and H. Boffin\inst{\ref{inst17}}$^{\href{https://orcid.org/0000-0002-9486-4840}{\includegraphics[scale=0.5]{orcid.jpg}}}$\and D. Cheryasov\inst{\ref{inst18}}$^{\href{https://orcid.org/0009-0003-4203-9667}{\includegraphics[scale=0.5]{orcid.jpg}}}$\and I. Strakhov\inst{\ref{inst18}}$^{\href{https://orcid.org/0000-0003-0647-6133}{\includegraphics[scale=0.5]{orcid.jpg}}}$\and F. Murgas\inst{\ref{inst19},\ref{inst20}}$^{\href{https://orcid.org/0000-0001-9087-1245}{\includegraphics[scale=0.5]{orcid.jpg}}}$\and M. D'Arpa\inst{\ref{inst3}}$^{\href{https://orcid.org/0009-0004-5914-7274}{\includegraphics[scale=0.5]{orcid.jpg}}}$\and K. Barkaoui\inst{\ref{inst21},\ref{inst22},\ref{inst19}}$^{\href{https://orcid.org/0000-0003-1464-9276}{\includegraphics[scale=0.5]{orcid.jpg}}}$\and P. Benni\inst{\ref{inst23}}$^{\href{https://orcid.org/0000-0001-6981-8722}{\includegraphics[scale=0.5]{orcid.jpg}}}$\and A. Bignamini\inst{\ref{inst4}}$^{\href{https://orcid.org/0000-0002-5606-6354}{\includegraphics[scale=0.5]{orcid.jpg}}}$\and A. Bonomo\inst{\ref{inst5}}$^{\href{https://orcid.org/0000-0002-6177-198X}{\includegraphics[scale=0.5]{orcid.jpg}}}$\and F. Borsa\inst{\ref{inst24}}$^{\href{https://orcid.org/0000-0003-4830-0590}{\includegraphics[scale=0.5]{orcid.jpg}}}$\and L. Cabona\inst{\ref{inst2}}$^{\href{https://orcid.org/0000-0002-5130-4827}{\includegraphics[scale=0.5]{orcid.jpg}}}$\and A. C. Cameron\inst{\ref{inst10}}$^{\href{https://orcid.org/0000-0002-8863-7828}{\includegraphics[scale=0.5]{orcid.jpg}}}$\and R. Claudi\inst{\ref{inst2}, \ref{inst41}}$^{\href{https://orcid.org/0000-0001-7707-5105}{\includegraphics[scale=0.5]{orcid.jpg}}}$\and W. Cochran\inst{\ref{inst25}}$^{\href{https://orcid.org/0000-0001-9662-3496}{\includegraphics[scale=0.5]{orcid.jpg}}}$\and K. A. Collins\inst{\ref{inst11}}$^{\href{https://orcid.org/0000-0001-6588-9574}{\includegraphics[scale=0.5]{orcid.jpg}}}$\and M. Damasso\inst{\ref{inst5}}$^{\href{https://orcid.org/0000-0001-9984-4278}{\includegraphics[scale=0.5]{orcid.jpg}}}$\and J. Dong\inst{\ref{inst26},\ref{inst27}}$^{\href{https://orcid.org/0000-0002-3610-6953}{\includegraphics[scale=0.5]{orcid.jpg}}}$\and M. Endl\inst{\ref{inst25}}$^{\href{https://orcid.org/0000-0002-7714-6310}{\includegraphics[scale=0.5]{orcid.jpg}}}$\and A. Fukui\inst{\ref{inst34},\ref{inst19}}$^{\href{https://orcid.org/0000-0002-4909-5763}{\includegraphics[scale=0.5]{orcid.jpg}}}$\and G. F{\H u}r{\' e}sz\inst{\ref{inst28}}$^{\href{https://orcid.org/0000-0001-8467-9767}{\includegraphics[scale=0.5]{orcid.jpg}}}$\and D. Gandolfi\inst{\ref{inst44}}$^{\href{https://orcid.org/0000-0001-8627-9628}{\includegraphics[scale=0.5]{orcid.jpg}}}$\and A. Ghedina\inst{\ref{inst29}}$^{\href{https://orcid.org/0000-0003-4702-5152}{\includegraphics[scale=0.5]{orcid.jpg}}}$\and J. Jenkins\inst{\ref{inst14}}$^{\href{https://orcid.org/0000-0002-4715-9460}{\includegraphics[scale=0.5]{orcid.jpg}}}$\and P. Kab\'{a}th\inst{\ref{inst31}}$^{\href{https://orcid.org/0000-0002-1623-5352}{\includegraphics[scale=0.5]{orcid.jpg}}}$\and D.~W.~Latham\inst{\ref{inst11}}$^{\href{https://orcid.org/0000-0001-9911-7388}{\includegraphics[scale=0.5]{orcid.jpg}}}$\and V. Lorenzi\inst{\ref{inst19},\ref{inst29}}$^{\href{https://orcid.org/0000-0002-1958-9930}{\includegraphics[scale=0.5]{orcid.jpg}}}$\and R. Luque\inst{\ref{inst32}}$^{\href{https://orcid.org/0000-0002-4671-2957}{\includegraphics[scale=0.5]{orcid.jpg}}}$\and J. Maldonado\inst{\ref{inst3}}$^{\href{https://orcid.org/0000-0002-2218-5689}{\includegraphics[scale=0.5]{orcid.jpg}}}$\and K. McLeod\inst{\ref{inst33}}$^{\href{https://orcid.org/0000-0001-9504-1486}{\includegraphics[scale=0.5]{orcid.jpg}}}$\and M. Molinaro\inst{\ref{inst4}}$^{\href{https://orcid.org/0000-0001-5028-6041}{\includegraphics[scale=0.5]{orcid.jpg}}}$\and N. Narita\inst{\ref{inst19},\ref{inst34},\ref{inst35}}$^{\href{https://orcid.org/0000-0001-8511-2981}{\includegraphics[scale=0.5]{orcid.jpg}}}$\and G. Nowak\inst{\ref{inst19},\ref{inst20},\ref{inst36}}$^{\href{https://orcid.org/0000-0002-7031-7754}{\includegraphics[scale=0.5]{orcid.jpg}}}$\and J. Orell-Miquel\inst{\ref{inst19},\ref{inst20}}$^{\href{https://orcid.org/0000-0003-2066-8959}{\includegraphics[scale=0.5]{orcid.jpg}}}$\and E. Pall{\' e}\inst{\ref{inst19},\ref{inst20}}$^{\href{https://orcid.org/0000-0003-0987-1593}{\includegraphics[scale=0.5]{orcid.jpg}}}$\and H. Parviainen\inst{\ref{inst19},\ref{inst20}}$^{\href{https://orcid.org/0000-0001-5519-1391}{\includegraphics[scale=0.5]{orcid.jpg}}}$\and M. Pedani\inst{\ref{inst29}}$^{\href{https://orcid.org/0000-0002-5752-6260}{\includegraphics[scale=0.5]{orcid.jpg}}}$\and S.~N.~Quinn\inst{\ref{inst11}}$^{\href{https://orcid.org/0000-0002-8964-8377}{\includegraphics[scale=0.5]{orcid.jpg}}}$\and H. Relles\inst{\ref{inst11}}$^{\href{https://orcid.org/0009-0009-5132-9520}{\includegraphics[scale=0.5]{orcid.jpg}}}$\and P. Rowden\inst{\ref{inst37}}$^{\href{https://orcid.org/0000-0002-4829-7101}{\includegraphics[scale=0.5]{orcid.jpg}}}$\and G. Scandariato\inst{\ref{inst15}}$^{\href{https://orcid.org/0000-0003-2029-0626}{\includegraphics[scale=0.5]{orcid.jpg}}}$\and R. Schwarz\inst{\ref{inst11}}$^{\href{https://orcid.org/0000-0001-8227-1020}{\includegraphics[scale=0.5]{orcid.jpg}}}$\and S. Seager\inst{\ref{inst28},\ref{inst38},\ref{inst39}}$^{\href{https://orcid.org/0000-0002-6892-6948}{\includegraphics[scale=0.5]{orcid.jpg}}}$\and A. Shporer\inst{\ref{inst38}}$^{\href{https://orcid.org/0000-0002-1836-3120}{\includegraphics[scale=0.5]{orcid.jpg}}}$\and A.~Vanderburg\inst{\ref{inst42}, \ref{inst43}}$^{\href{https://orcid.org/0000-0001-7246-5438}{\includegraphics[scale=0.5]{orcid.jpg}}}$\and T. G. Wilson\inst{\ref{inst40}}$^{\href{https://orcid.org/0000-0001-8749-1962}{\includegraphics[scale=0.5]{orcid.jpg}}}$
          }

   \institute{Dipartimento di Fisica e Astronomia ``Galileo Galilei'', Università di Padova, Vicolo dell'Osservatorio 3, IT-35122, Padova, Italy\\
              \email{giacomo.mantovan@phd.unipd.it}\label{inst1}
         \and
             INAF - Osservatorio Astronomico di Padova, Vicolo dell'Osservatorio 5, IT-35122, Padova, Italy\label{inst2}
\and INAF - Osservatorio Astronomico di Palermo, Piazza del Parlamento 1, I-90134, Palermo, Italy\label{inst3}
\and INAF - Osservatorio Astronomico di Trieste, Via Giambattista Tiepolo, 11, I-34131, Trieste (TS), Italy\label{inst4}
\and INAF - Osservatorio Astrofisico di Torino, via Osservatorio 20, I-10025, Pino Torinese, Italy\label{inst5}
\and Leibniz-Institut für Astrophysik Potsdam (AIP), An der Sternwarte 16, 14482 Potsdam, Germany\label{inst6}
\and INAF – Osservatorio Astronomico di Roma, Via Frascati 33, 00078, Monte Porzio Catone (Roma), Italy\label{inst7}
\and Department of Physics and Astronomy, Vanderbilt University, Nashville, TN 37235, USA\label{inst8}
\and Department of Physics, Fisk University, Nashville, TN 37208, USA\label{inst9}
\and SUPA, School of Physics \& Astronomy, University of St Andrews, North Haugh, St Andrews KY169SS, UK; Centre for Exoplanet Science, University of St Andrews, North Haugh, St Andrews KY169SS, UK\label{inst10}
\and Center for Astrophysics | Harvard \& Smithsonian, 60 Garden Street, Cambridge, MA, 02138, USA\label{inst11}
\and Space Science and Astrobiology Division, NASA Ames Research Center, MS 245-3, Moffett Field, CA 94035, USA\label{inst12}
\and SETI Institute, Mountain View, CA 94043 USA/NASA Ames Research Center, Moffett Field, CA 94035 USA\label{inst13}
\and NASA Ames Research Center, Moffett Field, CA  94035, USA\label{inst14}
\and INAF - Osservatorio Astrofisico di Catania, Oss. Astr. Catania, via S. Sofia 78, 95123 Catania Italy\label{inst15}
\and Department of Astrophysical Sciences, Princeton University, Princeton, NJ 08544, USA\label{inst16}
\and ESO, Karl-Schwarzschild-str. 2, 85748 Garching, Germany\label{inst17}
\and Sternberg Astronomical Institute Lomonosov Moscow State University\label{inst18}
\and Instituto de Astrof\'{i}sica de Canarias (IAC), 38205 La Laguna, Tenerife, Spain\label{inst19}
\and Departamento de Astrofísica, Universidad de La Laguna (ULL), Avd. Astrof\'isico Francisco S\'anchez s/n, E-38206 La Laguna, Tenerife, Spain\label{inst20}
\and Astrobiology Research Unit, Universit\'e de Li\`ege, 19C All\'ee du 6 Ao\^ut, 4000 Li\`ege, Belgium\label{inst21}
\and Department of Earth, Atmospheric and Planetary Science, Massachusetts Institute of Technology, 77 Massachusetts Avenue, Cambridge, MA 02139, USA\label{inst22}
\and Acton Sky Portal (private observatory), Acton, MA, USA\label{inst23}
\and INAF – Osservatorio Astronomico di Brera, Via E. Bianchi 46, 23807 Merate (LC), Italy\label{inst24}
\and McDonald Observatory and Center for Planetary Systems Habitability; The University of Texas, Austin Texas USA\label{inst25}
\and Flatiron Research Fellow\label{inst26}
\and Center for Computational Astrophysics, Flatiron Institute, 162 Fifth Avenue, New York, NY 10010, USA\label{inst27}
\and Department of Physics and Kavli Institute for Astrophysics and Space Research, Massachusetts Institute of Technology, Cambridge, MA 02139, USA\label{inst28}
\and Fundación Galileo Galilei – INAF, Rambla José Ana Fernandez Pérez 7, 38712 Breña Baja, TF, Spain\label{inst29}
\and NASA Goddard Space Flight Center, 8800 Greenbelt Rd, Greenbelt, MD 20771, USA\label{inst30}
\and Astronomical Institute, Czech Academy of Sciences, Fri\v{c}ova 298, 251 65, Ond\v{r}ejov, Czech Republic\label{inst31}
\and Department of Astronomy \& Astrophysics, University of Chicago, Chicago, IL 60637, USA\label{inst32}
\and Department of Astronomy, Wellesley College, Wellesley, MA 02481, USA\label{inst33}
\and Komaba Institute for Science, The University of Tokyo, 3-8-1 Komaba, Meguro, Tokyo 153-8902, Japan\label{inst34}
\and Astrobiology Center, 2-21-1 Osawa, Mitaka, Tokyo 181-8588, Japan\label{inst35}
\and Institute of Astronomy, Faculty of Physics, Astronomy and Informatics, Nicolaus Copernicus University, Grudzi\c{a}dzka 5, 87-100 Toru\'n, Poland\label{inst36}
\and Royal Astronomical Society, Burlington House, Piccadilly, London W1J 0BQ, UK\label{inst37}
\and Department of Physics and Kavli Institute for Astrophysics and Space Research, Massachusetts Institute of Technology, Cambridge, MA 02139, USA\label{inst38}
\and Department of Aeronautics and Astronautics, MIT, 77 Massachusetts Avenue, Cambridge, MA 02139, USA\label{inst39}
\and Department of Physics, University of Warwick, Gibbet Hill Road, Coventry CV4 7AL, UK\label{inst40}
\and Dipartimento di Matematica e Fisica, Universit\`a Roma Tre, Via della Vasca Navale 84, 00146 Roma, Italy\label{inst41}
\and Department of Astronomy, The University of Texas at Austin, Austin, TX 78712, USA\label{inst42}
\and NASA Sagan Fellow\label{inst43}
\and Dipartimento di Fisica, Universit\'a degli Studi di Torino, via Pietro Giuria 1, I-10125, Torino, Italy\label{inst44}
             }

   \date{Compiled: \today}

 
  \abstract
  {Short-period giant planets ($P$\,$\lesssim$\,10 days, $M_{\rm p}$\,>\,0.1 $M_{J}$) are frequently found to be solitary compared to other classes of exoplanets. Small inner companions to giant planets with $P$\,$\lesssim$\,15 days are known only in five compact systems: WASP-47, Kepler-730, WASP-132, TOI-1130, and TOI-2000. Here, we report the confirmation of TOI-5398, the youngest compact multi-planet system composed of a hot sub-Neptune (TOI-5398 c, $P_{\rm c}$\,=\,4.77271 days) orbiting interior to a short-period Saturn (TOI-5398 b, $P_{\rm b}$\,=\,10.590547 days) planet, both transiting around a 650\,$\pm$\,150 Myr G-type star. }
  {As part of the Global Architecture of Planetary System (GAPS) Young Object project, we confirmed and characterised this compact system, measuring the radius and mass of both planets, thus constraining their bulk composition. 
  }
  {Using multidimensional Gaussian processes, we simultaneously modelled stellar activity and planetary signals from \textit{TESS} Sector 48 light curve and our HARPS-N radial velocity time series. We have confirmed the planetary nature of both planets, TOI-5398 b and TOI-5398 c, alongside a precise estimation of stellar parameters.}
  {Through the use of astrometric, photometric, and spectroscopic observations, our findings indicate that TOI-5398 is a young, active G dwarf star (650\,$\pm$\,150 Myr), with a rotational period of $P_{\rm rot}$\,=\,7.34 days. The transit photometry and radial velocity measurements enabled us to measure both the radius and mass of planets b, $R_b$\,=\,10.30$\pm$0.40 $R_{\oplus}$, $M_b$\,=\,58.7$\pm$5.7 $M_{\oplus}$, and c, $R_c$\,=\,3.52\,$\pm$\,0.19 $R_{\oplus}$, $M_c$\,=\,11.8$\pm$4.8 $M_{\oplus}$. \textit{TESS} observed TOI-5398 during sector 48 and no further observations are planned in the current Extended Mission, making our ground-based light curves crucial for ephemeris improvement. With a Transmission Spectroscopy Metric value around 300, TOI-5398 b is the most amenable warm giant (10\,<\,$P$\,<\,100 days) for JWST atmospheric characterisation.}
  {}

   \keywords{ planetary systems -- planets and satellites: fundamental parameters -- stars: fundamental parameters -- stars: individual: BD+37 2118 -- techniques: photometric -- techniques: radial velocities -- planet-star interactions
               }

   \maketitle
%

\section{Introduction}

Multi-planet systems give the unique opportunity to investigate comparative planetary science and understand the interactions and processes between their planets (e.g., \citealt{Dragomir_2019,2019AJ....158...32K,2021MNRAS.501.4148L,2023AJ....165..179T}). By studying the relative planet sizes and orbital separations, the obliquity between the planetary orbital planes and the stellar rotation axis, and other parameters, we can constrain the formation and evolution processes of these systems (see, e.g., \citealt{2022A&A...664A.162M,2022A&A...668A.172G}). The precise measurement of the orbital architecture and the bulk compositions of the planets is essential to fully characterise similar systems. In this context, high-precision radial velocities (RVs) are critical to deriving masses and eccentricities, which, combined with the radii from transit, allow measuring precise inner bulk densities and, finally, exploring the differences in planetary structure and evolution.

The architecture of multi-planet systems exhibits a significant amount of diversity, with planets spanning the whole range of masses and grouped in a wide variety of dynamical configurations. However, short-period giant planets ($P$\,$\lesssim$\,10 days) 
are typically isolated planets compared to other classes of exoplanets \citep{Huang2016}, and usually, their companions, if present, are massive long-period planets with $P$ > 200 days \citep{2014ApJ...785..126K,2016ApJ...825...62S}. Compactness is a rare feature among known multi-planet systems with inner giant planets. Moreover, there is a scarcity of known compact systems composed of a close-orbit outer jovian-size planet and an inner orbit small-size planet (e.g., \citealt{2022AJ....164...13H}). Within this family of unique planetary systems, we have WASP-47 \citep{2012MNRAS.426..739H}, Kepler-730 \citep{Zhu_2018, 2019ApJ...870L..17C}, TOI-1130 \citep{Huang_2020,2023arXiv230515565K}, TOI-2000 \citep{2022arXiv220914396S}, and WASP-132 \citep{2022AJ....164...13H}.

Studying planets younger than 1 Gyr is crucial for understanding the mechanisms at play during the early stages of planetary formation and evolution, such as orbital migration, atmospheric evaporation, planetary impacts, and so on \citep[see, e.g.,][]{2007ApJ...654.1110T,2019NatAs...3..416B}. However, it is challenging to identify and model such systems, because of the magnetic activity of the host star. In fact, both the photometric and spectroscopic time series are affected by the variations in star flux caused by the numerous star spots, faculae, faculae network, and strong magnetic activity on the stellar surface. Despite the stellar activity that hides planetary signals, researchers have discovered planets orbiting members of young stellar clusters \citep{2016A&A...588A.118M,2019ApJ...880L..17N,2020AJ....160..179M,2020AJ....160..239B}, young stellar associations and moving groups \citep{2019A&A...630A..81B,2023A&A...672A.126D}, and even young field stars \citep{2022arXiv221007933D}. There is an increasing number of young exoplanets with well-constrained ages, radii, and masses. Within the Global Architecture of Planetary Systems (GAPS, \citealt{2013A&A...554A..28C,2020A&A...638A...5C}) Young Objects (YO) long-term program at Telescopio Nazionale Galileo (TNG), many systems with young transiting planets, first identified by space missions such as \textit{Kepler} \citep{2010Sci...327..977B}, \textit{K2} \citep{2014PASP..126..398H}, and \textit{TESS} (Transiting Exoplanet Survey Satellite, \citealt{2015JATIS...1a4003R}), are being characterised through intense RV monitoring (e.g., \citealt{2021A&A...645A..71C,2022A&A...664A.163N}). 

Among the systems under intensive scrutiny by GAPS, the moderately young ($\sim$ 650 Myr) solar-analogue star TOI-5398 (BD+37 2118) is of special interest. In this paper, we characterise the compact multi-planet system orbiting this star using a combination of \textit{TESS} and ground-based photometry, and RVs collected with the High Accuracy Radial velocity Planet Searcher (HARPS-N, \citealt{2012SPIE.8446E..1VC}) spectrograph (Section \ref{sec:obs}). The \textit{TESS} pipeline identified two candidate exoplanets: a transiting sub-Neptune (TOI-5398.02, $P \sim$ 4.77 d) and a giant planet ($P \sim$ 10.59 d) that has been thoroughly validated by \cite{2022MNRAS.516.4432M} and labelled TOI-5398 b. In Section \ref{sec:host-star}, we report the stellar properties determined using two independent methods. Section \ref{sec:analysis} reports the procedures used to identify and confirm the two planets in the system by outlining the detailed modelling of photometry and RV. In Section \ref{sec:discussion}, we discuss our results, provide suggestions for follow-up observations, highlight the rarity of our system, and call attention to the exquisite suitability of TOI-5398 b for future atmospheric characterisation. Concluding remarks are in Section \ref{sec:conclusions}.


\section{Observations}
\label{sec:obs}
\subsection{TESS Photometry}
\label{sec:tess}

The \textit{TESS} mission observed TOI-5398 (TIC 8260536) at 2 min cadence in Sector 48 from 28 January 2022 to 26 February 2022. In particular, we considered the \textit{TESS} Science Processing Operations Center (SPOC; \citealt{2016SPIE.9913E..3EJ}) Simple Aperture Photometry (SAP; \citealt{2010SPIE.7740E..23T,2020ksci.rept....6M}) flux light curve and corrected it for time-correlated instrumental signatures. We corrected the light curve using Cotrending Basis Vectors (CBVs) extracted through the algorithm by \cite{2020MNRAS.495.4924N} as well as all SAP light curves in the same Camera and CCD as TOI-5398. We did this rather than using the Presearch Data Conditioning Simple Aperture Photometry (PDCSAP) light curve from the SPOC because, with the latter, the target experienced numerous systematic effects \citep{2012PASP..124.1000S,2012PASP..124..985S,2014PASP..126..100S} as a result of over-corrections and/or injection of spurious signals. 

\begin{figure}
   \centering
   \includegraphics[width=\hsize]
   {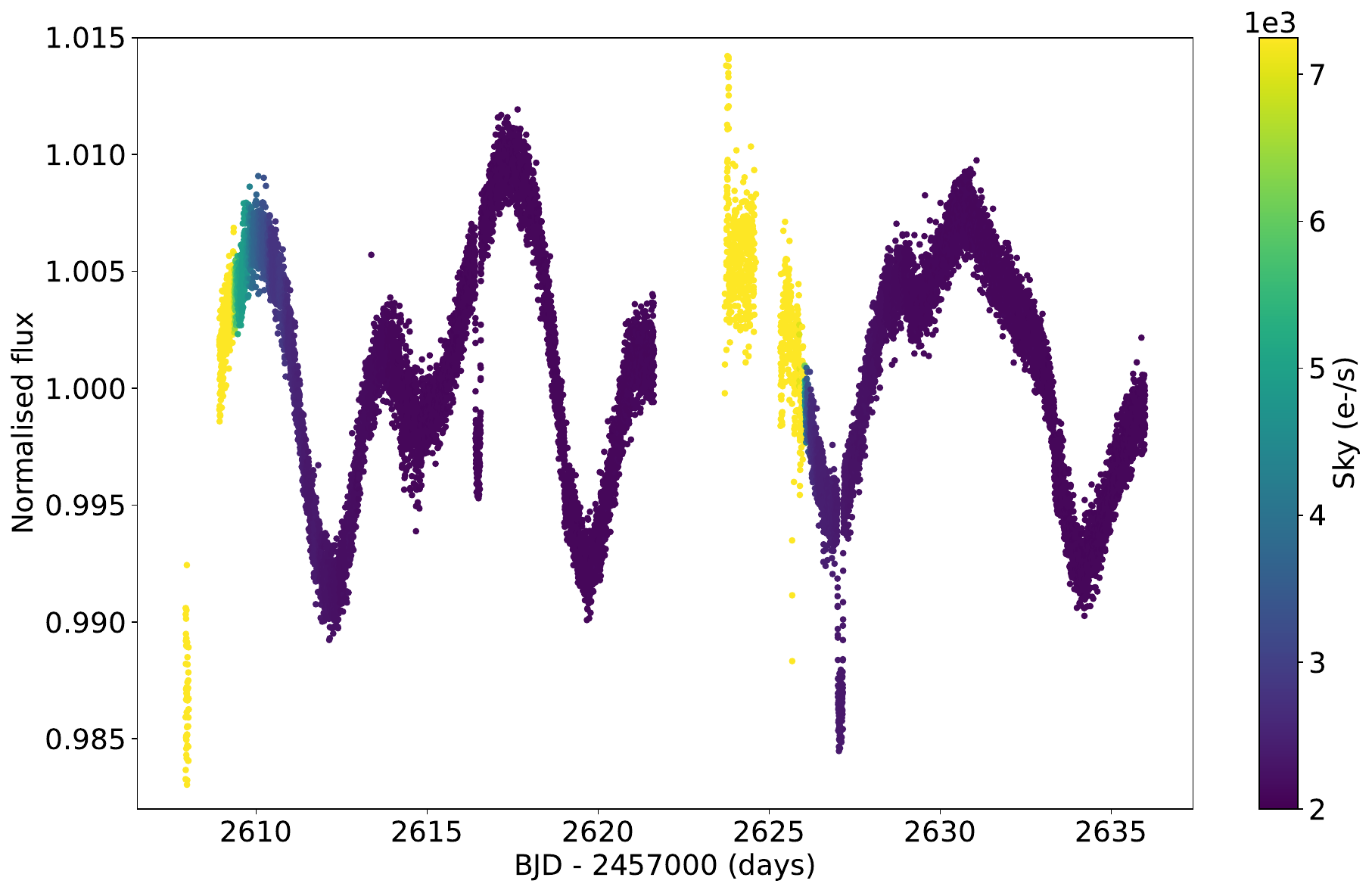}
   \caption{SAP-corrected light curve of TOI-5398, observed in Sector 48. The local background value is colour-coded. Yellow points -- with local background values $>$ 4$\sigma$ above its mean value along the light curve --  have been excluded from all analyses.}
   \label{fig:light}
\end{figure}

\begin{figure}
   \centering
   \includegraphics[width=\hsize,trim=4 4 4 4,clip]
   {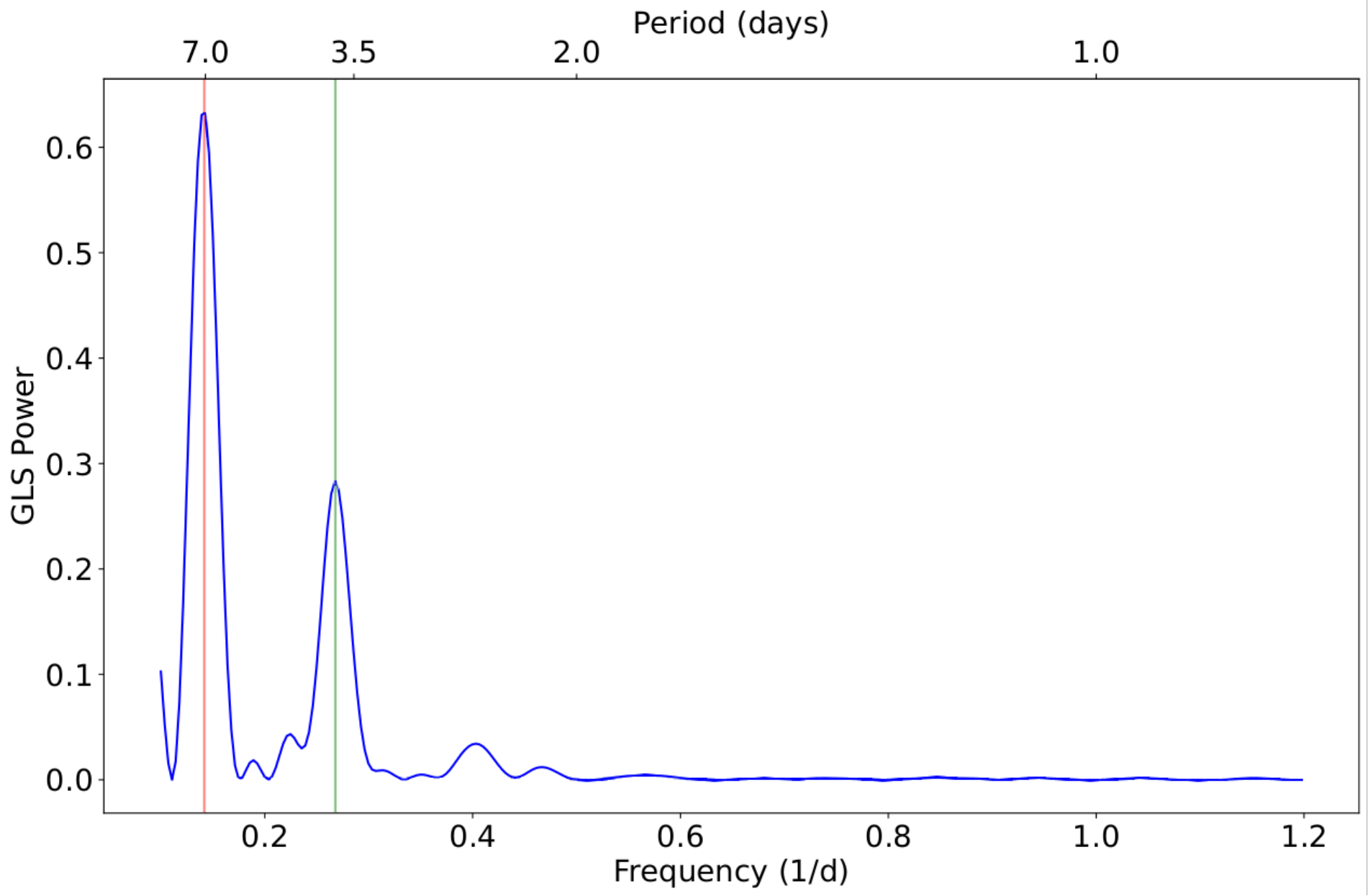}
   \caption{GLS periodogram of the \textit{TESS} SAP photometry corrected for time-correlated instrumental signatures. The vertical lines indicate the period at 7.18 days and its first harmonic. }
   \label{fig:gls_star}
\end{figure}

In this study, we used light curve points flagged with cadence quality flag values equal to 0 and associated with local background values < 4$\sigma$ above its mean\footnote{Mean of the sigma-clipped data, computed using \texttt{Astropy}.} value along the light curve. The only exceptions to this rule are the points before BTJD\footnote{BTJD = $\mathrm{BJD_{TDB}} - 2457000.0$} = 2609, which are flagged with a quality flag value equal to 32768\footnote{``Insufficient Targets for Error Correction Exclude'' \url{https://outerspace.stsci.edu/display/TESS/2.0+-+Data+Product+Overview}.} and had local background values up to 7$\sigma$ above the mean. We did this to preserve one of the TOI-5398.02's four genuine transits. The corrected light curve shown in Fig. \ref{fig:light} displays a clear modulation that we attribute to the stellar rotation. Therefore, to identify a rotation period of TOI-5398, we computed the Generalised Lomb-Scargle (GLS) periodogram \citep{2009A&A...496..577Z} of the aforementioned light curve. The periodogram reveals the most powerful peak at 7.18 $\pm$ 0.21 days, see Fig. \ref{fig:gls_star}. The link between this peak and the rotational period of the star will be discussed in Sec. \ref{sec:rotation}. The light curve also exhibits two clear dips at BTJD $\sim$\,2616 and 2627, which have been shown to be caused by a sub-stellar companion by \cite{2022MNRAS.516.4432M}.

\subsection{ASAS-SN}
\label{sec:asas}

\begin{figure}
   \centering
   \includegraphics[width=\hsize]
   {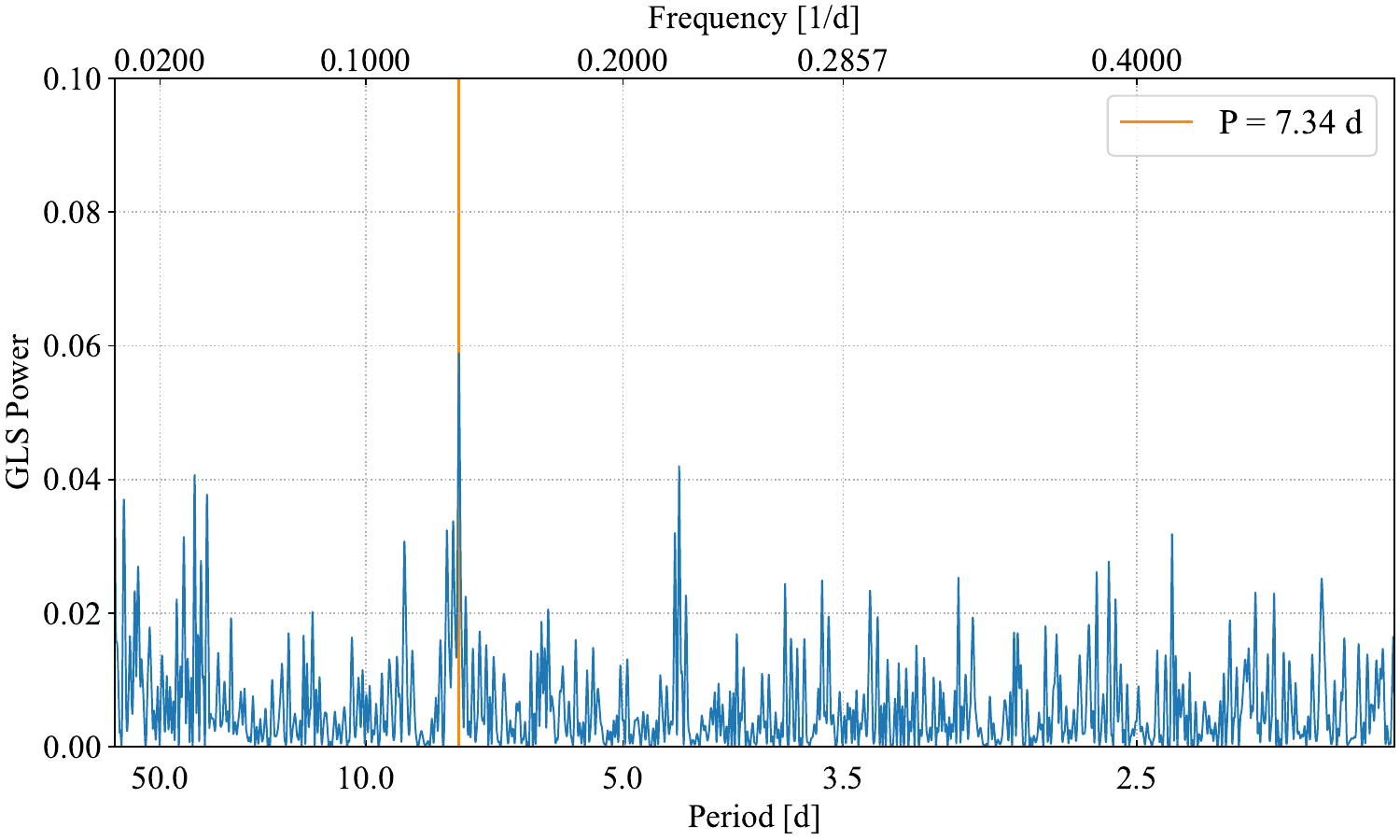}
   \caption{GLS periodogram of the \textit{ASAS-SN} photometry. The vertical line indicates the stellar rotation period at 7.34 days. }
   \label{fig:gls_asas}
\end{figure}

We downloaded almost three years of archival data from ASAS-SN \citep{2014ApJ...788...48S,2017PASP..129j4502K}, spanning from October 2020 to May 2023. The ASAS-SN images have a resolution of 8 arcsec/pixel ($\sim$15$\arcsec$ FWHM PSF), and the observations of TOI-5398 were conducted in the Sloan $g-$band. We checked the light curve to further constrain the stellar rotation period identified from the \textit{TESS} photometry. We extracted the GLS periodogram, which reveals the most powerful peak with a period of 7.34$\pm$0.01 days, see Fig. \ref{fig:gls_asas}.

\subsection{Photometric Follow-up (TASTE)}
A partial transit of TOI-5398 b was observed on January 31st, 2023 by The Asiago Search for Transit timing variations of Exoplanets (TASTE) programme, a long-term campaign to monitor transiting planets \citep{2011A&A...527A..85N}. To prevent saturation and increase the photometric accuracy, the AFOSC camera, mounted on the Copernico 1.82-m telescope at the Asiago Astrophysical Observatory in northern Italy, was purposely defocused up to $8\arcsec$~FWHM. A Sloan $r’$ filter was used to acquire 2916 frames with a constant exposure time of 5~s. A pre-selected set of suitable reference stars was always imaged in the same field of view in order to apply precise differential photometry. At the end of the series, the sky transparency was highly variable, and thick clouds passed by during the off-transit, leaving us with a subset of 2721 images with a good S/N ratio.

We reduced the Asiago data with the STARSKY code \citep{2013A&A...549A..30N}, a software pipeline specifically designed to perform differential photometry on defocused images. After a standard correction through bias and flat-field frames, the size of the circular apertures and the weights assigned to each reference star were automatically chosen by the code to minimise the photometric scatter of our target. The time stamps were consistently converted to the $\mathrm{BJD_{TDB}}$ standard following \cite{2010PASP..122..935E}, as done for the following photometric data sets as well. 

\subsection{Photometric Follow-up (TFOP)}

The \textit{TESS} pixel scale is $\sim 21\arcsec$/ pixel and photometric apertures typically extend out to roughly 1 arcminute, generally causing multiple stars to blend in the \textit{TESS} aperture. To determine the true source of transit signals in the \textit{TESS} data, improve the transit ephemerides, monitor for transit timing variations, and check the SPOC pipeline transit depth after accounting for the crowding metric, we conducted ground-based lightcurve follow-up observations of the field around TOI-5398 as part of the \textit{TESS} Follow-up Observing Program\footnote{https://tess.mit.edu/followup} Sub Group 1 \citep[TFOP;][]{collins:2019}. We used the {\tt TESS Transit Finder}, which is a customised version of the {\tt Tapir} software package \citep{Jensen:2013}, to schedule our transit observations. All the image data were calibrated, and all photometric data were extracted using {\tt AstroImageJ} unless otherwise noted. We used circular photometric apertures centred on TOI-5398, and checked also the flux from the nearest known neighbour in the {\it Gaia} DR3 \citep{2023A&A...674A...1G} and TICv8 catalogues (TIC 8260534), which is $\sim37\arcsec$ north-east of TOI-5398.

\subsubsection{LCOGT}

We observed four and two partial transit windows of TOI-5398 b and TOI-5398.02, respectively, in Pan-STARRS $z$-short band using the Las Cumbres Observatory Global Telescope \citep[LCOGT;][]{Brown:2013} 1.0\,m network nodes at Teide Observatory (TEID) on the island of Tenerife, McDonald Observatory (MCD) near Fort Davis, Texas, United States, and Cerro Tololo Inter-American Observatory in Chile (CTIO). We observed an ingress and egress of TOI-5398 b on UTC 2022 April 21 from TEID and MCD, respectively, an egress on UTC 2023 March 04 from TEID, and an ingress on UTC 2023 March 15 from MCD. We observed an egress of TOI-5398.02 on UTC 2022 April 02 from CTIO, and an ingress on UTC 2022 May 10 from MCD. The 1\,m telescopes are equipped with $4096\times4096$ Sinistro cameras having an image scale of $0.389\arcsec$/ pixel, resulting in a $26\arcmin\times26\arcmin$ field of view. The images were calibrated by the standard LCOGT {\tt BANZAI} pipeline \citep{McCully:2018}. We used circular photometric apertures having radii in the range $5.1\arcsec$ to $7.4\arcsec$, which excluded all of the flux from TIC 8260534.

\subsubsection{Acton Sky Portal}

A full transit window of TOI-5398 b was observed in Sloan $i'$ band on UTC 2022 April 21 from the Acton Sky Portal private observatory in Acton, MA, USA. The 0.36\,m telescope is equipped with a SBIG Alumna CCD4710 camera having an image scale of $1.0$ arcsec/ pixel, resulting in a $17.1\times17.1$ arcmin field of view. We used a circular photometric aperture with radius $9\arcsec$, which excluded all of the flux from TIC 8260534.

\subsubsection{Whitin}

A full transit window of TOI-5398 b was observed in Sloan $r'$ band on UTC 2022 April 21 using the Whitin observatory 0.7\,m telescope in Wellesley, MA, USA. The $2048\times2048$ FLI ProLine PL23042 detector has an image scale of $0.68$$\arcsec$/pixel, resulting in a $23.2\times23.2$ arcmin field of view. We used a circular photometric aperture with a radius of $8.2$ arcsec, which excluded all of the flux from TIC 8260534.

\subsubsection{KeplerCam}
We obtained photometric observations using the KeplerCam CCD on the 1.2m telescope at the Fred Lawrence Whipple Observatory (FLWO) at Mount Hopkins, Arizona to observe an egress of TOI-5398 b. Observations were taken in the Sloan z$^\prime$ band on UT April 21, 2022. KeplerCam is a $4096\times4096$ Fairchild detector which has a field of view of 23$\arcmin$x23$\arcmin$ and an image scale of 0.672$\arcsec$/pixel when binned by 2. We used circular photometric apertures with radius $6.7\arcsec$, which excluded all of the flux from TIC 8260534. 

\subsubsection{MuSCAT2}

TOI-5398 was observed by MuSCAT2 on the night of 30 March 2022. MuSCAT2 is a multi-band imager \citep{Narita2019} mounted on the Telescopio Carlos S\'{a}nchez (TCS, 1.52 m) at Teide Observatory, Spain. The instrument is capable of obtaining simultaneous images in Sloan $g'$, $r'$, $i'$, and $z_s$ bands with little readout time. Each camera has a field of view of $7.4\arcmin \times 7.4\arcmin$ and a pixel scale of 0.44$\arcsec$/ pixel.

The observation was made with the telescope defocused to avoid the saturation of the target. On the night of 30 March 2022, the $i'$-band camera presented technical issues and could not be used; the exposure times were set to 10, 5, and 10 seconds in $g'$, $r'$, and $z_s$, respectively. Standard data reduction, aperture photometry, and transit model fit including systematic noise was done using the MuSCAT2 pipeline \citep{Parviainen2015,Parviainen2019}.

\subsubsection{Results}

Due to their low S/N and to reduce the computational cost, we decided not to include light curves observed with LCOGT-CTIO and Whitin in our analysis. All other observations resulted in clear transit detections and the light curve data are included in the joint modelling in Sect. \ref{sec:phot_model} of this work. 

\subsection{HARPS-N Spectroscopic Follow-up}
\label{sec:obs-harpsn}
We collected observations of TOI-5398 with HARPS-N at TNG spanning the period between May 2022 and June 2023 and obtained a total of 86 spectra, with exposure times ranging from 900 to 1200 s. These spectra cover the wavelength range 383-693 nm with a resolving power of R $\sim$ 115 000. We performed the observations within the framework of the GAPS project and incorporated 10 hours coming from a time-sharing agreement (proposal A46TAC\_32, PI: G. Mantovan). Additionally, we included eight off-transit spectra from the DDT proposal A46DDT4 (PI: G. Mantovan) in our analysis, which we binned from 600 to 1200 s of exposure time. Six spectra taken in Spanish time (CAT22A\_48, PI: E. Pall\'e) are also included in a comprehensive analysis of the object. We excluded from our final analysis one of these six spectra as it was collected during the in-transit phase of planet b. We decided to proceed in this way after considering the expected amplitude of the Rossiter-McLaughlin \citep{Ohta2005} signature, which is further discussed in Sect. \ref{sec:form-evo}. We report the details about the observations and typical S/N in Table \ref{table:obs}.

\begin{table}
\caption{Observations from \textit{TESS} and HARPS-N summarised.}             
\label{table:obs}      
\centering                          
\begin{tabular}{c | c c}        
\hline\hline                 
 \multicolumn{3}{c}{TOI-5398} \rule{0pt}{2.5ex} \rule[-1ex]{0pt}{0pt} \\
\hline 
Dataset & Parameter & Value \rule{0pt}{2.5ex} \rule[-1ex]{0pt}{0pt} \\    
\hline                        
\multirow{4}{*}{\textit{TESS}} &  Sector & 48  \rule{0pt}{2.5ex} \rule[-1ex]{0pt}{0pt}\\ 
&  Camera & 1  \rule{0pt}{2.5ex} \rule[-1ex]{0pt}{0pt}\\
&  CCD & 1  \rule{0pt}{2.5ex} \rule[-1ex]{0pt}{0pt}\\
   \hline
\multirow{7}{*}{HARPN-N}    & N$^{\circ}$ spectra    & 86 \rule{0pt}{2.5ex} \rule[-1ex]{0pt}{0pt} \\
    & Time-span (days)     & 439 \rule{0pt}{2.5ex} \rule[-1ex]{0pt}{0pt} \\
    & $\sigma_{\rm RV}$ (m s$^{-1}$)   & 29 \rule{0pt}{2.5ex} \rule[-1ex]{0pt}{0pt} \\
    & $\langle \rm RV_{\rm err} \rangle$ (m s$^{-1}$)   & 3.5 \rule{0pt}{2.5ex} \rule[-1ex]{0pt}{0pt} \\
    & $\langle \rm S/N \rangle_{5460\AA}$    & 64 \rule{0pt}{2.5ex} \rule[-1.2ex]{0pt}{0pt} \\ 
\hline                                   
\end{tabular}
\end{table}

We reduced the data collected using the HARPS-N Data Reduction Software (DRS 3.7.0
), and we computed the RV through the Cross-Correlation Function (CCF) method (\citealt{2002A&A...388..632P} and references therein). With this method, the scientific spectra are cross-correlated with a binary mask describing the typical features of a star with a chosen spectral type. We used a G2 mask for TOI-5398. The resulting CCFs provide us with a representation of the mean line profile of each spectrum. However, the high levels of stellar activity might distort the core of the average line profile, and the stellar rotation broadens the line. Therefore, we needed to proceed with care in selecting the half-window for the evaluation of the CCF and use a width large enough to include the continuum when fitting the CCF profile (see \citealt{2020A&A...642A.133D}). We decided to use the G2 mask with a half-window of 40 km s$^{-1}$ (instead of the default value of 20 km s$^{-1}$) and reprocessed our data using the DRS version implemented through the YABI workflow interface \citep{Hunter2012} at the Italian Center for Astronomical Archives\footnote{\url{https://www.ia2.inaf.it/}}. We obtained RVs with a dispersion of a few tens of m\,s$^{-1}$ ($\sim$29 m\,s$^{-1}$), while their internal errors are approximately a few m s$^{-1}$ ($\sim$3.5 m\,s$^{-1}$).

\begin{figure*}
   \centering
   \includegraphics[width=\hsize]{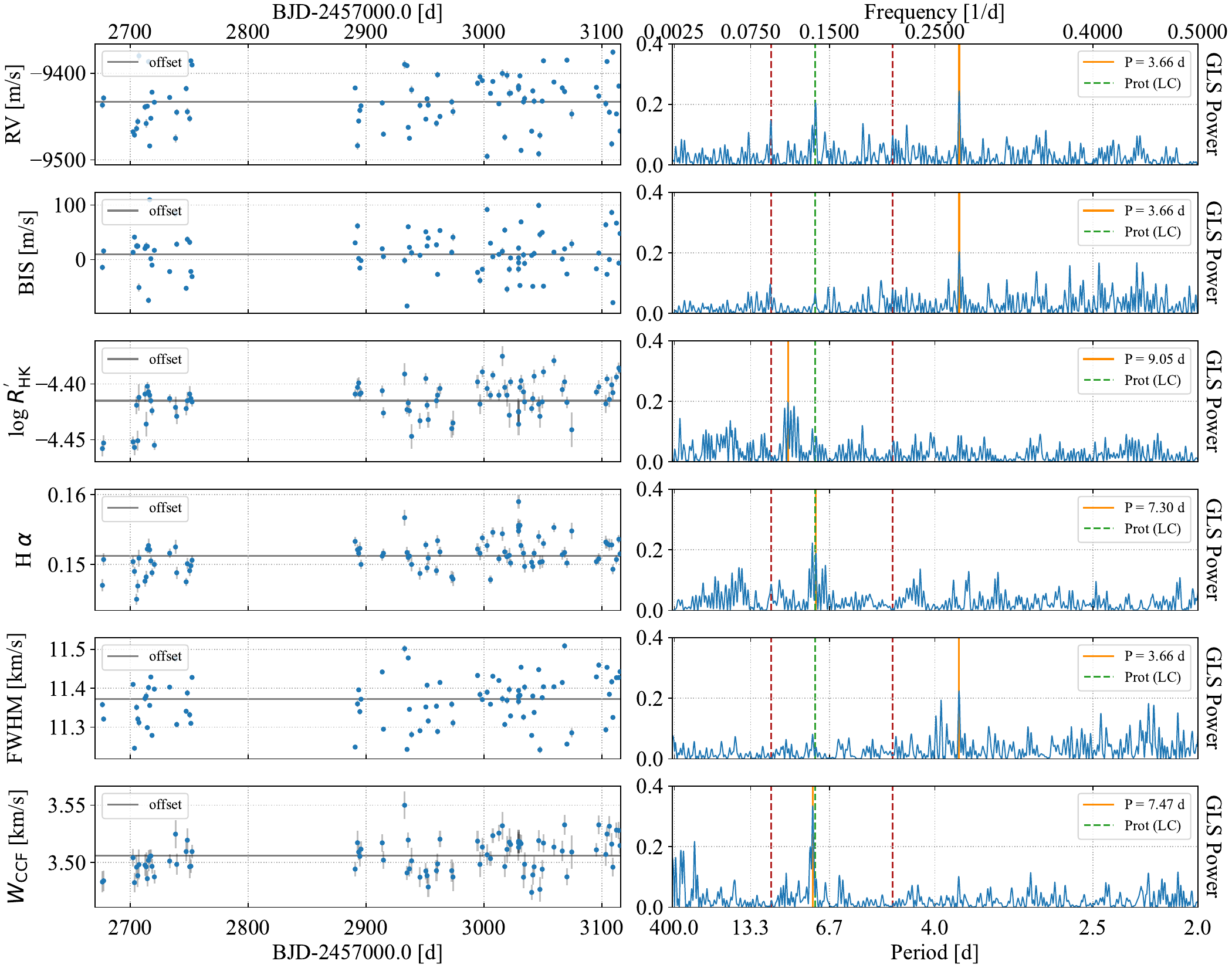}
   \caption{\textit{Left:} HARPS-N spectroscopic time series used in this work. The time series for RV, BIS, $\log R^{'}_{\rm HK}$, H$\alpha$, FWHM, and $W_{\rm CCF}$ are shown in the panels in order from top to bottom. \textit{Right:} GLS periodogram of the RVs and the spectroscopic activity indicators under analysis. The primary peak of each periodogram is indicated by a vertical orange line. The dotted green lines represent the stellar rotation period described in Sect. \ref{sec:asas}. The signals along the red dotted vertical lines correspond to the transit-like signals with periods 4.77 and 10.59 d. }
   \label{fig:gls}
\end{figure*}

To assess the jitter in the RV series caused by the stellar activity, we additionally extracted a set of activity indices. The value of the CCF bisector span (BIS), as well as the CCF's full width at half-maximum (FWHM) depth and the CCF's equivalent width ($W_{\rm CCF}$, see \citealt{2019MNRAS.487.1082C} for further details), are provided by the HARPS-N DRS, while the log $R'_{HK}$ index from the Ca II H\&K lines was obtained using a method available on YABI (based on the prescriptions of \citealt{2011arXiv1107.5325L} and references therein) and using the $(B-V)_0$ colour index quoted in Sect. \ref{sec:host-star}. Finally, we extracted the H$\alpha$ index using the ACTIN 2 Code\footnote{\url{https://github.com/gomesdasilva/ACTIN2}} (\citealt{2018JOSS....3..667G, 2021A&A...646A..77G}). Figure \ref{fig:gls} displays the spectroscopic time series.

We also obtained medium precise RVs with the Tull spectrograph at the Harlan J. Smith 2.7m telescope at McDonald Observatory. They can be found in the Appendix.


\subsection{High angular resolution data}
\label{sec:highres_obs}

\begin{figure}
   \centering
   \includegraphics[width=\hsize]{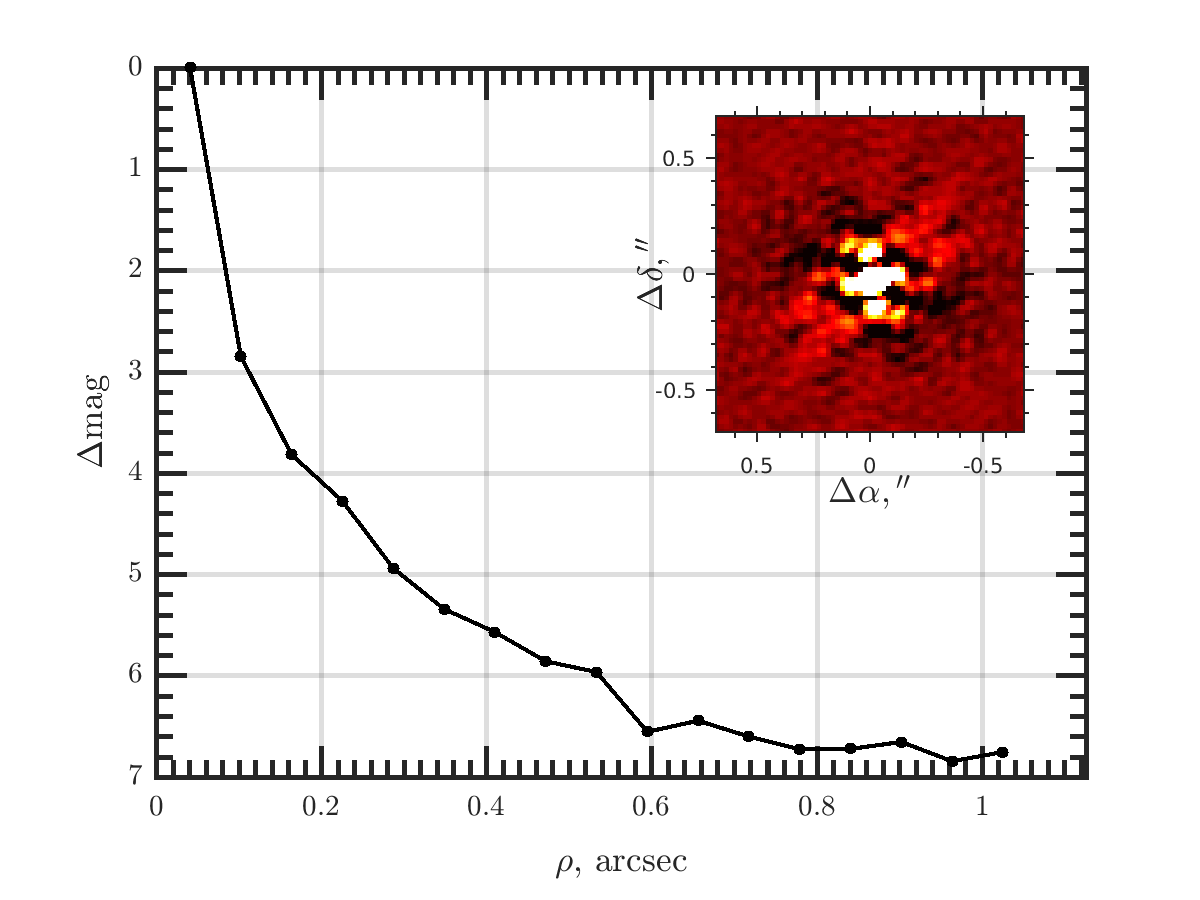}
   \caption{High angular resolution speckle imaging of TOI-
5398 in $I_c$ filter using SAI-2.5 m telescope. }
   \label{fig:high-res}
\end{figure}

TOI-5398 was also observed on 2 December 2022 with the speckle polarimeter on the 2.5 m telescope at the Caucasian Observatory of Sternberg Astronomical Institute (SAI) of Lomonosov Moscow State University. Speckle polarimeter uses high--speed low--noise CMOS detector Hamamatsu ORCA--quest \citep{Strakhov2023}. The atmospheric dispersion compensator was active, which allowed using the $I_c$ band. The respective angular resolution is $0.083\arcsec$, while the long--exposure atmospheric seeing was $0.6\arcsec$. We did not detect any stellar companions brighter than $\Delta I_c = 4.0$ and 6.8 mag at $\rho=0.25\arcsec$ and $1.0\arcsec$, respectively, where $\rho$ is the separation between the source and the potential companion (Fig. \ref{fig:high-res}).

\section{The host star}
\label{sec:host-star}

\subsection{Atmospheric parameters and metallicity}
\label{sec:atmo_param}

Given the relatively young age of TOI-5398, we analysed the HARPS-N combined spectrum following the same methodology as, for example, in \cite{2022A&A...664A.163N} and \cite{2023A&A...672A.126D}. 
Specifically, we adopted an innovative approach to derive the stellar parameters with the equivalent width (EW) method using a combination of iron (Fe) and titanium (Ti) lines. In this way, we avoid issues related to the effect of the stellar activity that can shape the stellar spectrum at young ages \citep[for a detailed explanation we refer the reader to][]{2020baratella_gaps,2020baratella_ges}.
Our initial guesses of the atmospheric parameters were estimated with \textit{Gaia} DR3 \citep{2016A&A...595A...1G} and 2MASS photometry \citep{2003yCat.2246....0C} using the tool \texttt{colte} \citep{2021casagrande} and adopting $E(B-V)=0.008\pm0.016$ \citep{2014A&A...561A..91L,2017A&A...606A..65C}. The photometric estimates vary from $5978\pm65$\,K in $(G_{\rm BP}-J)$ to $6039\pm34$\,K in $(G_{\rm RP}-H)$. 
From these initial values and taking the \textit{Gaia} parallax into account, we derived $4.44\pm0.09$ dex as input surface gravity, while an initial microturbulence $\xi=1.07\pm0.05$ km\,s$^{-1}$ was estimated from the relation by \cite{2016dutra}. 

The final stellar parameters ($T_{\mathrm{eff}}$, $\log g$, $\xi$, [Fe/H]) were then derived through the MOOG code (\citealt{sneden1973}) and adopting the ATLAS9 grid of model atmospheres with new opacities (\citealt{castellikurucz2003}). 
Our spectroscopic analysis provides as final atmospheric parameters $T_{\mathrm{eff}}=6000\pm75$\,K, $\log g=4.44\pm0.10$\,dex and $\xi=1.12\pm0.12$ km\,s$^{-1}$, in excellent agreement with the initial guesses. The derived iron abundance (computed with respect to the solar Fe abundance as in \citealt{2020baratella_gaps}) is [Fe/H]$=+0.09\pm0.06$, while the titanium abundance is [Ti/H]$=+0.08\pm0.05$, where the errors include the scatter due to the EWs measurements and the uncertainties in the stellar parameters. Despite the error bars, there is an indication of a slightly super-solar metallicity.

\subsection{Projected rotational velocity}
\label{sec:vsini}

With the same code and model atmospheres described in Sect.\,\ref{sec:atmo_param}, we adopted the atmospheric parameters ($T_{\mathrm{eff}}$, $\log g$, $\xi$, [Fe/H]) previously derived to measure the stellar projected rotational velocity ($v\sin{i_{\star}}$). In particular, fixing the macroturbulence velocity to the value of 3.8\,km s$^{-1}$ from the relationship by \cite{Doyleetal2014}, we applied the spectral synthesis method within MOOG for three spectral regions around 5400, 6200, and 6700\,\AA. Our final value of $v\sin{i_{\star}}$ is $7.5\pm0.6$\,km s$^{-1}$. We refer the readers to \cite{Biazzoetal2022} for the procedure based on spectral synthesis and the description of the uncertainty measurement. 

\subsection{Lithium Abundance}
\label{sec:lithium}

Since the lithium line at $\lambda$ 6707.8\,\AA\, in the co-added spectrum of TOI-5398 resulted to be blended with the iron line at $\lambda$ 6707.4\,\AA, we applied the spectral synthesis technique as done in Sect.\,\ref{sec:vsini}. Using MOOG and the \cite{castellikurucz2003} model atmospheres, we fixed the stellar parameters and $v\sin{i_{\star}}$ to the values derived in the previous steps. Then, adopting two different line lists by \cite{Carlbergetal2012} and Chris Sneden (priv. comm.) and applying the non-LTE calculations of \cite{Lindetal2009}, we obtained a lithium abundance of $\log n$(Li)$_{\rm NLTE}=2.82\pm0.11$, where the error bar considers uncertainties in the line list, in stellar parameters and in the definition of the continuum position around the Li line. With its effective temperature and lithium abundance, the position of our target in the $\log n$({\rm Li})$-T_{\rm eff}$ appears to be in between that of the M35 cluster ($\sim$ 200 Myr) and that of the Hyades cluster ($\sim$ 650 Myr, see, e.g., \citealt{SestitoRandich2005}, \citealt{Cummingsetal2017}).

\subsection{Chromospheric activity}
\label{sec:activity}

The lower chromosphere Ca II H\&K emission was measured on HARPS-N spectra using YABI (see Sect. \ref{sec:obs-harpsn}). The average value of the S-index, calibrated in the Mt. Wilson scale \citep{1995ApJ...438..269B}, is 0.325, corresponding to $\log R'_{\rm HK}$ = 
-4.43 $\pm$ 0.02 (arithmetic average and standard deviation).

For the $\log R'_{\rm HK}$ determination, we adopted $(B-V)_0$ = 0.58 mag, which was derived from $T_{\rm eff}$ through \cite{casagrande2006} calibration, as the observed values from Tycho2 and APASS suffer from large uncertainties. Our determination of $\log R'_{\rm HK}$ implies a stellar age of 370 Myr using the activity-age calibration and an expected rotation period of 4.25 d (corresponding to a gyrochronology age of 260 Myr), through \cite{mamajek2008} relationships. The difference with the observed photometric period (see Sect. \ref{sec:rotation}) is not significant considering observational errors and intrinsic scatter of magnetic activity (see e.g. Fig. 7 of \citealt{mamajek2008}). It is possible that the star is caught by chance at a level of chromospheric activity higher than the average one or that the activity is somewhat enhanced by the presence of planetary companions.

The upper chromosphere H$\alpha$ line, extracted following Sect. \ref{sec:obs-harpsn}, has a value of 
0.151 $\pm$ 0.002 (arithmetic average and standard deviation).

The tidal evolution of the stellar rotation has a timescale much longer than the age of the system, even considering an extremely strong tidal interaction with a stellar modified tidal quality factor $Q^{\prime}_{\star}=10^{5}$ \citep[e.g.,][]{2002ApJ...573..829M}. Therefore, we do not expect that tides can affect the estimate of the stellar age based on gyrochronology.

\subsection{Rotation period}
\label{sec:rotation}

The extensive datasets we collected allowed us to obtain several estimates of the rotation period of the star, a key parameter for age determination.

As mentioned in Sect. \ref{sec:tess} and \ref{sec:asas}, from the analysis of \textit{TESS} and ASAS-SN light curves we obtained photometric periodicities of 7.18$\pm$0.21 days and 7.34$\pm$0.01 days, respectively. The larger error of the determination based on \textit{TESS} data is due to the short baseline of the monitoring (single sector).

We also exploited the HARPS-N time series of RVs and several spectroscopic indicators (Sect. \ref{sec:obs-harpsn}). Figure \ref{fig:gls} displays the GLS periodograms, computed for the frequency range 0.0001 - 0.5 days$^{-1}$, or 2 - 10000 days. We removed a linear trend from $\log R'_{\rm HK}$ and H$\alpha$ time series before running the GLS. 
The RVs, BIS, and FWHM periodograms show a clear peak (orange line, normalised GLS power $>$ 0.2) close to the first harmonic of the stellar rotation period detected using the ASAS-SN photometry (7.34 days, dotted green lines). As the period is compatible with the modulation observed in the photometry and consistently recovered in both the radial velocity and the activity indexes, we associated this signal to the stellar activity. This result could be a sign that RV variations are dominated by dark spots rather than the quenching of convective motions in the magnetised regions of the photosphere, as explained in the study by \cite{2010A&A...520A..53L}. Moreover, the periodograms of H$\alpha$ and $W_{\rm CCF}$ show peaks consistent with the photometric period. On the other hand, the second-highest peak in the RVs periodogram reveals the signal of the validated gas giant planet. We also add that the detailed modelling of the RV time series with Gaussian processing regression (Sect. \ref{sec:rv_analysis}) yields a value of 7.37$\pm$0.03 days for the rotation period.

In short, the available photometric and spectroscopic time series consistently indicate a periodicity of 7.34$\pm$0.05 days, which we interpret as the rotation period of the star. The shorter periodicity seen in some spectroscopic indicators is fully compatible with being the first harmonic of the true periodicity, as observed in several other cases (see, e.g., TOI-1807 \citealt{2022A&A...664A.163N}). Conversely, we think it is unlikely that the true period is the shorter one (3.66 d), with this periodicity arising from the presence of active regions of comparable areas on opposite stellar hemispheres, especially considering the long-time baseline of the ASAS-SN time series.



\subsection{Kinematics}
\label{sec:kinematics}

Kinematic space velocities were derived adopting {\it Gaia} kinematic parameters
and using the formalism by \citet{uvw}.
The resulting $U$, $V$, and $W$ (Table \ref{tab:star_param}) are at the boundary
of the kinematic space of young stars \citep[age younger than about 500 Myr, ][]{montes2001,2010A&A...521A..12M} in agreement with the other age diagnostics.

TOI-5398 is not a member of known young moving groups and a dedicated search of comoving objects within a few degrees in the \textit{Gaia} DR3 catalogue does not yield convincing candidates.

\subsection{Age, mass and radius}
\label{sec:age}

The indirect methods discussed above point towards an age of a few hundred Myr, similar to the Hyades and Praesepe open clusters, which have also super-solar metallicity (Hyades with +0.15 dex, \citealt{Cummingsetal2017}, and Praesepe with +0.21 dex, \citealt{2020A&A...633A..38D}). In the case of a G dwarf star with a few hundred million years, the most robust age indicator is the rotation period. An age of 680 Myr is obtained with the \cite{mamajek2008} calibration for our adopted rotation period. Using different photometric colours and calibrations \citep[e.g., G-K][]{messina2022} yields similar results. The Lithium abundance is intermediate between members of Hyades ($\sim$ 650 Myr) and M35 ($\sim$ 200 Myr) of similar colours, pointing to a somewhat younger age. The level of chromospheric activity also suggests a younger age than the gyrochronology, but the discrepancy of both lithium and $\log R^{'}_{\rm HK}$ with the expectations for an age similar to that of the Hyades cluster is marginal. Finally, kinematic parameters are fully consistent with an age of a few hundred Myr and the lack of comoving objects prevents a more precise age estimate. We further add that isochrone fitting, performed using the \texttt{param}\footnote{ \url{http://stev.oapd.inaf.it/cgi-bin/param_1.3}} tool \citep{param} does not add further relevant information (nominal age 1.6$\pm$1.6 Gyr). We then adopt a system age of 650$\pm$150 Myr from the indirect indicators.

Through the \texttt{param} tool and by imposing the age range allowed by indirect methods to avoid the inclusion of solutions not compatible with the above results \citep{desidera2015}, we derived the stellar mass and radius. We obtained in this way a stellar mass of 1.146$\pm$0.013  $M_{\odot}$ and a stellar radius $R=$ 1.051$\pm$0.013 $R_{\odot}$, where the uncertainties are those provided by the \texttt{param} interface and do not include possible systematics of the adopted stellar models. 

From the combination of $R_{\star}$, $P_{\rm rot}$ and $v \sin{i_{\star}}$, we infer a system orientation fully compatible with edge-on. Indeed, for the nominal parameters, $\sin i_{\star}$ is just below unity, and taking error bars into account we estimate $i_{\star} \ge 69$ deg.

The stellar parameters outlined in this and the preceding subsections serve as the reference for this study. We present them in Table \ref{tab:star_param} for reference.

\begin{table}[!htb]
   \caption[]{Stellar properties of TOI-5398}
     \label{tab:star_param}
     \small
     \centering
       \begin{tabular}{lcc}
         \hline
         \noalign{\smallskip}
         Parameter   &  \object{TOI-5398} & Ref  \\
         \noalign{\smallskip}
         \hline
         \noalign{\smallskip}
$\alpha$ (J2000)          &   10 47 31.09	& {\it Gaia} DR3    \\
$\delta$ (J2000)          &   +36 19 45.86  & {\it Gaia} DR3  \\
$\mu_{\alpha}$ (mas yr$^{-1}$)  &    1.360$\pm$0.014  & {\it Gaia} DR3  \\
$\mu_{\delta}$ (mas yr$^{-1}$)  &    7.003$\pm$0.012  & {\it Gaia} DR3  \\
RV     (km s$^{-1}$)            &    -9.95$\pm$0.30   & {\it Gaia} DR3   \\
$\pi$  (mas)             &    7.6190$\pm$0.0143 & {\it Gaia} DR3  \\
$U$   (km s$^{-1}$)             &  	4.07$\pm$0.14        & This paper (Sect. \ref{sec:kinematics}) \\
$V$   (km s$^{-1}$)             &      4.90$\pm$0.02        & This paper (Sect. \ref{sec:kinematics}) \\
$W$   (km s$^{-1}$)             &     -8.83$\pm$0.27       & This paper (Sect. \ref{sec:kinematics})  \\
\noalign{\medskip}
V (mag)                  &    10.06$\pm$0.03     &   Tycho-2 \citep{2000AA...355L..27H}   \\
$B-V$ (mag)                &    0.58  & This paper (Sect. \ref{sec:activity})  \\
$G$ (mag)                  &    9.9875$\pm$0.0004  & {\it Gaia} DR3  \\
$G_{BP}-G_{RP}$ (mag)              &         0.7582         & {\it Gaia} DR3  \\
$J_{\rm 2MASS}$ (mag)    &   9.026$\pm$0.021  & 2MASS  \\
$H_{\rm 2MASS}$ (mag)    &   8.772$\pm$0.023  & 2MASS  \\
$K_{\rm 2MASS}$ (mag)    &   8.713$\pm$0.018  & 2MASS  \\
\noalign{\medskip}
$T_{\rm eff}$ (K)        &  6000$\pm$75       & This paper (spec) (Sect. \ref{sec:atmo_param}) \\  
$\log g$                 &  4.44$\pm$0.10     & This paper (Sect. \ref{sec:atmo_param}) \\ 
${\rm [Fe/H]}$ (dex)     &  +0.09$\pm$0.06     & This paper (Sect. \ref{sec:atmo_param}) \\ 
$E(B-V)$ (mag)                   &  $\leq$ 0.024\tablefootmark{a}     & PIC \citep{pic} \\ 
\noalign{\medskip}
$S_{\rm MW}$             &  0.325 $\pm$0.008   & This paper (Sect. \ref{sec:activity})\\
$\log R^{'}_{\rm HK}$    &     -4.43$\pm$0.02 &  This paper (Sect. \ref{sec:activity}) \\  
$v\sin{i_{\star}}$ (km s$^{-1}$)      &   7.5$\pm$0.6   & This paper (Sect. \ref{sec:vsini}) \\  
$P_{\rm rot}$ (d)  &   7.34$\pm$0.05  &   This paper (Sect. \ref{sec:rotation}) \\
A(Li)                    &  2.82$\pm$0.11  &  This paper  (Sect. \ref{sec:lithium}) \\
\noalign{\medskip}
Mass ($M_{\odot}$)       &    1.146$\pm$0.013   & This paper (Sect. \ref{sec:age}) \\
Radius ($R_{\odot}$)     &    1.051$\pm$0.013   & This paper (Sect. \ref{sec:age}) \\
Age  (Myr)               &    650$\pm$150 & This paper (Sect. \ref{sec:age}) \\
$i_{\star}$ (deg)     &    $\ge$69    & This paper (Sect. \ref{sec:age}) \\
         \noalign{\smallskip}
         \hline
      \end{tabular}

\tablefoot{\tablefoottext{a}{84th percentile.}}
\end{table}

\subsection{Spectral Energy Distribution}
\label{sec:sed}

\begin{figure}
    \centering
    \includegraphics[width=\linewidth,trim=80 70 60 60,clip]{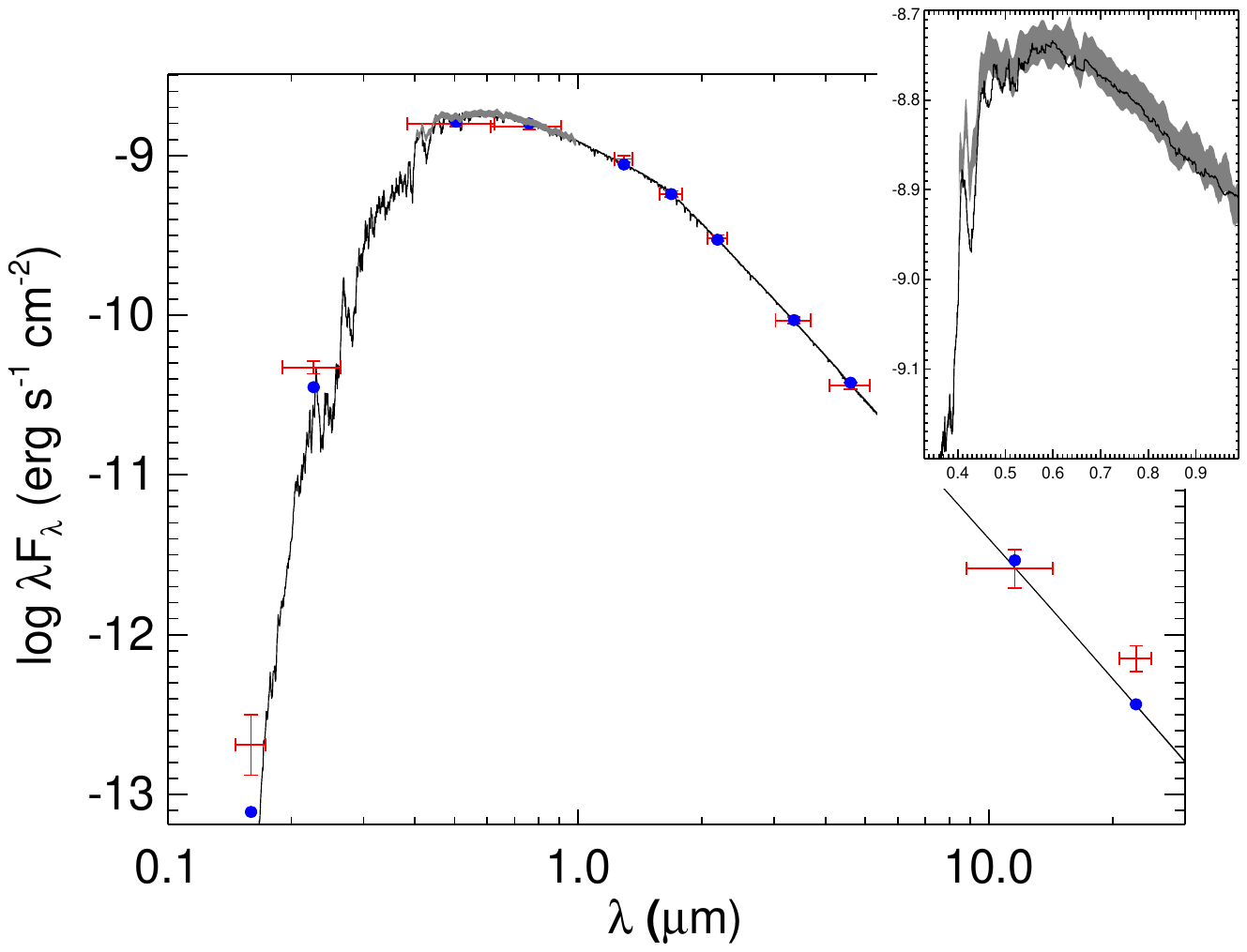}
    \caption{Spectral energy distribution of TOI-5398. Red symbols represent the observed photometric measurements, where the horizontal bars represent the effective width of the passband. Blue symbols are the model fluxes from the best-fit PHOENIX atmosphere model (black). The {\it Gaia\/} spectrum is overlaid as a grey swathe and shown in closer detail in the inset plot.}
    \label{fig:sed}
\end{figure}

As an independent determination of the basic stellar parameters, we performed an analysis of the broadband spectral energy distribution (SED) of the star together with the {\it Gaia\/} DR3 parallax \citep[see, e.g.,][]{StassunTorres:2021}, in order to determine an empirical measurement of the stellar radius, following the procedures described in \citet{Stassun:2016,Stassun:2017,Stassun:2018}. We pulled the $JHK_S$ magnitudes from {\it 2MASS}, the W1--W4 magnitudes from {\it WISE} \citep{2010AJ....140.1868W}, and the $G_{\rm BP}\,G_{\rm RP}$ magnitudes from {\it Gaia} \citep{2016A&A...595A...1G}, and the FUV and NUV magnitudes from {\it GALEX} \citep{2005ApJ...619L...1M}. Together, the available photometry spans the full stellar SED over the wavelength range 0.2--22~$\mu$m (see Fig.~\ref{fig:sed}). 

We performed a fit using PHOENIX stellar atmosphere models \citep{2013A&A...553A...6H}, with the free parameters being the effective temperature ($T_{\rm eff}$) and metallicity ([Fe/H]), as well as the extinction $A_V$, which we limited to maximum line-of-sight value from the Galactic dust maps of \citet{Schlegel:1998}. The resulting fit (Fig.~\ref{fig:sed}) has a best-fit $A_V = 0.05 \pm 0.02$ mag, $T_{\rm eff} = 6025 \pm 100$~K, [Fe/H] = $0.0 \pm 0.5$, with a reduced $\chi^2$ of 1.3 (excluding the FUV and W4 fluxes, which suggest modest excesses in both the UV and mid-IR). Integrating the (unreddened) model SED gives the bolometric flux at Earth, $F_{\rm bol} = 2.470 \pm 0.029 \times 10^{-9}$ erg~s$^{-1}$~cm$^{-2}$. Taking the $F_{\rm bol}$ and $T_{\rm eff}$ together with the {\it Gaia\/} parallax, gives the stellar radius, $R_\star = 1.059 \pm 0.036$~R$_\odot$. In addition, we can estimate the stellar mass from the empirical relations of \citet{Torres:2010}, obtaining $M_\star = 1.12 \pm 0.07$~M$_\odot$. 

Finally, we may estimate the stellar rotation period from the above radius together with the spectroscopically measured $v\sin{i_{\star}}$, giving $P_{\rm rot}/\sin i_{\star} = 7.1 \pm 0.6$~d. 

These results are fully compatible with those presented in Sects. \ref{sec:atmo_param} and \ref{sec:age}.

\section{Analysis}
\label{sec:analysis}
\subsection{Planets detection and vetting tests}

Two candidate exoplanets orbiting TOI-5398 were identified in Sector 48 light curves in both SPOC \citep{2016SPIE.9913E..3EJ} and QLP \citep{2020RNAAS...4..204H} pipeline: one giant and one sub-Neptune ($P$ $\sim$ 10.59 d and $P$ $\sim$ 4.77 d, respectively). The SPOC performed a transit search with an adaptive, noise-compensating matched filter (\citealt{2002ApJ...575..493J,2010SPIE.7740E..0DJ,2020ksci.rept....9J}), producing Threshold Crossing Events (TCEs) for which an initial limb-darkened transit model was fitted \citep{2019PASP..131b4506L} and diagnostic tests were conducted to help assess the planetary nature of the signal \citep{2018PASP..130f4502T}. The QLP performed its transit search with the Box Least Squares Algorithm \citep{2002A&A...391..369K}. The two transit signatures passed all \textit{TESS} data validation diagnostic tests, and the \textit{TESS} Science Office issued alerts for TOI-5398.01 (10.59\,days) and TOI-5398.02 (4.77\,days) on 24 March 2022. The SPOC difference image centroid offsets \citep{2018PASP..130f4502T} localised the transit source for TOI 5398.01 within 0.41 $\pm$ 2.55 arcsec and for TOI 5398.02 within 2.69 $\pm$ 2.67 arcsec; all TIC v8 \citep{2019AJ....158..138S} objects other than TOI-5398 were excluded as the source of each transit signature. 

\cite{2022MNRAS.516.4432M} thoroughly validated the planetary nature of the giant, which they label TOI-5398 b, by ruling out any false positive (FP) scenarios capable of mimicking the observed transit signal. In fact, mainly due to the low spatial resolution of \textit{TESS} cameras ($\approx$ 21 arcsec/pixel), some objects initially identified as sub-stellar candidates might be FPs. As a result, vetting and validation tests are critical. Therefore, in order to better understand also the nature of the sub-Neptune candidate, TOI-5398.02, and to exclude FP scenarios, we followed the procedure adopted in \cite{2022MNRAS.516.4432M}. The latter approach, further described in Appendix \ref{app:stat}, considers the major concerns reported in \cite{2023RNAAS...7..107M} and ensures reliable results when using VESPA \citep{Morton_2012,2015ascl.soft03010M}. This procedure allowed us to statistically validate the sub-Neptune exoplanet and label it as TOI-5398 c.

\subsection{Photometry time-series analysis}
\label{sec:phot_model}
To characterise the properties of TOI-5398 b and c, we investigated all the ground-based photometry simultaneously with the two transits of TOI-5398 b and the four transits of TOI-5398 c observed by \textit{TESS} in a Bayesian framework using \texttt{PyORBIT}\footnote{\url{https://github.com/LucaMalavolta/PyORBIT}.} (\citealt{2016A&A...588A.118M}, \citealt{2018AJ....155..107M}), a Python package for modelling planetary transits and radial velocities while simultaneously accounting for stellar activity effects. The ground-based observations rule out false positive scenarios caused by BEBs and further confirm the transits of two planetary companions orbiting TOI-5398.

In the present case, we took our \textit{TESS} corrected light curve (see Sect. \ref{sec:tess}) and carefully considered the influence of stellar contamination from neighbour stars. We verified the stellar dilution by measuring a dilution factor, which defines the total flux from contaminants that fall into the photometric aperture divided by the flux contribution of the target star. Following Sect. 2.2.2 of \cite{2022MNRAS.516.4432M}, we determined the dilution factor (and its associated error) by calculating the contribution of the flux that falls into the \textit{TESS} aperture for each star. We determined its value to be 0.00735 $\pm$ 0.00005, and we imposed it as a Gaussian prior in the modelling. Then, we selected each space-based transit event from the corrected light curve -- and an out-of-transit part as long as the corresponding transit duration (both before the ingress and after the egress) -- and created a mask that flags each transit and cuts the corresponding portions of the light curve. The transits of the two planets do not overlap. 

We simultaneously modelled each transit (ground- and space-based) using the code \texttt{BATMAN} \citep{2015PASP..127.1161K}, fitting the following parameters: the central time of transit ($T_0$), the planetary-to-star radius ratio ($R_p/R_\star$), the impact parameter, $b$, the stellar density ($\rho_{\star}$, in solar units), the quadratic limb-darkening (LD) coefficients $u_1$ and $u_2$ adopting the LD parametrization ($q_1$ and $q_2$) introduced by \cite{2013MNRAS.435.2152K}, a second-order polynomial trend to take into account the local stellar variability (with c$_0$ as the intercept, $c_1$ as the linear coefficient, and $c_2$ as the quadratic coefficient), and a jitter term to be added in quadrature to the errors of the photometry to account for any effects that were not included in our model (e.g., short-term stellar activity) or any underestimation of the error bars. We applied an airmass detrending technique to each ground-based light curve, and we estimated $u_1$ and $u_2$ using \texttt{PyLDTk}\footnote{\url{https://github.com/hpparvi/ldtk}} \citep{2013A&A...553A...6H, 2015MNRAS.453.3821P} and applying the specific filters used for the observations. We imposed a Gaussian prior on the stellar density, whereas we imposed uniform priors on the period and $T_0$. We then imposed a Gaussian prior on the eccentricities following \cite{2019AJ....157...61V}. 

\begin{figure}
   \centering
   \includegraphics[width=\hsize]{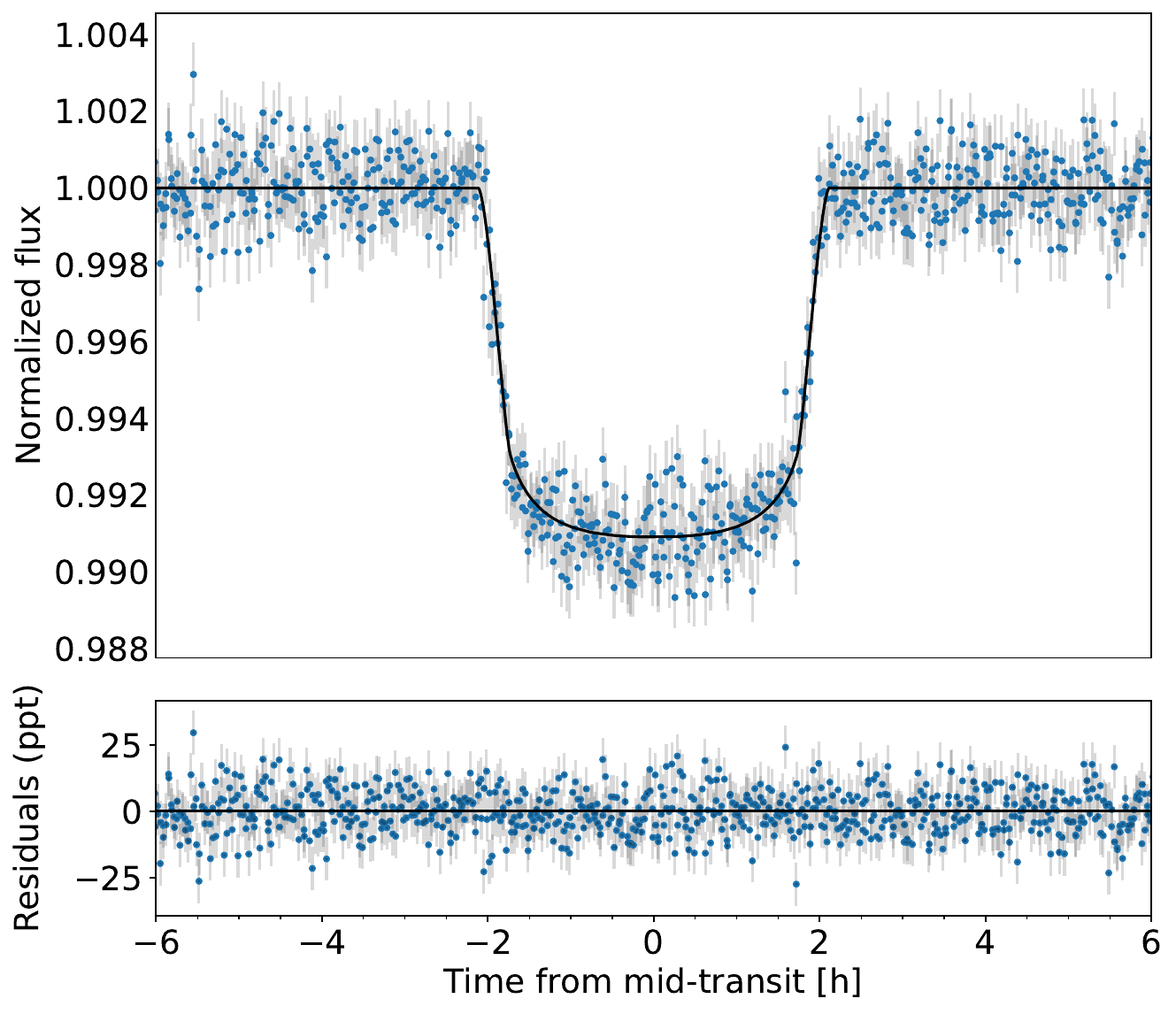}
   \caption{Photometric modelling of TOI-5398 b planetary signal. In the top panel, we display the \textit{TESS} phase-folded transits after normalisation along with the transit model (black line). In the panel below, we show the residuals. }
   \label{fig:pl_b}
\end{figure}

\begin{figure}
   \centering
   \includegraphics[width=\hsize]{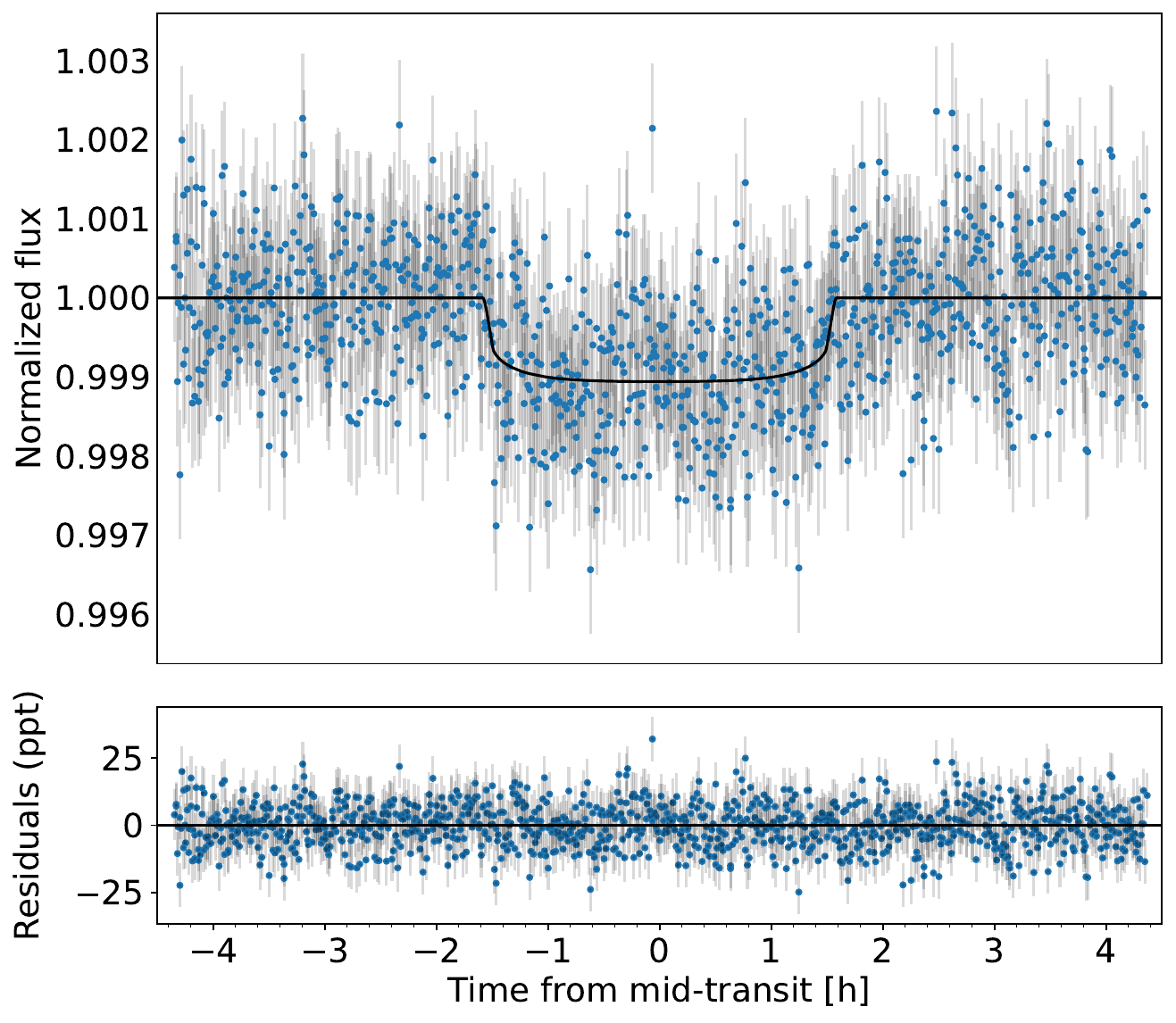}
   \caption{As in Fig. \ref{fig:pl_b}, but for TOI-5398 c.}
   \label{fig:pl_c}
\end{figure}

\begin{figure*}
   \centering
   \includegraphics[width=\hsize]{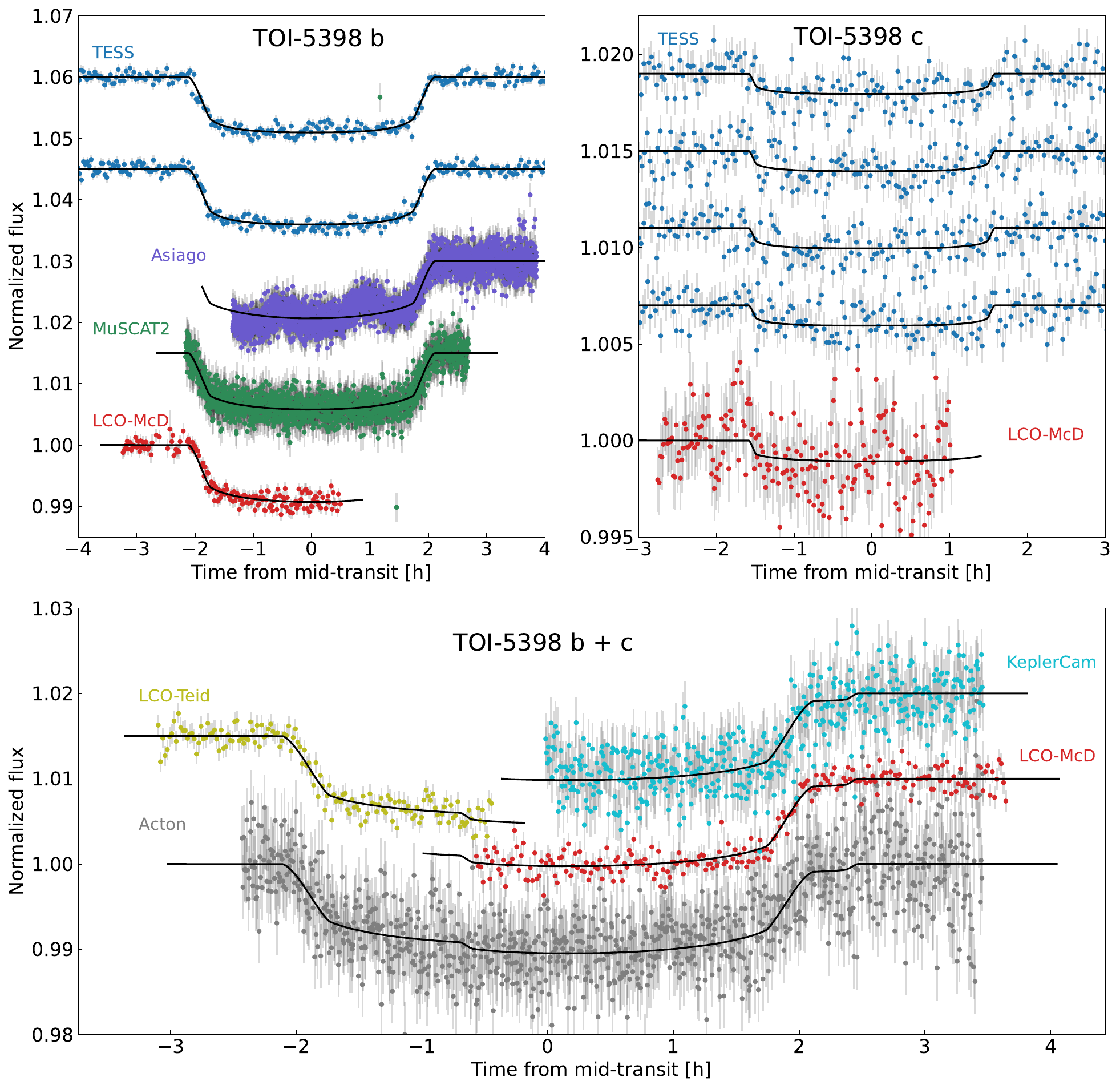}
   \caption{Ground-based photometric data, simultaneously modelled with \textit{TESS} transits. Different datasets are represented by distinct colours. \textit{Top}: Individual transits of TOI-5398 b on the left and TOI-5398 c on the right. \textit{Bottom}: Simultaneous transit of TOI-5398 b and c, observed during the night of 21 April 2022.}
   \label{fig:complete_lc}
\end{figure*}

We performed a global optimisation of the parameters by executing a differential evolution algorithm (\citealt{Storn1997}, \texttt{PyDE}\footnote{\url{https://github.com/hpparvi/PyDE}}) and performing a Bayesian analysis of each selected light curve around each transit. The latter was achieved using the affine-invariant ensemble sampler (\citealt{2010CAMCS...5...65G}) for Markov Chain Monte Carlo (MCMC) implemented within the package \textsc{emcee} (\citealt{2013PASP..125..306F}). We used $4n_{dim}$ walkers (with $n_{dim}$ being the dimensionality of the model) for 50\,000 generations with \texttt{PyDE} and then with 100 000 steps with \textsc{emcee} – where we applied a thinning factor of 200 to reduce the effect of the chain auto-correlation. We discarded the first 25\,000 steps (burn-in) after checking the convergence of the chains with the Gelman–Rubin (GR) statistics (Gelman \& Rubin 1992, threshold value \^{R} = 1.01). Unless specifically mentioned otherwise, the same sampling configuration and process have been used throughout all occurrences of \texttt{PyDE} and \textsc{emcee}. Figures \ref{fig:pl_b}, \ref{fig:pl_c}, \ref{fig:complete_lc}, and Table \ref{table:model_pl} present the results of the modelling.

\begin{table*}
\caption{Priors and outcomes of the model of planet b and c from the analysis of the photometric time series. In Appendix \ref{app:table_extra}, we show the second part of this Table, moved for improved readability of the main text.}             
\label{table:model_pl}      
\centering          
\begin{tabular}{l c c c}     
\hline\hline     

 \multicolumn{4}{c}{Photometric time-series fit} \rule{0pt}{2ex} \rule[-0.9ex]{0pt}{0pt} \\ 
\hline    
Parameter & Unit & Prior & Value \rule{0pt}{2.2ex} \rule[-0.9ex]{0pt}{0pt}\\ 
\hline    
   Stellar density ($\rho_{\star}$) & $\rho_{\sun}$ & $\mathcal{U}$(0.75, 10.25) & 0.971$^{+0.042}_{-0.049}$ \rule{0pt}{2.2ex} \rule[-1.2ex]{0pt}{0pt}\\
   
\hline
\multicolumn{4}{c}{Planet b} \rule{0pt}{2.2ex} \rule[-0.9ex]{0pt}{0pt}\\ 
\hline    
Parameter & Unit & Prior & Value \rule{0pt}{2.2ex} \rule[-0.9ex]{0pt}{0pt}\\ 
\hline 
   Orbital period ($P_{\rm b}$) & days & $\mathcal{U}$(10.57, 10.61) & 10.590547$^{+0.000012}_{-0.000011}$ \rule{0pt}{2.2ex} \rule[-0.9ex]{0pt}{0pt}\\
   Central time of the first transit ($T_{\rm 0,b}$) & BTJD & $\mathcal{U}$(2616.4, 2616.6) & 2616.49232$^{+0.00022}_{-0.00021}$ \rule{0pt}{2.2ex} \rule[-0.9ex]{0pt}{0pt}\\
   Semi-major axis to stellar radius ratio ($a_{\rm b}/R_{\star}$) &  & ... & 20.0$\pm$0.6 \rule{0pt}{2.2ex} \rule[-0.9ex]{0pt}{0pt}\\
   Orbital semi-major axis ($a_{\rm b}$) & au & ... & 0.098$\pm$0.005 \rule{0pt}{2.2ex} \rule[-0.9ex]{0pt}{0pt}\\
   Orbital inclination ($i$) & deg & ... & 89.21$^{+0.31}_{-0.21}$ \rule{0pt}{2.2ex} \rule[-0.9ex]{0pt}{0pt}\\
   Orbital eccentricity ($e_{\rm b}$) &  & $\mathcal{N}$(0, 0.098) & $\leqslant$ 0.094\tablefootmark{a} \rule{0pt}{2.2ex} \rule[-0.9ex]{0pt}{0pt}\\
   Impact parameter ($b$) &  & $\mathcal{U}$(0, 1) & 0.272$^{+0.069}_{-0.110}$ \rule{0pt}{2.2ex} \rule[-0.9ex]{0pt}{0pt}\\
   Transit duration ($T_{14}$) & days & ... & 0.1774$^{+0.0062}_{-0.0043}$ \rule{0pt}{2.2ex} \rule[-0.9ex]{0pt}{0pt}\\
   Planetary radius ($R_{\rm b}$) & $R_{\oplus}$ & ... & 10.30$\pm$0.40 \rule{0pt}{2.2ex} \rule[-0.9ex]{0pt}{0pt}\\
\hline
\multicolumn{4}{c}{Planet c} \rule{0pt}{2.2ex} \rule[-0.9ex]{0pt}{0pt}\\ 
\hline    
Parameter & Unit & Prior & Value \rule{0pt}{2.2ex} \rule[-0.9ex]{0pt}{0pt}\\ 
\hline 
   Orbital period ($P_{\rm c}$) & days & $\mathcal{U}$(4.770, 4.776) & 4.77271$^{+0.00016}_{-0.00014}$ \rule{0pt}{2.2ex} \rule[-0.9ex]{0pt}{0pt}\\
   Central time of the first transit ($T_{\rm 0,c}$) & BTJD & $\mathcal{U}$(2628.5, 2628.7) & 2628.61781$^{+0.00090}_{-0.00086}$ \rule{0pt}{2.2ex} \rule[-0.9ex]{0pt}{0pt}\\
   Semi-major axis to stellar radius ratio ($a_{\rm c}/R_{\star}$) &  & ... & 11.8$\pm$0.4 \rule{0pt}{2.2ex} \rule[-0.9ex]{0pt}{0pt}\\
   Orbital semi-major axis ($a_{\rm c}$) & au & ... & 0.057$\pm$0.003 \rule{0pt}{2.2ex} \rule[-0.9ex]{0pt}{0pt}\\
   Orbital inclination ($i$) & deg & ... & $\geqslant$ 88.40\tablefootmark{b} \rule{0pt}{2.2ex} \rule[-0.9ex]{0pt}{0pt}\\
   Orbital eccentricity ($e_{\rm c}$) &  & $\mathcal{N}$(0, 0.098) & $\leqslant$ 0.117\tablefootmark{a} \rule{0pt}{2.2ex} \rule[-0.9ex]{0pt}{0pt}\\
   Impact parameter ($b$) &  & $\mathcal{U}$(0, 1) & $\leqslant$ 0.34\tablefootmark{a} \rule{0pt}{2.2ex} \rule[-0.9ex]{0pt}{0pt}\\
   Transit duration ($T_{14}$) & days & ... & 0.1303$^{+0.0042}_{-0.0053}$ \rule{0pt}{2.2ex} \rule[-0.9ex]{0pt}{0pt}\\
   Planetary radius ($R_{\rm c}$) & $R_{\oplus}$ & ... & 3.52$\pm$0.19 \rule{0pt}{2.2ex} \rule[-0.9ex]{0pt}{0pt}\\
\hline
\end{tabular}
\tablefoot{\tablefoottext{a}{84th percentile.}\tablefoottext{b}{16th percentile.} }
\end{table*}

\subsection{RV time-series analysis}
\label{sec:rv_analysis}

Using \texttt{PyORBIT}, we investigated the RV time series data in a Bayesian framework. We tried various approaches to model the stellar activity through the use of Gaussian processes (GP, \citealt{rasmussen2006gaussian}, \citealt{2014MNRAS.443.2517H}). We experimented with a number of data set combinations to constrain the GP hyper-parameters and consider various planetary system architectures (one, two, or more planets). Here we outline the three most notable cases. 

Due to the high computational cost of the Gaussian Processes, we only modelled the spectroscopic time series in each test. However, we included the inclination from the photometry model to determine the true masses.

\subsubsection{Case 1: \textit{Two-planet system and activity modelling trained on spectroscopy -- uni-dimensional GP}}
In the first case, we tested a multi-planet system model formed by TOI-5398 b and TOI-5398 c. 

\texttt{PyORBIT} simultaneously modelled the stellar activity and the signals of the two planets in the RV series. In the model, we included the inclination measured from photometry and the stellar mass $M_{\star}$ as derived in Sect. \ref{sec:host-star} to determine the true masses of the planets. We then imposed a Gaussian prior on the eccentricities following \cite{2019AJ....157...61V} and included the RV semi-amplitudes $K_{\rm b}$ and $K_{\rm c}$ in the RV modelling. Moreover, we used Gaussian priors on both orbital periods ($P_{\rm b}$, $P_{\rm c}$) and central time of the first transits ($T_{0,{\rm b}}$, $T_{0,{\rm c}}$) by considering the parameters outlined in Sect. \ref{sec:phot_model}. 

We modelled the stellar activity in the RV, BIS, and $\log{R^{'}_{HK}}$ series simultaneously, through a Gaussian process (GP) regression. We used a quasi-periodic kernel as defined by \cite{2015ApJ...808..127G}. As part of this modelling, we set the stellar rotation period $P_{\rm rot}$ (Gaussian prior, as defined in Sec. \ref{sec:asas}), the characteristics decay time scale $P_{\rm dec}$, and the coherence scale $w$. In accordance with \cite{2013PASP..125...83E}, we fitted the periods and semi-amplitudes of the RV signals in the linear space, and we determined the eccentricity $e$ and the argument of periastron $\omega$ by fitting $\sqrt{e} \cos{\omega}$ and $\sqrt{e} \sin{\omega}$. We performed a global optimisation of the parameters by running \texttt{PyDE} and performing a Bayesian analysis of the planetary signals and activity in the RV time series using \textsc{emcee}. The results of this analysis will be discussed after presenting each respective Case and in Table \ref{table:diff}.

\subsubsection{Case 2: \textit{Two-planet system and activity modelling trained on spectroscopy -- multidimensional GP}}
In the present case, we modelled the stellar activity of TOI-5398 using the multidimensional GP framework developed by \cite{2015MNRAS.452.2269R} and re-implemented in \texttt{PyORBIT} in accordance with the prescriptions in the paper (see also \citealt{barragan2022pyaneti}). Again, we relied on the quasi-periodic kernel and its derivatives. We modelled RV, BIS, and $\log{R^{'}_{HK}}$ spectroscopic time series. The three-dimensional GP model is the following:
\begin{align}
\begin{split}
\label{eq:model}
    & \Delta RV = V_{\rm c}\,G(t) + V_{\rm r}\,\dot{G}(t),\\
    & \log R'_{\rm HK} = L_{\rm c}\,G(t),\\
    & BIS = B_{\rm c}\,G(t) + B_{\rm r}\,\dot{G}(t),
\end{split}
\end{align} where $G(t)$ is the GP and $\dot{G}(t)$ its time derivative. The constants denoted by subscripts $r$ and $c$ represent free parameters linking the individual time series to $G(t)$ and $\dot{G}(t)$ \citep{2023MNRAS.522.3458B}.

To perform the modelling with \texttt{PyORBIT}, we assigned the same priors described in Case 1 to planets and stellar parameters. Then, we run a global optimisation of the parameters with \texttt{PyDE} and a Bayesian analysis of the planetary signals and activity in the RV time series with \textsc{emcee}. The results of this analysis will be discussed after presenting each respective Case and in Table \ref{table:diff}.



\subsubsection{Case 3: \textit{Extended list of activity indexes and search for a third planet}}
To better disentangle planetary and stellar signals, we extended the list of activity indexes that \texttt{PyORBIT} simultaneously models with the two keplerian signals (described in Case 2). In particular, we included the chromospheric activity indicator H$\alpha$ \citep{2011A&A...534A..30G} as well as the two CCF asymmetry diagnostics FWHM and equivalent width $W_{\rm CCF}$. 

On the one hand, the H$\alpha$ line complements the lower chromosphere indicator $\log{R^{'}_{\rm HK}}$ by providing information about the conditions in the upper chromosphere of a star \citep{2011A&A...534A..30G}. In fact, H$\alpha$ and Ca H \& K are emitted from different depths -- and formed at different temperatures -- in the chromosphere \citep{2013ApJ...764....3R,2014A&A...566A..66G}. Therefore, it is helpful to study the H$\alpha$ and $\log{R^{'}_{\rm HK}}$ indices simultaneously \citep{2011A&A...534A..30G,2014A&A...566A..66G} to learn more about the presence of distinct activity-related features and disentangle their signals from keplerian ones in the RV time series. On the other hand, according to three years of RV monitoring of the Sun \citep{2019MNRAS.487.1082C}, the line-shape parameters of the CCF appear to respond to different components of the active regions. Moreover, they help to track global temperature changes in the photosphere (see also \citealt{2017MNRAS.469.3965M}). 

\begin{figure}
   \centering
   \includegraphics[width=\hsize]{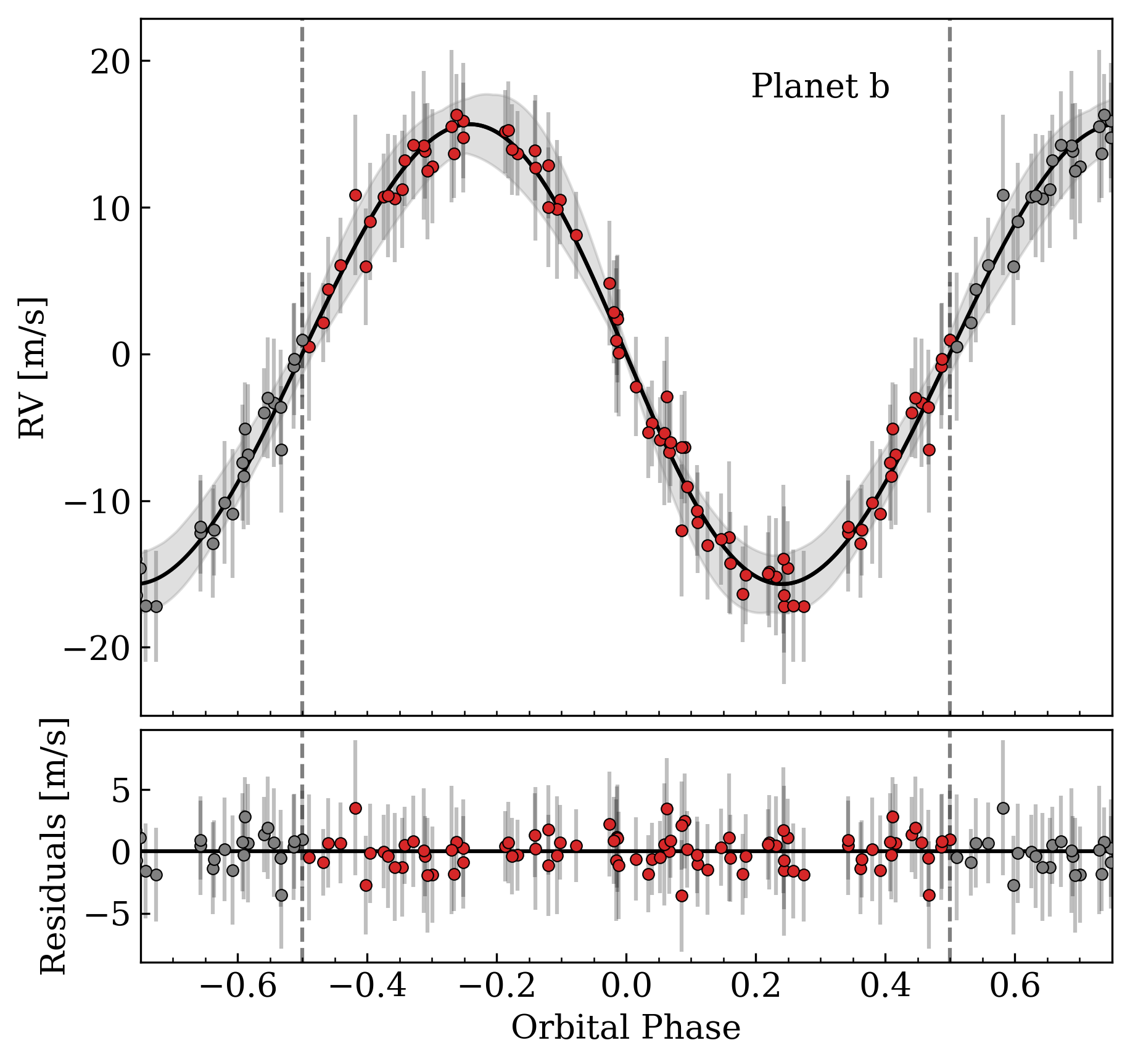}
   \caption{Phase-folded RV fit of TOI-5398 b planetary signal. The reported error bars include the jitter term, added in quadrature. The shaded area represents the RV model's $\pm$1$\sigma$ uncertainties. The bottom panel displays the residuals of the fit.}
   \label{fig:rvb_best}
\end{figure}

\begin{figure}
   \centering
   \includegraphics[width=\hsize]{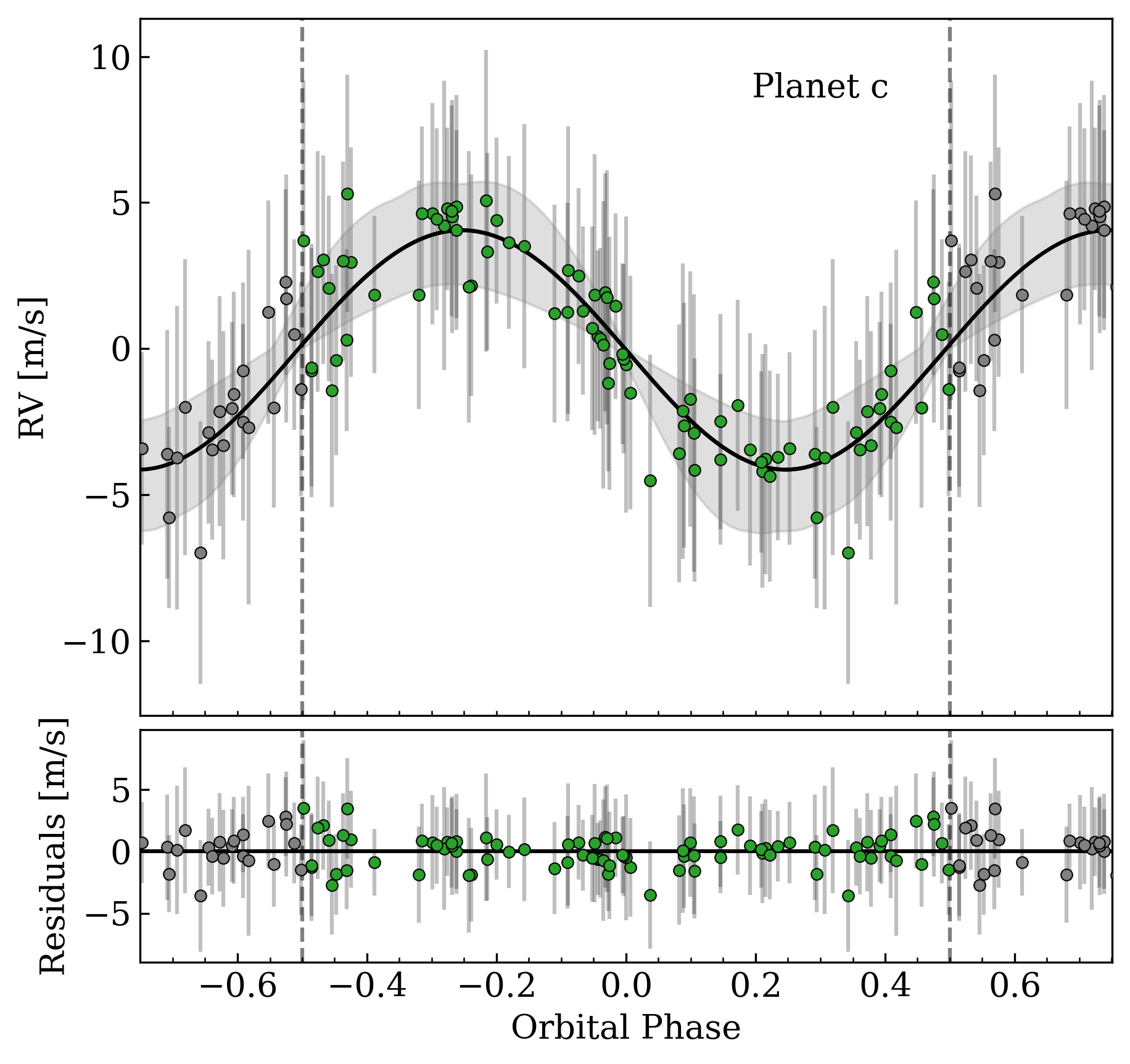}
   \caption{As in Fig. \ref{fig:rvb_best}, but for planet c.}
   \label{fig:rvc_best}
\end{figure}

For the reasons listed, we decided to include in the modelling both the chromospheric indicators H$\alpha$ and $\log{R^{'}_{\rm HK}}$, as well as the three CCF asymmetry diagnostics BIS, FWHM, and $W_{\rm CCF}$. We followed the multidimensional GP formalism introduced by \cite{2015MNRAS.452.2269R} to examine the RV and BIS time series, using the first derivative of the GP. Conversely, we did not use the first derivative for the remaining four time series, as suggested also in \cite{2023MNRAS.522.3458B}. The six-dimensional GP model is an extension of Eq. \ref{eq:model}, with the addition of the following supplementary terms:
\begin{align}
\begin{split}
\label{eq:model2}
    & {\rm H}\alpha = L2_{\rm c}\,G(t)\\
    & {\rm FWHM} = L3_{\rm c}\,G(t),\\
    & W_{\rm CCF} = L4_{\rm c}\,G(t).
\end{split}
\end{align}
 We performed the modelling with \texttt{PyORBIT} in the same way as described in the previous cases, and we show the outcomes in Figs. \ref{fig:rvb_best}, \ref{fig:rvc_best}, and Table \ref{table:model-rv-best}. 

We emphasise that the inclusion of the additional activity indicators reduces the uncertainties in the keplerian signals and the RV jitter term with respect to Case 1 and Case 2. Notably, the orbital parameters remain consistent across all cases. We used the Bayesian Information Criterion (BIC, \citealt{1978AnSta...6..461S}) to compare the first two cases. Our analysis revealed a strong preference for Case 2 over Case 1, with a substantial $\Delta$BIC$_{12}$ value of 148 \citep{doi:10.1080/01621459.1995.10476572}.
In general, the BIC may not be the optimal estimator for the Bayesian evidence; in this specific case, however, we believe that the extreme difference between the BIC values ($\Delta$BIC = 148) largely overcomes any possible bias, thus favouring Case 2 over Case 1. As for Case 3, a direct BIC comparison is not possible due to the different datasets employed in the analysis. Instead, based on logical grounds, we selected Case 3 as our reference model. Each additional activity indicator provides valuable information on specific aspects of stellar activity, as evidenced in previous paragraphs, and their inclusion is justified by the amplitude parameter of the covariance matrix being significantly different from zero for every extra activity indicator, i.e. those activity indicators further constrain the activity model independently from the observed radial velocities. The adopted masses for planet b and c are $58.7^{+5.7}_{-5.6}$ and $11.8^{+4.8}_{-4.7}\, M_{\oplus}$, respectively. 
We include the final parameters of planets b and c in Table \ref{table:param}, while Table \ref{table:diff} lists the differences between this model and the other two. The significantly reduced jitter observed in Case 2 and 3, compared to Case 1, may be attributed to the limited capability of the uni-dimensional GP framework to model the different periodicities present in the spectroscopic time series. In particular, the RV and chromospheric indexes exhibit different periodicities (see also Sec. \ref{sec:rotation} and Fig. \ref{fig:gls}) when the RV variations are dominated by dark spots. In contrast, Case 2 and 3 use the formalism outlined in \cite{2015MNRAS.452.2269R}, which effectively handles variations in periodicity. 

\begin{table}
\caption{Comparison of the three different models
}             
\label{table:diff}      
\centering          
\begin{tabular}{l c c c}     
\hline     
Parameter & Case 1 & Case 2 & Case 3 \rule{0pt}{2.3ex} \rule[-1ex]{0pt}{0pt}\\ 
\hline    

   $K_{\rm b}$ (m s$^{-1}$)  & 14.0$\pm$2.7 & 14.8$\pm$1.7 & 15.7$\pm$1.5 \rule{0pt}{2.3ex} \rule[-1ex]{0pt}{0pt}\\
   $K_{\rm c}$ (m s$^{-1}$) & 3.9$^{+2.7}_{-2.4}$ & 3.1$^{+1.7}_{-1.6}$ & 4.1$^{+1.6}_{-1.7}$ \rule{0pt}{2.3ex} \rule[-1ex]{0pt}{0pt}\\
   $\sigma^{RV}_{jitter}$ (m s$^{-1}$) & 8.3$^{+7.3}_{-6.2}$ & 1.8$^{+2.0}_{-1.2}$ & 1.5$^{+1.6}_{-1.1}$ \rule{0pt}{2.3ex} \rule[-1ex]{0pt}{0pt}\\
   
\hline
\end{tabular}
\end{table}

In addition to the modelling mentioned above, we looked for the presence of a third planet in our RV dataset, by applying wide uniform priors on its period $P$ and RV semi-amplitude $K$. While we did not impose the orbit of the third planet to be circular, we applied a Gaussian prior on the eccentricity following \cite{2019AJ....157...61V}. Initially, the global optimization algorithm and the first ten thousand steps of the MCMC exploration suggested a significant detection of a third planet. However, the chains of the planet's orbital period diverged quite soon, preventing us from claiming a third planet detection. We underline that the solutions for planets b and c in the system show little variation, which further strengthens the validity of their detections.

\begin{table*}
\caption{Priors and outcomes of the model of planet b and c from analysing spectroscopic series with a multidimensional GP framework (Case 3).}             
\label{table:model-rv-best}      
\centering          
\begin{tabular}{l c c c}     
\hline\hline     

 \multicolumn{4}{c}{GP framework parameters} \rule{0pt}{2ex} \rule[-0.9ex]{0pt}{0pt} \\ 
\hline    
Parameter & Unit & Prior & Value \rule{0pt}{2.2ex} \rule[-0.9ex]{0pt}{0pt}\\ 
\hline    
   Uncorrelated RV jitter ($\sigma^{\rm RV}_{\rm jitter,0}$) & m s$^{-1}$ & ... & 1.5$^{+1.6}_{-1.1}$ \rule{0pt}{2.2ex} \rule[-0.8ex]{0pt}{0pt}\\  
   RV offset ($\gamma^{\rm RV}_{0}$) & m s$^{-1}$ & ... & -9432.9$^{+2.9}_{-2.8}$ \rule{0pt}{2.2ex} \rule[-0.8ex]{0pt}{0pt}\\
   Uncorrelated BIS jitter ($\sigma^{\rm BIS}_{\rm jitter,0}$) & m s$^{-1}$ & ... & 17.1$^{+1.7}_{-1.6}$ \rule{0pt}{2.2ex} \rule[-0.8ex]{0pt}{0pt}\\
   BIS offset ($\gamma^{\rm BIS}_{0}$) & m s$^{-1}$ & ... & 10.6$^{+2.6}_{-2.6}$ \rule{0pt}{2.2ex} \rule[-0.8ex]{0pt}{0pt}\\
   Uncorrelated log$R^{'}_{HK}$ jitter ($\sigma^{\rm logR^{'}_{HK}}_{\rm jitter,0}$) &  & ... & 0.0127$^{+0.0014}_{-0.0013}$ \rule{0pt}{2.3ex} \rule[-0.8ex]{0pt}{0pt}\\
   log$R^{'}_{HK}$ offset ($\gamma^{\rm logR^{'}_{HK}}_{0}$) &  & ... & -4.4137$\pm $0.0030 \rule{0pt}{2.2ex} \rule[-0.8ex]{0pt}{0pt}\\
   Uncorrelated H${\alpha}$ jitter ($\sigma^{\rm H{\alpha}}_{\rm jitter,0}$) &  & ... & 0.0017$\pm $0.0002 \rule{0pt}{2.2ex} \rule[-0.8ex]{0pt}{0pt}\\
   H${\alpha}$ offset ($\gamma^{\rm H\alpha}_{0}$) &  & ... & 0.1513$\pm $0.0004 \rule{0pt}{2.2ex} \rule[-0.8ex]{0pt}{0pt}\\
   Uncorrelated FWHM jitter ($\sigma^{\rm FWHM}_{\rm jitter,0}$) & km s$^{-1}$ & ... & 0.046$\pm $0.004 \rule{0pt}{2.2ex} \rule[-0.8ex]{0pt}{0pt}\\
   FWHM offset ($\gamma^{\rm FWHM}_{0}$) & km s$^{-1}$ & ... & 11.37$\pm $0.01 \rule{0pt}{2.2ex} \rule[-0.8ex]{0pt}{0pt}\\
   Uncorrelated $W_{\rm CCF}$ jitter ($\sigma^{W_{\rm CCF}}_{\rm jitter,0}$) & km s$^{-1}$ & ... & 0.0013$^{+0.0013}_{-0.0009}$ \rule{0pt}{2.2ex} \rule[-0.8ex]{0pt}{0pt}\\
   $W_{\rm CCF}$ offset ($\gamma^{W_{\rm CCF}}_{0}$) & km s$^{-1}$ & ... & 3.505$\pm $0.003 \rule{0pt}{2.2ex} \rule[-0.8ex]{0pt}{0pt}\\
   \cline{0-1} 
   \textit{Multidimensional GP parameters \citep{2015MNRAS.452.2269R}} & & & \rule{0pt}{2.2ex} \rule[-0.8ex]{0pt}{0pt}\\
   $V_c$ & m s$^{-1}$ & $\mathcal{U}$(-100.0, 100.0) & -11.2$^{+2.3}_{-2.8}$ \rule{0pt}{2.2ex} \rule[-0.8ex]{0pt}{0pt}\\ 
   $V_r$ & m s$^{-1}$ & $\mathcal{U}$(0.0, 100.0) & 27.1$^{+4.3}_{-3.6}$ \rule{0pt}{2.2ex} \rule[-0.8ex]{0pt}{0pt}\\ 
   $B_c$ & m s$^{-1}$ & $\mathcal{U}$(-100.0, 100.0) & 4.3$^{+3.3}_{-3.2}$ \rule{0pt}{2.2ex} \rule[-0.8ex]{0pt}{0pt}\\ 
   $B_r$ & m s$^{-1}$ & $\mathcal{U}$(-100.0, 100.0) & -35.6$^{+4.8}_{-6.0}$ \rule{0pt}{2.2ex} \rule[-0.8ex]{0pt}{0pt}\\ 
   $L_c$ (log$R^{'}_{HK}$) &  & $\mathcal{U}$(-0.1, 0.1) &  -0.011$\pm$0.002 \rule{0pt}{2.2ex} \rule[-0.8ex]{0pt}{0pt}\\ 
   $L2_c$ (H${\alpha}$) &  & $\mathcal{U}$(-0.1, 0.1) & -0.0015$\pm$0.0003 \rule{0pt}{2.2ex} \rule[-0.9ex]{0pt}{0pt}\\ 
   $L3_c$ (FWHM) & km s$^{-1}$ & $\mathcal{U}$(-0.5, 0.5) & -0.043$^{+0.007}_{-0.008}$ \rule{0pt}{2.2ex} \rule[-0.9ex]{0pt}{0pt}\\ 
   $L4_c$ ($W_{\rm CCF}$) & km s$^{-1}$ & $\mathcal{U}$(-0.02, 0.02) & -0.014$\pm$0.002 \rule{0pt}{2.2ex} \rule[-0.9ex]{0pt}{0pt}\\ 
\hline         
\multicolumn{4}{c}{Stellar activity (RV + activity indexes)} \rule{0pt}{2.3ex} \rule[-0.9ex]{0pt}{0pt}\\ 
\hline    
Parameter & Unit & Prior & Value \rule{0pt}{2.2ex} \rule[-0.9ex]{0pt}{0pt}\\ 
\hline 
   Rotational period ($P_{\rm rot}$) & days & $\mathcal{N}$(7.34, 0.15) & 7.37$\pm$0.03 \rule{0pt}{2.2ex} \rule[-0.9ex]{0pt}{0pt}\\
   Decay Timescale of activity ($P_{\rm dec}$) & days & $\mathcal{U}$(10.0, 2000.0) & 26.1$^{+3.3}_{-3.0}$ \rule{0pt}{2.2ex} \rule[-0.9ex]{0pt}{0pt}\\
   Coherence scale ($w$) &  & $\mathcal{U}$(0.01, 0.60) & 0.36$\pm$0.03 \rule{0pt}{2.2ex} \rule[-0.9ex]{0pt}{0pt}\\
\hline
\multicolumn{4}{c}{Planet b} \rule{0pt}{2.2ex} \rule[-0.9ex]{0pt}{0pt}\\ 
\hline    
Parameter & Unit & Prior & Value \rule{0pt}{2.2ex} \rule[-0.9ex]{0pt}{0pt}\\ 
\hline 
   Orbital period ($P_{\rm b}$) & days & $\mathcal{N}$(10.590547, 0.000012) & 10.590547$\pm$0.000012 \rule{0pt}{2.2ex} \rule[-0.9ex]{0pt}{0pt}\\
   Central time of the first transit ($T_{\rm 0,b}$) & BTJD & $\mathcal{N}$(2616.49232, 0.00022) & 2616.49232$\pm$0.00022 \rule{0pt}{2.2ex} \rule[-0.9ex]{0pt}{0pt}\\
   Orbital eccentricity ($e_{\rm b}$) &  & $\mathcal{N}$(0.00, 0.098) & $\leqslant$ 0.13 \rule{0pt}{2.2ex} \rule[-0.9ex]{0pt}{0pt}\\
   Argument of periastron ($\omega_{\rm b}$) & deg & ... & 92$^{+82}_{-45}$ \rule{0pt}{2.2ex} \rule[-0.9ex]{0pt}{0pt}\\
   Semi-major axis to stellar radius ratio ($a_{\rm b}/R_{\star}$) &  & ... & 20.2$\pm$0.3 \rule{0pt}{2.2ex} \rule[-0.9ex]{0pt}{0pt}\\
   Orbital semi-major axis ($a_{\rm b}$) & au & ... & 0.0988$\pm$0.0004 \rule{0pt}{2.2ex} \rule[-0.9ex]{0pt}{0pt}\\
   RV semi-amplitude ($K_{\rm b}$) & m s$^{-1}$ & $\mathcal{U}$(0.01, 100.0) & 15.7$^{+1.5}_{-1.5}$ \rule{0pt}{2.2ex} \rule[-0.9ex]{0pt}{0pt}\\
   Planetary mass ($M_{\rm p,b}$) & $M_{\oplus}$ & ... & 58.7$^{+5.7}_{-5.6}$ \rule{0pt}{2.2ex} \rule[-1.5ex]{0pt}{0pt}\\
\hline
\multicolumn{4}{c}{Planet c} \rule{0pt}{2.2ex} \rule[-0.9ex]{0pt}{0pt}\\ 
\hline    
Parameter & Unit & Prior & Value \rule{0pt}{2.2ex} \rule[-0.9ex]{0pt}{0pt}\\ 
\hline 
   Orbital period ($P_{\rm c}$) & days & $\mathcal{N}$(4.77271, 0.00016) & 4.77270$\pm$0.00016 \rule{0pt}{2.2ex} \rule[-0.9ex]{0pt}{0pt}\\
   Central time of the first transit ($T_{\rm 0,c}$) & BTJD & $\mathcal{N}$(2628.6178, 0.0009) & 2628.6178$\pm$0.0009 \rule{0pt}{2.2ex} \rule[-0.9ex]{0pt}{0pt}\\
   Orbital eccentricity ($e_{\rm c}$) &  & $\mathcal{N}$(0.00, 0.098) & $\leqslant$ 0.14 \rule{0pt}{2.2ex} \rule[-0.9ex]{0pt}{0pt}\\
   Argument of periastron ($\omega_{\rm c}$) & deg & ... & 172$^{+79}_{-107}$ \rule{0pt}{2.2ex} \rule[-1ex]{0pt}{0pt}\\
   Semi-major axis to stellar radius ratio ($a_{\rm c}/R_{\star}$) &  & ... & 11.9$\pm$0.2 \rule{0pt}{2.2ex} \rule[-0.9ex]{0pt}{0pt}\\
   Orbital semi-major axis ($a_{\rm c}$) & au & ... & 0.0581$\pm$0.0002 \rule{0pt}{2.2ex} \rule[-0.9ex]{0pt}{0pt}\\
   RV semi-amplitude ($K_{\rm c}$) & m s$^{-1}$ & $\mathcal{U}$(0.01, 100.0) & 4.1$^{+1.7}_{-1.6}$ \rule{0pt}{2.2ex} \rule[-0.9ex]{0pt}{0pt}\\
   Planetary mass ($M_{\rm p,c}$) & $M_{\oplus}$ & ... & 11.8$^{+4.8}_{-4.7}$ \rule{0pt}{2.2ex} \rule[-1.5ex]{0pt}{0pt}\\
\hline
\end{tabular}
\end{table*}

\subsection{Search for TTVs}
To investigate the potential presence of dynamical interactions between TOI-5398 b and c
we performed a search for Transit Timing Variations 

\citep[TTVs; e.g.,][]{Agol2005MNRAS.359..567A, HolmanMurray2005Sci...307.1288H, borsato2014A&A...571A..38B, borsato2019MNRAS.484.3233B, borsato2021MNRAS.506.3810B}
of planet b and c. 



\begin{figure}
   \centering
   \includegraphics[width=\columnwidth]{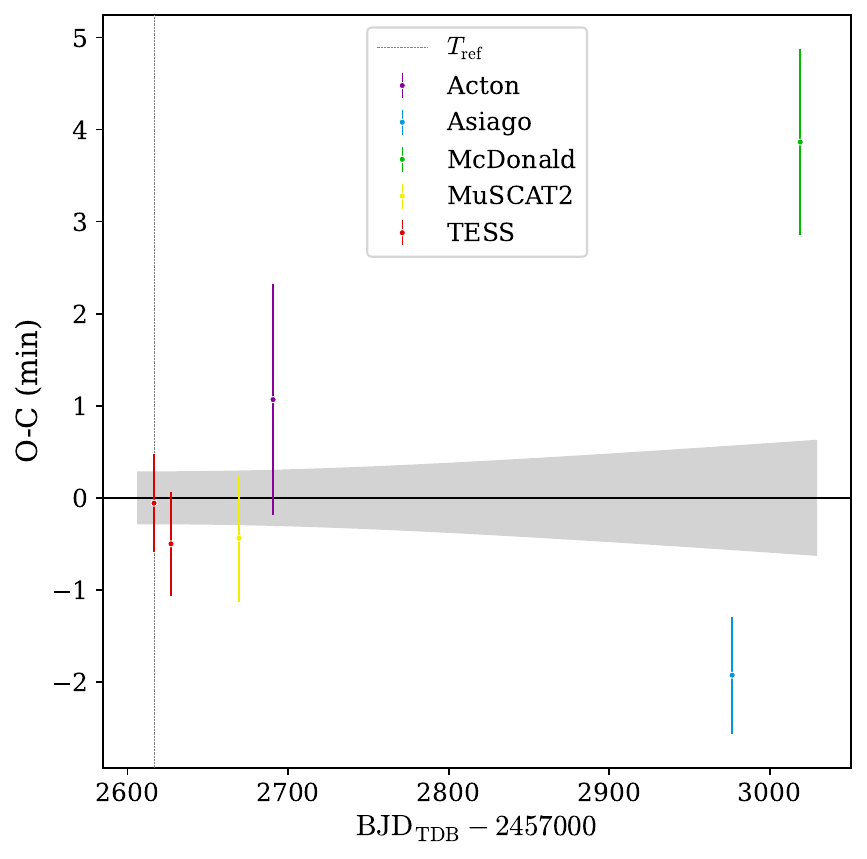}
   \caption{O-C plot representing the observed (O) and calculated (C) transit times for 
   the linear ephemeris
   of TOI-5398 b (see Table~\ref{table:param}).
   Each dataset is shown in a distinct colour.}
   \label{fig:ttv}
\end{figure}

\begin{figure}
   \centering
   \includegraphics[width=\columnwidth]{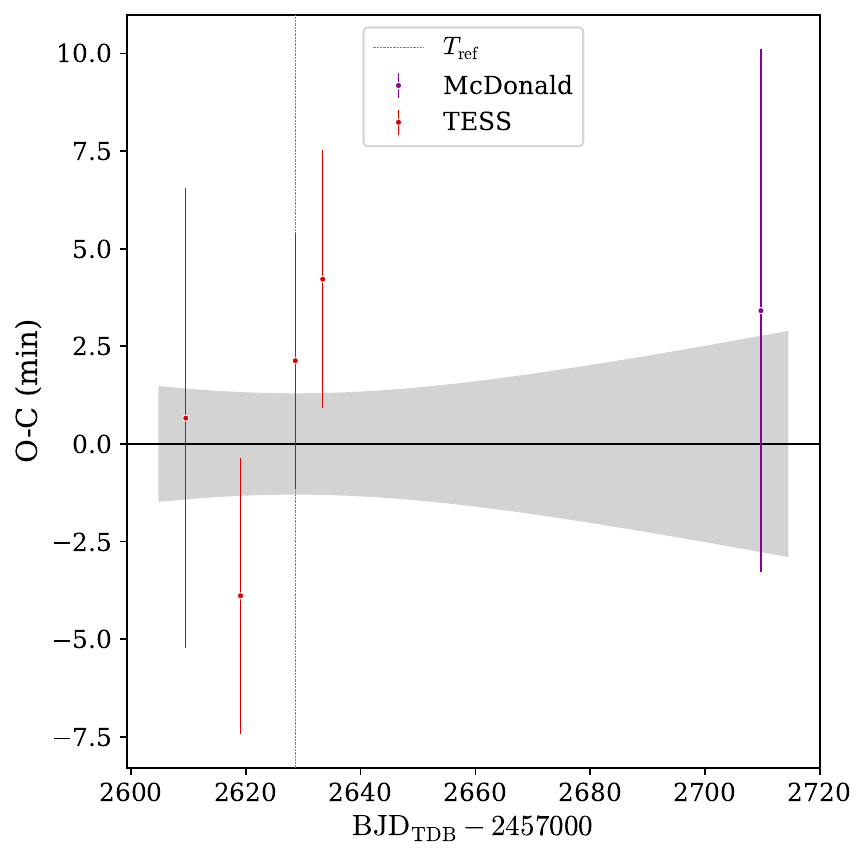}
   \caption{As in Fig. \ref{fig:ttv}, but for TOI-5398 c.}
   \label{fig:ttv_c}
\end{figure}

We computed the observed (O) - calculated (C) diagrams for both planets removing the linear ephemeris (in Table~\ref{table:param}) to each transit times. See the O-C diagrams in Fig.~\ref{fig:ttv} and \ref{fig:ttv_c} for planets b and c, respectively.
The possible TTV amplitude ($A_\mathrm{TTV}$), computed as the semi-amplitude of the O-C
, is of $2.9_{-1.0}^{+1.2}$ minutes for planet b and $4_{-2}^{+7}$ minutes for planet c. The associated error is derived from the subtraction of $A_\mathrm{TTV}$ from the high-density interval at 95\% of a Monte-Carlo sampling of 10\,000 repetitions.

\begin{figure}
   \centering
   \includegraphics[width=\hsize]{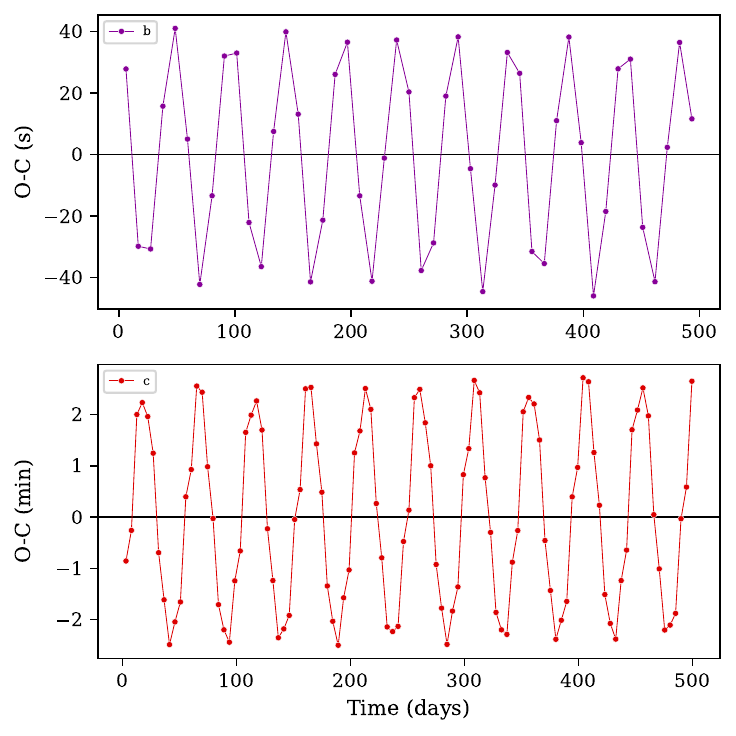}
   \caption{
   Synthetic O-C diagrams (top-panel planet b, bottom-panel planet c) computed from the dynamical simulation with \texttt{TRADES} and with parameters from Table~\ref{table:model-rv-best}.
   }
   \label{fig:ttv_sim}
\end{figure}

Although our data do not cover a sufficient portion of the super-period to offer direct evidence, they do suggest a possible TTV due to the gravitational interaction of planets b and c (see Fig. \ref{fig:ttv}). 
The sparse sampling of the TTV signals did not allow us to run a dynamical fit, so we decided to run a forward dynamical model with \texttt{TRADES}\footnote{\url{https://github.com/lucaborsato/trades}}\citep{borsato2014A&A...571A..38B, 2019MNRAS.484.3233B, borsato2021MNRAS.506.3810B}, similar to what has been done in \citet{tuson2023MNRAS.523.3090T}. We took the masses and the orbital parameters from Tables~\ref{table:model-rv-best} and \ref{table:param} and we integrated the orbits for about 500 days and computed the transit times of each planet. We computed the O-C diagrams (see Fig.~\ref{fig:ttv_sim}) and we found that the simulated 
$A_\mathrm{TTV}$
is of the order of $\sim40$ seconds and of $\sim2$ minutes for planet b and c, respectively.

We used the orbital parameters from Table~\ref{table:param}
as the starting point of the dynamical simulation,
i.e., we assumed that the value we obtained from our global fit spanning about 500 days
represents a specific configuration in time, which may not be true.
The amplitude of the simulated TTVs may also be dependent on the eccentricity of the two planets.
Nevertheless, the outcome of the simulation is compatible 
(at $2\sigma$ for b and at $1\sigma$ for c) with the measured TTVs.
The scope of this analysis is to show that TTVs can indeed be present in this system,
while a full dynamical analysis is outside the scope of this paper and it is left to future works.

\section{Discussion}
\label{sec:discussion}
\begin{table}
\caption{Final parameters of TOI-5398 multi-planet system.}             
\label{table:param}      
\centering          
\begin{tabular}{l c c}     
\hline     
Parameter & TOI-5398 b & TOI-5398 c \rule{0pt}{2.3ex} \rule[-1ex]{0pt}{0pt}\\ 
\hline    

   $P$ (days)  & 10.590547$^{+0.000012}_{-0.000011}$ & 4.77271$^{+0.00016}_{-0.00014}$ \rule{0pt}{2.3ex} \rule[-1ex]{0pt}{0pt}\\
   $T_{\rm 0}$ (BTJD)  & 2616.49232$^{+0.00022}_{-0.00021}$ & 2628.61781$^{+0.00090}_{-0.00086}$ \rule{0pt}{2.3ex} \rule[-1ex]{0pt}{0pt}\\
   $a/R_{\star}$ & 20.0$\pm$0.6 & 11.8$\pm$0.4 \rule{0pt}{2.3ex} \rule[-1ex]{0pt}{0pt}\\
   $a$ (au) & 0.098$\pm$0.005 & 0.057$\pm$0.003 \rule{0pt}{2.3ex} \rule[-1ex]{0pt}{0pt}\\
   $R_{\rm p}/R_{\star}$  & 0.0899$^{+0.0007}_{-0.0006}$ & 0.0308$^{+0.0011}_{-0.0012}$ \rule{0pt}{2.5ex} \rule[-1ex]{0pt}{0pt}\\
   $R_{\rm p}$ ($R_{\oplus}$) & 10.30$\pm$0.40 & 3.52$\pm$0.19 \rule{0pt}{2.5ex} \rule[-1.5ex]{0pt}{0pt}\\
   $b$  & 0.272$^{+0.069}_{-0.110}$ & $\leqslant$ 0.34\tablefootmark{a}
   \rule{0pt}{2.5ex} \rule[-1.5ex]{0pt}{0pt}\\
   $i$ (deg) & 89.21$^{+0.31}_{-0.21}$ & $\geqslant$ 88.4\tablefootmark{b} 
   \rule{0pt}{2.5ex} \rule[-1ex]{0pt}{0pt}\\
   $T_{14}$ (hour) & 4.258$^{+0.149}_{-0.103}$ & 3.127$^{+0.108}_{-0.127}$ \rule{0pt}{2.5ex} \rule[-1ex]{0pt}{0pt}\\
   $e$ & 
   $\leqslant$ 0.13\tablefootmark{a} & 
   $\leqslant$ 0.14\tablefootmark{a}\rule{0pt}{2.3ex} \rule[-1ex]{0pt}{0pt}\\
   $K$ (m\,s$^{-1}$) & 15.7$\pm$1.5 & 4.1$^{+1.7}_{-1.6}$ \rule{0pt}{2.3ex} \rule[-1ex]{0pt}{0pt}\\
   $M_{\rm p}$ ($M_{\oplus}$) & 58.7$^{+5.7}_{-5.6}$ & 11.8$^{+4.8}_{-4.7}$ \rule{0pt}{2.3ex} \rule[-1.5ex]{0pt}{0pt}\\
   $\rho_{\rm p}$ (g\,cm$^{-3}$) & 0.29$\pm$0.05 & 1.50$\pm$0.68 \rule{0pt}{2.3ex} \rule[-1.5ex]{0pt}{0pt}\\
   $g$ (m\,s$^{-2}$) & 5.36$\pm$0.86 & 8.99$\pm$3.76 \rule{0pt}{2.3ex} \rule[-1.5ex]{0pt}{0pt}\\
   $T_{\rm eq}$ (K) & 947$\pm$28 & 1242$\pm$37\rule{0pt}{2.3ex} \rule[-1.5ex]{0pt}{0pt}\\

\hline
\end{tabular}
\tablefoot{\tablefoottext{a}{84th percentile.}\tablefoottext{b}{16th percentile.} }
\end{table}

\subsection{Peculiar architecture}
TOI-5398 is a compact multi-planet system composed of a warm giant (TOI-5398 b, $P_{\rm orb}$ $\sim$ 10.59 d) and a hot sub-Neptune planet (TOI-5398 c, $P_{\rm orb}$ $\sim$ 4.77 d) orbiting a moderately young solar-analogue star. The peculiarity of this system comes from its compactness and planetary architecture, which is uncommon among known multi-planet systems with short-period giant planets. In fact, the sub-Neptune is the closest planet to the host star. There are currently very few notable examples of compact systems consisting of an inner orbit small-size planet and an outer short-period giant companion (e.g., \citealt{2022AJ....164...13H}). 
The most famous multi-planet system with similar characteristics is WASP-47 \citep{2012MNRAS.426..739H, 2015ApJ...812L..18B, Nascimbeni2023A&A...673A..42N}.
Within this family of peculiar planetary systems, we also have Kepler-730 \citep{Zhu_2018, 2019ApJ...870L..17C}, TOI-1130 \citep{Huang_2020,2023arXiv230515565K}, TOI-2000 \citep{2022arXiv220914396S}, and WASP-132 \citep{2022AJ....164...13H}. 
We list them in Table \ref{table:compact}, together with a list of available datasets and a few notable parameters that we want to highlight. 

Among these compact systems, TOI-5398 stands out, along with WASP-47 and TOI-2000, due to its precise transit photometry and radial velocity measurements. These observations enable the precise measurement of planetary bulk densities and make it an extremely appealing target for continued monitoring with follow-up observations and surveys, including PLATO and Ariel, along with telescopes like ELTs.  

Moreover, TOI-5398 is the youngest compact system with a gas giant ever confirmed with a quite good age estimation. 

\begin{table*}
\caption{Confirmed compact multi-planet systems, sorted by age.}             
\label{table:compact}      
\centering          
\begin{tabular}{l c c c c c c l}     
\hline     
\hline  
System & n$^{\circ}$ known & Transiting & PRV\tablefootmark{a} & TTV\tablefootmark{b} 
& TSM\tablefootmark{c} & Age & Reference \rule{0pt}{2.3ex} \rule[-1ex]{0pt}{0pt}\\ 
 & planets & planets &  &  
 &  & (Gyr) &  \rule{0pt}{2.3ex} \rule[-1ex]{0pt}{0pt}\\ 
\hline    

   TOI-5398 & 2  & True, 2/2 & True & Potential & 
   288 & 0.65$\pm$0.15 & This paper (Sect. \ref{sec:age}) \rule{0pt}{2.3ex} \rule[-1ex]{0pt}{0pt}\\
   WASP-132 & 2  & True, 2/2 & True, 1/2 & False & 
   106 & 3.2$\pm$0.5 & \cite{2022AJ....164...13H} \rule{0pt}{2.3ex} \rule[-1ex]{0pt}{0pt}\\
   TOI-2000 & 2 &  True, 2/2 & True & Potential & 
   68 & 5.3$\pm$2.7 & \cite{2022arXiv220914396S} \rule{0pt}{2.3ex} \rule[-1ex]{0pt}{0pt}\\
   WASP-47 & 4 & True, 4/4 & True & True & 
   47 & 6.5$^{+2.6}_{-1.2}$ & \cite{2012MNRAS.426..739H} \rule{0pt}{2.3ex} \rule[-1ex]{0pt}{0pt}\\
   TOI-1130 & 2  & True, 2/2 & True & True & 
   345 & 8.2$^{+3.8}_{-4.9}$ & \cite{Huang_2020} \rule{0pt}{2.3ex} \rule[-1ex]{0pt}{0pt}\\
   Kepler-730 & 2  & True, 2/2 & False & False & 
   25 & 9.5$^{+2.5}_{-2.7}$ & \cite{Zhu_2018} \rule{0pt}{2.3ex} \rule[-1ex]{0pt}{0pt}\\
   

\hline
\end{tabular}
\tablefoot{\tablefoottext{a}{Precise Radial Velocity.} \tablefoottext{b}{Transit Time Variations.} \tablefoottext{c}{Transmission Spectroscopy Metric of the giant planet in the system.}}
\end{table*}


\subsection{Variety among compact multi-planet systems}
\begin{figure}
   \centering
   \includegraphics[width=\hsize]
   {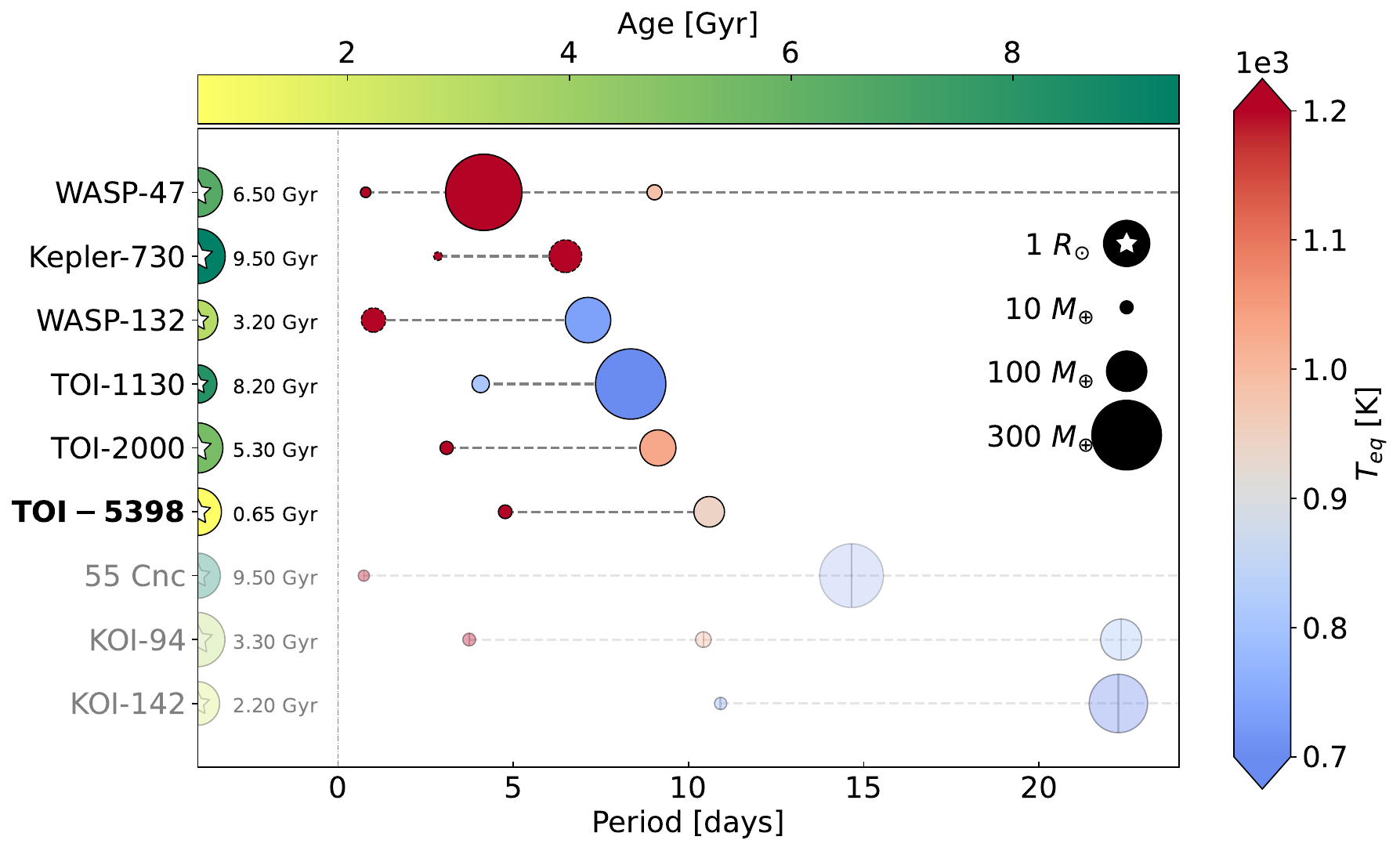}
   \caption{Architecture of compact multi-planet systems hosting small-size planets orbiting inner to short-period gas giants ($P$ $\lesssim$ 10 d). 
   Each row represents one planetary system ($y$-axis) and the planetary orbital periods ($x$-axis). The sizes of the dots correspond to the planet masses, and the colours of the points to the equilibrium temperatures (see colorbar to the right). From top to bottom, the systems are sorted in ascending order of the period of the giant. Shaded dots represent the next three systems with gas giants on $P$ < 25 d orbits. Dots with a vertical line represent planets whose mass is multiplied by $\sin(i)$. The semi-circular dots filled with a star shape stand for the host stars, colour-coded by their age (see colorbar to the top), while the size represents their radii.}
   \label{fig:peas}
\end{figure}

In Fig. \ref{fig:peas}, we show the architecture of compact multi-planet systems with small-size planets orbiting interior to short-period giant planets ($P$ $\lesssim$ 10 d). For comparison, we show the next three systems with giants whose orbital periods are $P$ < 25 d \citep{1997ApJ...474L.115B, 2018A&A...619A...1B,2013ApJ...768...14W,2013ApJ...777....3N}. The semicircular dots represent the host stars, colour-coded by their age, while the size represents their radii. Planets, on the other hand, are colour-coded by their equilibrium temperature $T_{\rm eq}$, and their size reflects their planetary masses. The inner planet of WASP-132 and the Kepler-730 planets do not yet have mass measurements, so we extracted them following \cite{2016ApJ...825...19W} or we showed upper limits \citep{2022AJ....164...13H}.

\begin{figure}
   \centering
   \includegraphics[width=\hsize]{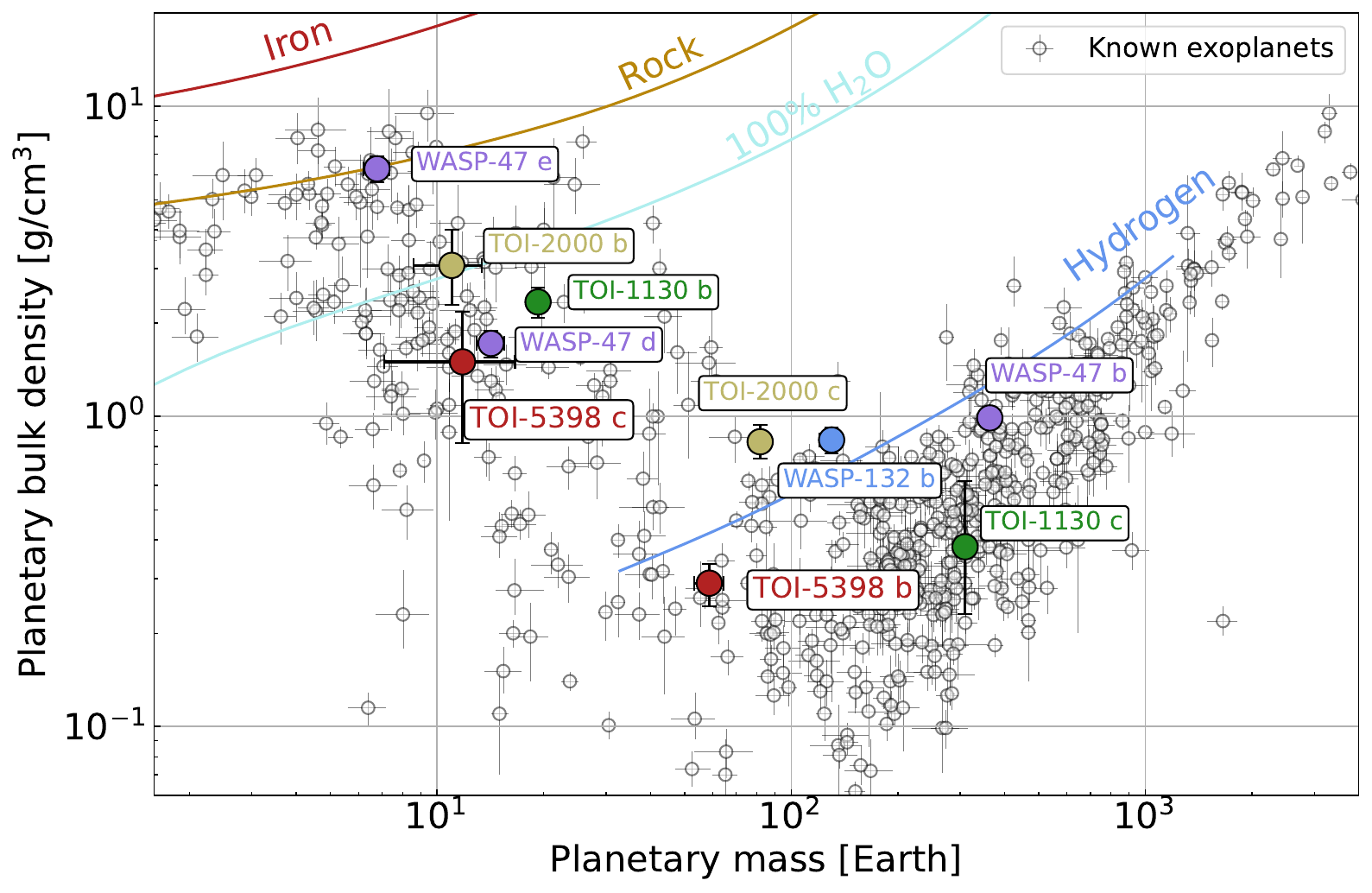}
   \caption{Mass-density distribution of all confirmed planets from the NASA Exoplanet Archive with mass and radius determination better than 20\%. The red dots represent TOI-5398 b and c, while the remaining planets mentioned in Table \ref{table:compact} are represented by dots of different colours. We also plot theoretical mass-radius curves for planets of various pure compositions from \citet[online at \url{https://lweb.cfa.harvard.edu/~lzeng/planetmodels.html}]{2019PNAS..116.9723Z}: solid red indicates a pure-Iron core, brown an Earth-like rocky core (32.5\% Fe and 67.5\% MgSiO3), light-blue a 100\% Water world at 1000K, and blue a 100\% cold-Hydrogen world.}
   \label{fig:mass-density}
\end{figure}

TOI-5398 hosts the youngest gas giant planet with $P$ < 25$\,$d and $M_p$ > 1/2 $M_{\rm Saturn}$ that is known to have an inner companion. The gas giant TOI-5398 b has a radius similar to Jupiter and a mass close to 2/3 of Saturn, which is the smallest mass among giants in compact systems. As a result, its bulk density is around half that of Saturn and roughly equal to that of TOI-1130 c (0.38 g $cm^{-3}$, \citealt{Huang_2020}). These properties, as well as its orbital period, make TOI-5398 b quite similar to the hot-Saturn planet TOI-2000 c, while they are in contrast with the properties of the larger WASP-47 b, Kepler-730 b, and TOI-1130 c, which all have radii $\sim$ 1.1 $R_J$ and masses $\sim$ 1 $M_J$ (apart from Kepler-730 b that does not have a mass measurement yet). However, the larger radius and smaller mass of TOI-5398 b compared to those of TOI-2000 c -- which results in having a relatively low density compared to giant planets of similar mass (see Fig. \ref{fig:mass-density} and cf., for example, \citealt{2022AJ....164...70Y}) -- make the giant planet under study more like a \textit{puffy} Saturn \citep{2022A&A...667A...8N}. Moreover, we calculated the equilibrium temperature $T_{\rm eq}$ (cf. Eq. 4 from \citealt{2011ApJ...729...54C}) of TOI-5398 b, assuming zero albedo and full day-night heat redistribution following
\begin{equation}
    T_{\rm eq} = T_{\rm eff} \sqrt{\frac{R_\star}{a}}\left(\frac{1}{4}\right)^{1/4},
\end{equation} where $a$ is the orbital semi-major axis given in the same units as $R_\star$. We obtained a value of 947$\pm$28 K, which indicates that TOI-5398 b is unlikely to be affected by the hot-Jupiter anomalous radius inflation mechanism \citep{2016ApJ...831...64T}.
By contrast, the $T_{\rm eq}$ of the hot-Saturn TOI-2000 c is slightly above 1000 K \citep{2022arXiv220914396S}; therefore, considering the $T_{\rm eq}$ = 1000 K threshold and its orbital period exceeding 10$\,$days, we describe TOI-5398 b as a \textit{warm}-Saturn planet. 

In addition, TOI-5398 b is located in the Neptunian ``savanna'' \citep{2023A&A...669A..63B}, a light deficit of planets close to the Hot-Neptune desert \citep{2007A&A...461.1185L,2016A&A...589A..75M}. 

\begin{figure}
   \centering
   \includegraphics[width=\hsize]{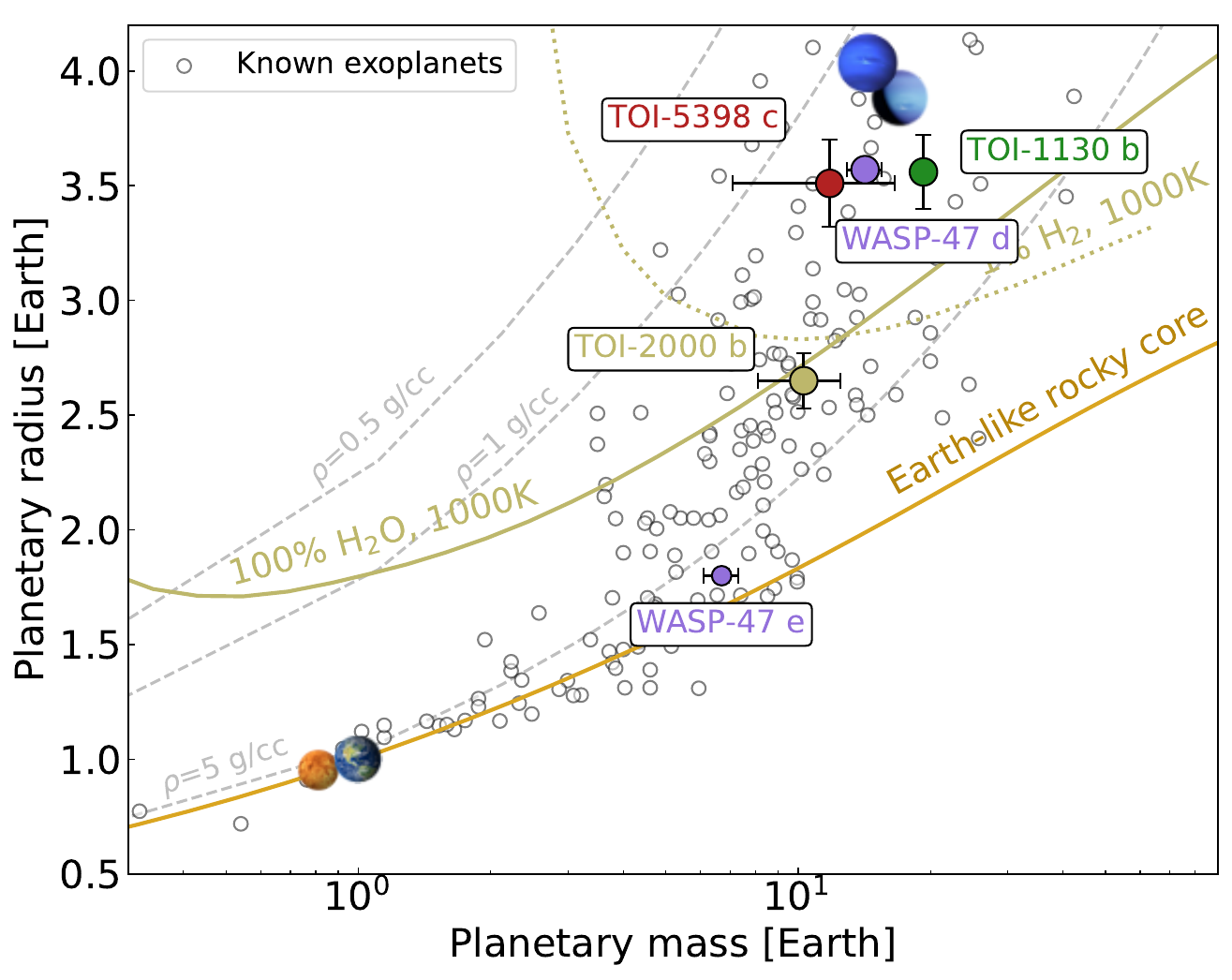}
   \caption{Mass-radius distribution of all confirmed planets with R$_{\rm p}$ $<$ 4 R$_{\oplus}$ from the NASA Exoplanet Archive with mass and radius determination better than 20\%. The red dot represents TOI-5398 c, the beige one TOI-2000 b, the green one TOI-1130 c, and the violet dots are respectively WASP-47 d and e. We include the Solar System planets Earth, Venus, Uranus, and Neptune. We add theoretical mass-radius curves from \cite{2019PNAS..116.9723Z}: solid brown indicates an Earth-like rocky core (32.5\% Fe and 67.5\% MgSiO3), beige a 100\% water world at 1000K, and dotted beige a 1\% hydrogen envelope and 99\% Earth-like rocky core at the same temperature. Grey dashed curves represent densities $\rho$ = 0.5, 1, and 5 g/cm$^3$, respectively.}
   \label{fig:mass_rad}
\end{figure}

The sub-Neptune TOI-5398 c is one of the very few inner companions to short-period gas giants with precise values of both mass and radius. In fact, only four small planets with $R_{\rm p} <$ 4 $R_{\oplus}$ in compact systems (P $\lesssim$ 15 d) have mass measurements: WASP-47 d, WASP-47 e \citep{2017AJ....154..237V}, TOI-1130 c \citep{2023arXiv230515565K}, and TOI-2000 b. It is worth noting that they have quite different bulk densities (see Fig. \ref{fig:mass_rad}), ranging from being composed of rocky cores to having masses and radii similar to those of Neptune and Uranus. The latter composition is true for TOI-5398 c, which shares a bulk density similar to that of the inner companion planet WASP-47 d. As a result, we provide additional evidence that inner companions to transiting giant planets tend to have the same density diversity as other small planets \citep{2022arXiv220914396S}. 

\subsection{Ephemeris improvements}

\begin{figure}
   \centering
   \includegraphics[width=\hsize]{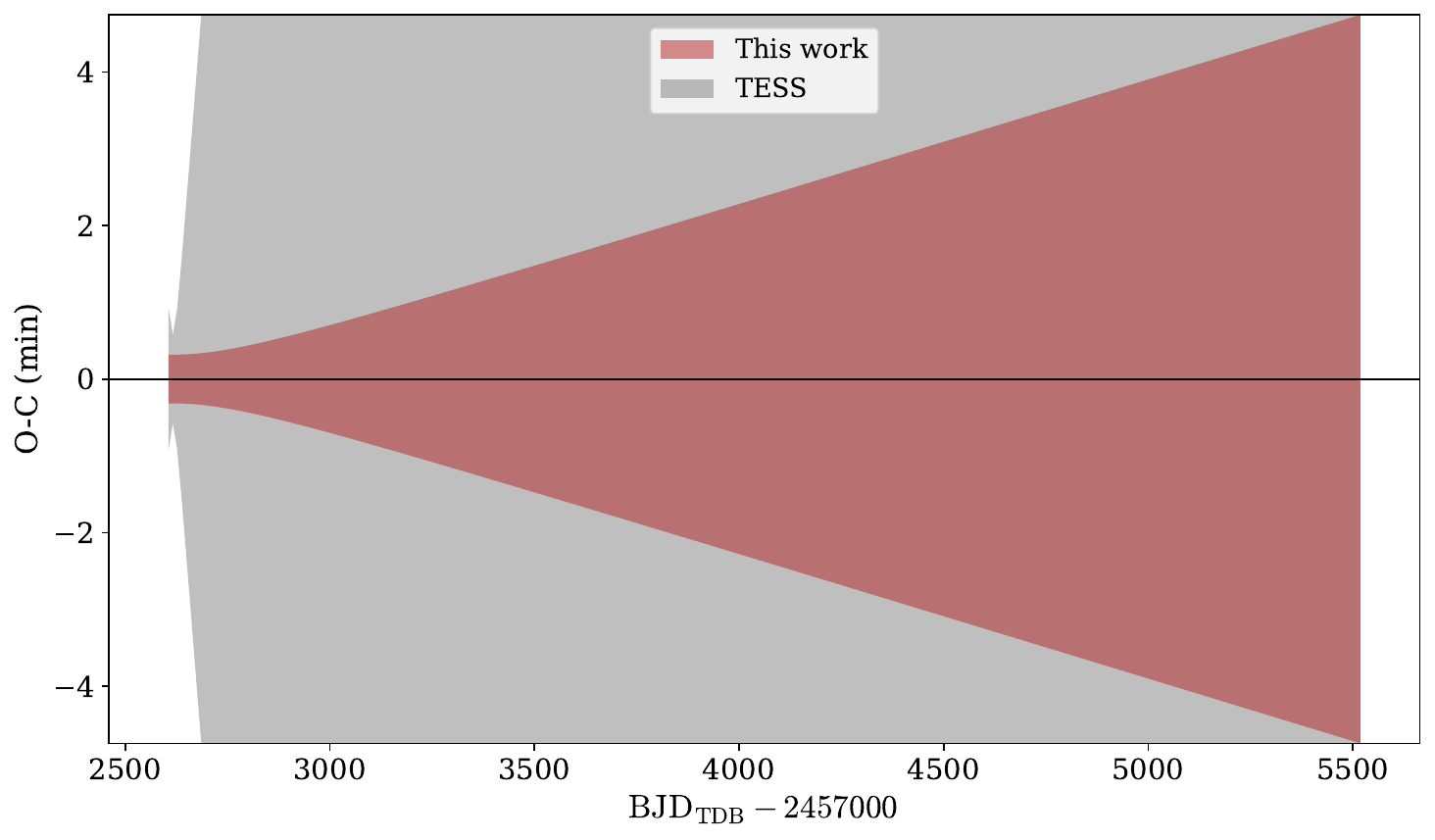}
   \caption{Ephemeris uncertainty of TOI-5398 b propagated until the start of 2030. On the $x$-axis, we show the date, while on the $y$-axis the error bars in minutes. In grey, we show the uncertainty propagation considering \textit{TESS} data alone. In red, we show the improvement considering also Ground-based photometry data.  }
   \label{fig:err_prop}
\end{figure}

\begin{figure}
   \centering
   \includegraphics[width=\hsize]{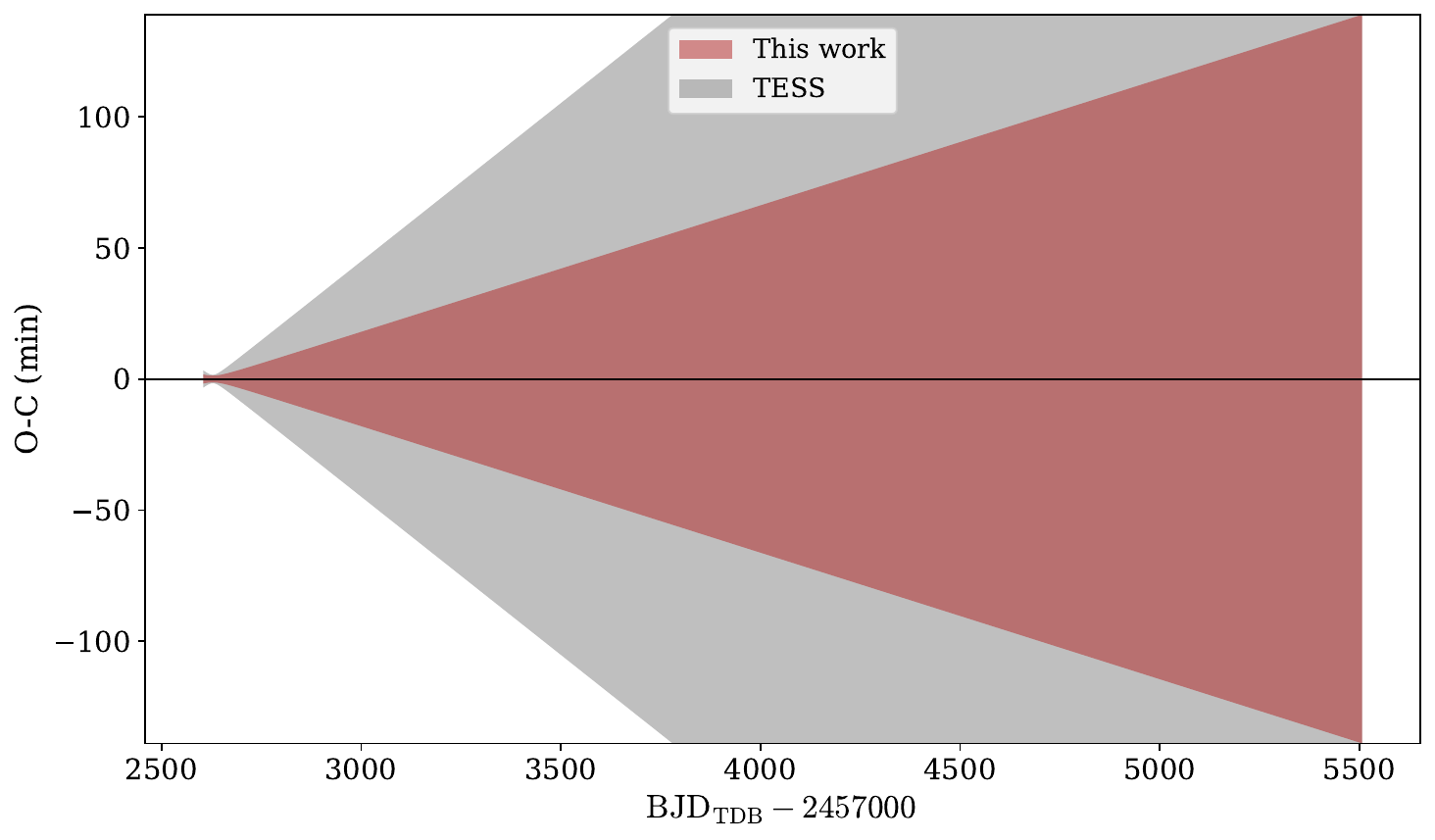}
   \caption{As in Fig. \ref{fig:err_prop}, but for TOI-5398 c.  }
   \label{fig:err_propc}
\end{figure}

An important step of our analysis is the derivation of new and updated mean ephemeris for TOI-5398 b and c. Our best-fit relation for the warm Saturn and the sub-Neptune are:

\begin{equation}
\begin{split}
    T_{0, \rm b} &= 2459616.49232 \pm 0.00022 \, \mathrm{BJD_{TDB}} \\
    &\quad +N \times (10.590547 \pm 0.000012),
\end{split}
\end{equation} 
\begin{equation}
\begin{split}
    T_{0, \rm c} &= 2459628.6178 \pm 0.0009 \, \mathrm{BJD_{TDB}} \\
    &\quad +N \times (4.77271 \pm 0.00016),
\end{split}
\end{equation}
where the variable $N$ is an integer number commonly referred to as the ``epoch'' and arbitrarily set to zero at our reference transit time $T_{\rm ref}$. We emphasise that if we propagate the new ephemeris at 2030-01-01 (see Fig. \ref{fig:err_prop}, and \ref{fig:err_propc}), the level of uncertainty is significantly reduced to $\sim$ 5 minutes compared to the previous $\sim$ 197 minutes for TOI-5398 b when only \textit{TESS} photometry was available. This means that when also the ground-based photometry is taken into account, the error bar for TOI-5398 b is 98\% smaller compared to using \textit{TESS} data alone. For TOI-5398 c, the error bar is 60\% smaller than when we rely solely on \textit{TESS} data. Accurately identifying the transit windows is crucial for upcoming space-based observations, given the significant investment in observing time and the time-critical nature of such observations. It is crucial to note that no further observations of TOI-5398 are planned in the current \textit{TESS} Extended Mission\footnote{As it results from the Web \textit{TESS} Viewing Tool \url{https://heasarc.gsfc.nasa.gov/cgi-bin/tess/webtess/wtv.py}}. 




\subsection{Planetary system formation and evolution}
\label{sec:form-evo}
The obliquity between the planetary orbital plane and the stellar rotation axis is a key diagnostic for the mechanisms of formation and orbital migration of exoplanets (e.g., \citealt{Naoz2011}). This can be detected with in-transit RVs through the Rossiter-McLaughlin effect (RM, \citealt{Ohta2005,1924ApJ....60...15R,1924ApJ....60...22M,2000A&A...359L..13Q}). Short-period giant planets are thought to form in situ close to the final orbit, or in the outer regions and migrate inward \citep{Dawson2018}. Different mechanisms, such as dynamical interactions (high-eccentricity migration) through planet-planet scattering \citep{Marzari2006} or the Kozai mechanism \citep{Wu2003}, and disc-planet interaction \citep{Lin1996}, can shrink their orbits. These mechanisms are expected to imprint different signatures in the planets' obliquity. Scattering encounters should randomise the alignments of the orbital planes, while migration through disc-planet interactions should keep the planetary orbits roughly co-planar throughout the entire process. 

TOI-5398's uncommon architecture and moderately young age make it particularly promising for measuring the obliquity between the orbital plane of the giant and the spin axis of the star. First, we can access the original configuration when observing systems young enough to have avoided tidal alterations of the obliquity. Then, unlike ordinary short-period giants, we can rule out the high-eccentricity migration scenario \citep{Mustill2015} for compact systems such as TOI-5398 and test the other formation models through detailed atmospheric characterisation.

Following Eq. 40 from \cite{winn2010}, we determined the expected amplitude of the radial velocity variation produced by the RM effect when planet b transits (58 m s$^{-1}$) or when planet c transits (7.0 m s$^{-1}$). Given the activity level of our target (typical RV dispersion: 27 m s$^{-1}$) and considering a planetary transit time scale, that is, when the activity level is significantly lower than its typical amplitude, we predict a perfectly suitable detection of the RM effect caused by TOI-5398 b and a possible detection for TOI-5398 c.


\subsubsection{Circularisation time scale and $\alpha$ parameter}
We calculated the circularisation time scale, denoted as $\tau_{\rm circ}$, for TOI-5398 b, using Eq. 6 from \cite{Matsumura2008}. Assuming a circular orbit and a modified tidal quality factor $Q$ of $10^5$ \citep{2014ARA&A..52..171O}, the calculated value for $\tau_{\rm circ}$ is 2.57 $\pm$ 0.88 Gyr, which significantly exceeds the age of the system. This suggests that the near-circular orbit of TOI-5398 b may well be primordial, indicating favourable conditions for preserving its close planetary companion. 

Following the methodology described in \cite{Bonomo2017}, we computed the parameter $\alpha$, which represents the ratio of the planet's semi-major axis to its Roche limit. \cite{Bonomo2017} conclude that planets with $\alpha$ > 5 and circular orbits are unlikely to undergo high eccentricity migration. With an $\alpha$ value of 5.6, TOI-5398 b falls in the middle $\alpha$ range (4.55 - 7.44) for short-period giants with close companions. This finding favours the notion that short-period giants with close companions should not be distinguished between \textit{hot} (P < 10 days) and \textit{warm} (P > 10 days) planets, that is, they belong to the same population of exoplanets, distinct in turn from the typical giants that experience the high-eccentricity migration scenario. 

\subsubsection{Atmospheric mass-loss}
\label{sec:mass-loss}
We investigated how the planetary masses, radii, and atmospheric mass-loss rates change with time due to photoevaporation and internal heating.

We evaluated the mass-loss rate of the two planets' atmospheres using the hydro-based approximation developed by \cite{kuby+2018a,kuby+2018b}, coupled with the planetary core-envelope model by \cite{LopFor14} and the MESA Stellar Tracks (MIST; \citealt{choi+2016}). For the stellar X-ray emission at different ages, we adopted the analytic description by \cite{Penz08a}, with the current X-ray luminosity, $L_{\rm x} = 10^{29}$ erg s$^{-1}$ in the 5--100 $\AA$ band. This value was derived from the rotation-activity relationships by \cite{Pizz03}, and closely resembles the median X-ray luminosity of Hyades stars. The stellar EUV luminosity (100--920\,\AA) was computed at any given time using the scaling law by \cite{SF22}. The subsequent paragraphs present the outcomes of our forward-in-time simulations, which include the evolution of the XUV irradiation and the planetary structure in response to stellar behaviour. More details on our modelling of atmospheric evaporation are provided in \citet{maggio2022} and \citet{2023A&A...672A.126D}.

We performed several simulations assuming different possible values for the planetary core radius at the current age. The test cases were selected by comparing them with the grids of planetary internal structures by \cite{Fortney2007}, assuming cores composed of 50\% rocks and 50\% ices. For planet b, we explored core radii ranging from 2.4 to 7 $R_{\rm \oplus}$, corresponding to core masses of 10 to 25 $M_\oplus$. As for planet c, due to its smaller size, we limited our analysis to core radii between 1 and 2.5 $R_{\rm \oplus}$, with core masses of $\sim 10 M_\oplus$. In Table \ref{tab:photevaporation}, we show the results of the simulations.

\begin{table*}
    \caption{Atmospheric mass-loss simulations}
    \label{tab:photevaporation}
    \centering
    \begin{tabular}{c| c c |c| c c c c}
        \hline \rule{0pt}{2.3ex} \rule[-1.2ex]{0pt}{0pt}
      Core Radius  & \multicolumn{2}{c|}{Current age} & Evaporation & \multicolumn{4}{c}{$t = 5$\,Gyr}\\
        \cline{2-3}
        \cline{5-8}
	 ($R_\oplus$) & $f_{\rm atm}$(\%) & Mass loss rate (g/s)  & time scale$^a$ & Mass ($M_\oplus$) & Radius ($R_\oplus$) & $f_{\rm atm}$(\%) &  Mass loss rate (g/s) \rule{0pt}{2.3ex} \rule[-1.2ex]{0pt}{0pt} \\         
        \hline
	\multicolumn{8}{c}{Planet b} \rule{0pt}{2.3ex} \rule[-1.2ex]{0pt}{0pt} \\
        \hline \rule{0pt}{2.3ex} \rule[-1.2ex]{0pt}{0pt}
	2.4  & 98.9 &  5.3$\times 10^{11}$ & $> 5$\,Gyr  & 56.7 & 8.9 & 98.8 & 1.6 $\times 1.6^{10}$ \\
	5    & 50.4 &  5.3$\times 10^{11}$ & $> 5$\,Gyr  & 56.5 & 9.3 & 48.5 & 2.0 $\times 1.6^{10}$ \\
	7    & 22.7 &  5.3$\times 10^{11}$ & $> 5$\,Gyr  & 56.4 & 9.5 & 19.6 & 2.2 $\times 1.6^{10}$ \rule{0pt}{2.3ex} \rule[-1.2ex]{0pt}{0pt} \\
        \hline                       
	\multicolumn{8}{c}{Planet c} \rule{0pt}{2.3ex} \rule[-1.2ex]{0pt}{0pt}\\
        \hline		\rule{0pt}{2.3ex} \rule[-1.2ex]{0pt}{0pt}
        1    & 7.0  & 3.5$\times 10^{12}$  & 96\,Myr & 9.7   & 1.4 & 0.4 & 5.2$\times 10^{8}$\\
	1.8  & 3.5  & 3.5$\times 10^{12}$  & 27\,Myr & 10.3 & 1.8 & 0.0 & 0.0 \\
	2.5  & 1.5  & 3.5$\times 10^{12}$  & 8\,Myr & 10.2 & 2.5 & 0.0 & 0.0 \rule{0pt}{2.3ex} \rule[-1.2ex]{0pt}{0pt}\\ 
	\hline      
  \end{tabular}
  \tablefoot{\tablefoottext{a}{The evaporation time scale is equal to the time required for losing half of the atmospheric mass.}}
\end{table*} 

In our reference model for planet b, the core has a radius of $R_{\rm core} = 5 R_{\rm \oplus}$ and a core mass of $M_{\rm core} \sim 29 M_\oplus$, resulting in an atmospheric mass fraction $f_{\rm env}$ of $\sim 50\%$. The current photo-evaporation rate is $\sim 5.3 \times 10^{11}$\,g s$^{-1}$, and the planet will maintain a large envelope mass fraction throughout its main-sequence lifetime, with $f_{\rm env} \sim 48\% M_{\rm p}$ at time $t \sim 5$\,Gyr. Its radius will only be reduced by $\sim 10$\,\%. In the range explored, these results depend little on the assumed characteristics of the core.

Conversely, the evolution of the inner planet is very different due to the smaller distance from the host star, higher equilibrium temperature, and higher high-energy irradiation. Our reference model has a core radius $R_{\rm core} = 1.8 R_{\rm \oplus}$ and a core mass $M_{\rm core} \sim 10 M_\oplus$, resulting in an atmospheric mass fraction $f_{\rm env}$ of $\sim 3.5\% M_{\rm p}$. The current photo-evaporation rate is $\sim 3.5 \times 10^{12}$\,g s$^{-1}$, and the planet will lose its entire envelope in $\lesssim 200$\,Myr from now. The planetary size will decrease to match that of the core. However, a larger core radius implies a smaller atmospheric mass fraction and shorter evaporation time scales. For example, for a core radius $R_{\rm core} = 1 R_{\rm \oplus}$, the planet would keep a residual atmospheric envelope even at $t = 5$\,Gyr.

\subsubsection{TOI-5398's global formation history}
The different masses of the two planets could be either primordial or, as suggested by their different evaporation rates discussed in Sect. \ref{sec:mass-loss}, the result of distinct photoevaporation histories. However, these two scenarios have different implications regarding formation regions and bulk compositions. Our preliminary exploration of the formation tracks of the two planets using the methodology from \cite{2019A&A...622A.202J} highlights the following possibilities. 

For the mass of TOI-5398 c to be primordial or close to the original one, the planet should not have captured significant quantities of disk gas after reaching its pebble isolation mass. This condition is satisfied if the planet had started its formation beyond about ten au and comparatively late ($\sim$ 1 Myr) in the life of its native disk. However, in this scenario, TOI-5398 b should still be able to capture its present gaseous envelope. For this to occur, planet b should have started its formation at an earlier time ($\sim$ 0.1 Myr) than planet c: as a result, planet b would already be close to its current orbit while planet c is forming and migrating, and the two planets would have to cross orbits to reach their current architecture. 
Such an encounter would likely result in a planet-planet scattering event: this scenario would result in higher eccentricities and inclinations than the currently observed ones. Even in the case that the dynamical excitation created by the planet-planet scattering event is removed by the interactions of the two planets with the disk gas, the density of planet c appears too low for a realistic mixture of rock and ice resulting from its growth track  (as a comparison, the density of the ice-rich dwarf planet Pluto is 2 g cm$^{-3}$). 

The other possible scenario results from the two planets having started forming early in the lifetime of their native disk at a few au from their host star. In this case, the simulated growth tracks favour planet c to have possessed an extended primordial atmosphere of comparable mass to that of planet b. In this scenario, the present-day mass disparity between the two planets would result from their different photoevaporation histories, with planet c experiencing a much higher mass loss than its outer counterpart. Their gas accretion phases would have occurred close to their final orbits, in the innermost and hottest regions of the native disk, which suggests that their atmospheres could exhibit stellar composition.

\subsection{Planetary bulk composition prediction}
Accurate data on mass, eccentricity, and radius allow for measuring precise inner bulk densities and exploring the differences in planetary structure and evolution, from inflated Hot Jupiters (HJ) to ``over-dense'' Warm Jupiters (WJs,\citealt{Fortney2021}). While the prediction of hotter interiors and larger radii for HJs \citep{Guillot_1996}, compared to Jupiter, has been proven true, the mechanism(s) behind some HJs having anomalously large radii remain a challenge to explain \citep{Thorngren2018}. For non-inflated giant planets ($F_\star$ < 2 $\times$ 10$^8$ erg\,s$^{-1}$\,cm$^{-2}$), \cite{2016ApJ...831...64T} found a relation between planet mass and bulk metallicity, which confirms a key prediction of the core-accretion planet formation model \citep{Mordasini2014}. A recent study \citep{2023arXiv230412782M} presents the current knowledge of mass-metallicity trends for warm giant exoplanets \citep{Teske2019,2020ApJ...903..147M,2023A&A...669A..24M}, and raises some doubts about its extent and existence. \cite{2023arXiv230412782M} link this ambiguity to theoretical uncertainties on the assumed models and the need for accurate stellar age and atmospheric measurements. 

Understanding this relationship and open questions regarding giants require characterising planets and host stars, focusing on the metal enrichment of planetary atmospheres \citep{2011ApJ...736L..29M}. It is particularly true for warm giants -- scarce among confirmed planets\footnote{Only 20 warm giants have precise bulk densities (density determination better than 20\%) in the NASA Exoplanet Archive} -- unaffected by the radius inflation mechanism, as we can reasonably constrain their bulk metal enrichment and interpret atmospheric features more safely \citep{2016ApJ...831...64T}.

\begin{figure*}
   \centering
   \includegraphics[width=\hsize]{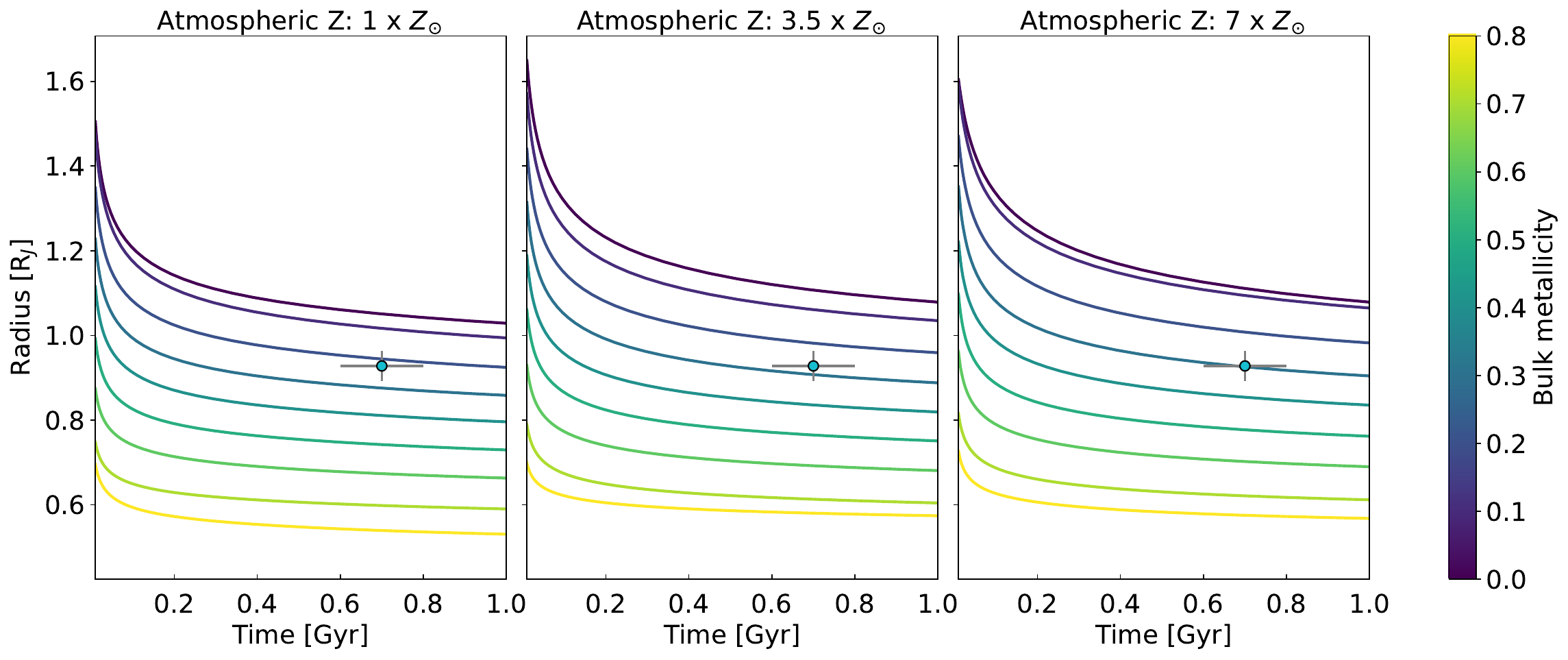}
   \caption{Radius evolution in time ($x$-axis) for various bulk metallicities (coloured lines, in units of planet b masses). The three plots adopt three different atmospheric heavy-element mass fractions. The light-blue dot represents TOI-5398 b.  }
   \label{fig:bulk}
\end{figure*}

TOI-5398 b is a perfect case study if we consider both its low insolation $F_\star$ and its good age estimation. Therefore, we estimated its planetary bulk heavy-element mass fraction (or bulk metallicity), using evolution models\footnote{\url{https://github.com/tiny-hippo/planetsynth/blob/main}} from \cite{2021MNRAS.507.2094M}. Figure \ref{fig:bulk} shows the radius evolution for various bulk metallicities (coloured lines) for three different atmospheric heavy-element mass fractions. The bulk heavy-element mass fraction of TOI-5398 b, depending on the adopted atmospheric metallicity, varies between 20 and 30 per cent of its total mass. Our result follows the mass-metallicity trend from \cite{2016ApJ...831...64T}, but the heavy-element mass appears to be slightly lower than expected. Therefore, we may infer that the trend from \cite{2023A&A...669A..24M}, which predicts a lower heavy-element mass for a given planetary mass compared to \cite{2016ApJ...831...64T}, might offer a more plausible explanation of our result.


\subsection{Future atmospheric characterisation}

To break the degeneracy in determining the planetary bulk composition, it is crucial to perform atmospheric measurements and to get information on metal enrichment. \cite{2023arXiv230412782M} show that atmospheric measurements by JWST and Ariel can significantly reduce this degeneracy and that this is particularly promising for warm giant planets \citep{2023A&A...669A..24M}. The precise characterisation of the TOI-5398 b atmosphere will be crucial to validate or disprove formation and evolution theories.

\begin{figure*}
   \centering
   \includegraphics[width=\hsize]{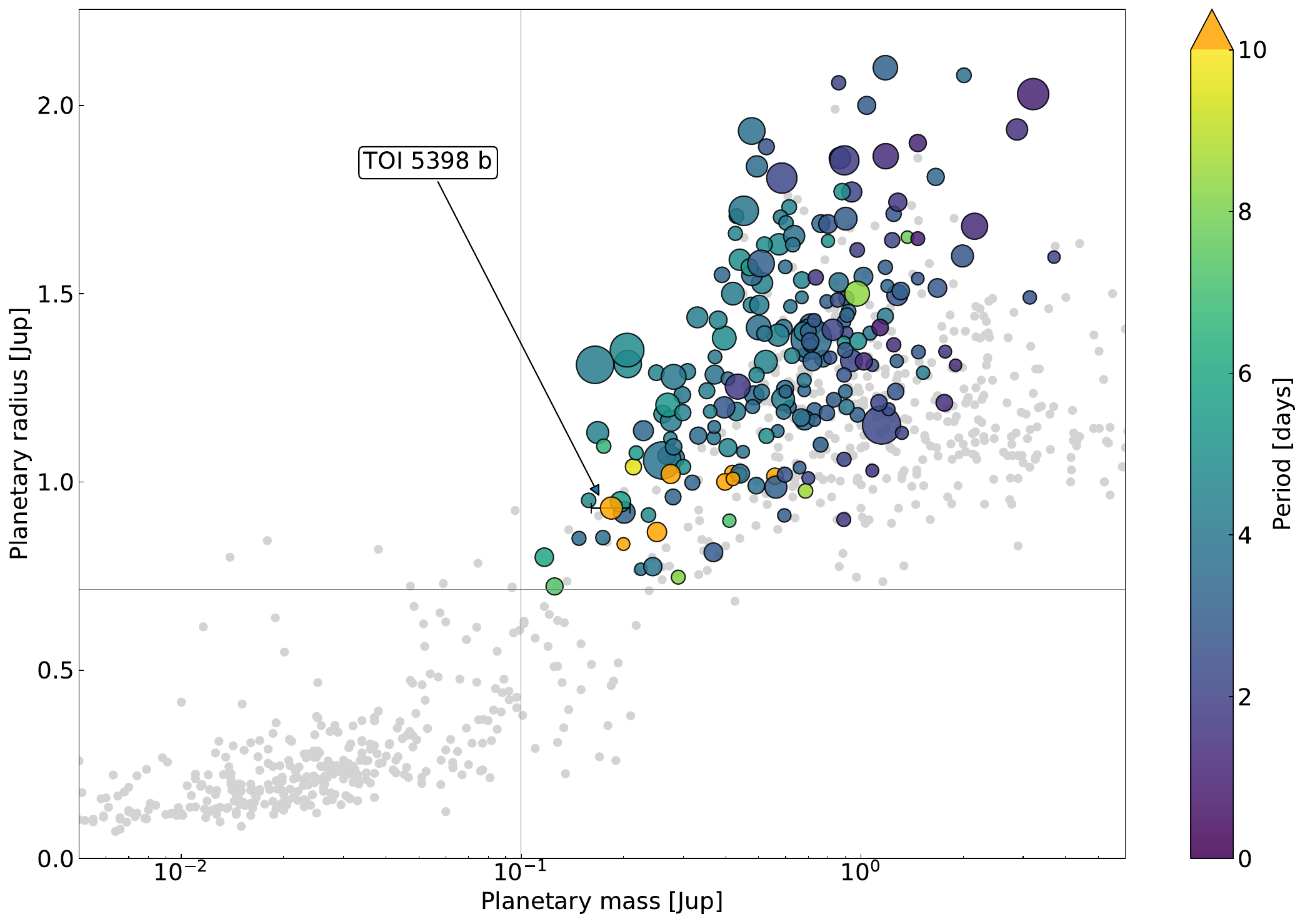}
   \caption{Mass radius distribution of all confirmed planets that are present in the TEPCat catalogue \citep{2011MNRAS.417.2166S}. The orbital period of a planet is colour-coded when it has TSM > 90, $M_{\rm p}$ > 0.1 $M_{\rm J}$, and $R_{\rm p}$ > 8 $R_\oplus$. Planets without these characteristics are coloured grey. The dot size tracks the TSM. The planetary mass is shown on a logarithmic scale. }
   \label{fig:tsm}
\end{figure*}

TOI-5398 is a fascinating compact system, as planet b has the highest Transmission Spectroscopy Metric (TSM, \citealt{2018PASP..130k4401K}) value ($\sim$\,300) among warm giant planets (10 < $P$ < 100, $M_{\rm p}$ > 0.1 $M_{J}$) currently known, making it ideal for atmospheric characterisation by JWST. Indeed, the TSM parameter quantifies the expected signal-to-noise in transmission spectroscopy for a given planet, and according to \cite{2018PASP..130k4401K}, a giant planet's atmosphere is considered amenable to JWST observations when its TSM value is greater than 90. In Fig. \ref{fig:tsm}, we include all confirmed planets, with colour-coding only for those with a TSM > 90, and $M_{\rm p}$ > 0.1 $M_{\rm J}$. In this plot, we show the planetary masses versus radii, where planets are colour-coded by their orbital period. We colour-code planets with periods longer than 10 days in orange to highlight the warm-giant planets, defined as giant planets with periods exceeding 10 days (e.g., \citealt{2021AJ....162..240Y,2023AJ....165...17G} and references therein). Instead, the size of the dots tracks the TSM. For comparison, WASP-47 b has only a modest TSM value of $\sim$ 47 \citep{2022AJ....163..197B}.

\subsection{Characterisation with JWST/NIRSpec}

TOI-5398 b is an ideal candidate for precise transmission spectroscopy and atmospheric characterisation, with a focus on the carbon-to-oxygen (C/O) ratio. This ratio is essential for understanding planetary formation mechanisms and grasping planetary atmosphere composition, shedding light on volatile content and atmospheric chemistry. 
The availability of carbon and oxygen determines different chemical reactions as well as the stability of molecules of planetary atmospheres. Understanding the C/O ratio helps in predicting the composition and behaviour of atmospheric constituents like carbon dioxide (CO2), carbon monoxide (CO), methane (CH4), and water (H2O) \citep{keyte2023}.

To test the feasibility of atmospheric characterisation using JWST, we investigated three different atmospheric scenarios for TOI-5398 b. We assumed equilibrium chemistry as a function of temperature and pressure using FastChem \citep{stock2018} and three different C/O ratios: 0.5, 1.0 and 1.5. We used FastChem within TauREx3 \citep{alrefaie2021} using the \texttt{taurex-fastchem}\footnote{\url{https://pypi.org/project/taurex-fastchem}} plugin. TauREx is a retrieval code that uses a Bayesian approach to infer atmospheric properties from observed data, utilising a forward model to generate synthetic spectra by solving the radiative transfer equation throughout the atmosphere. We used all the possible gases contributions within FastChem and the active absorption contribution given by K, Na, HCN, H2CO, CH4, CO, CO2, C2H2, C2H4, H2O, NH3, SiO, TiO, VO and SO2 opacities.

\begin{figure*}
    \centering
    \includegraphics[width=\hsize]{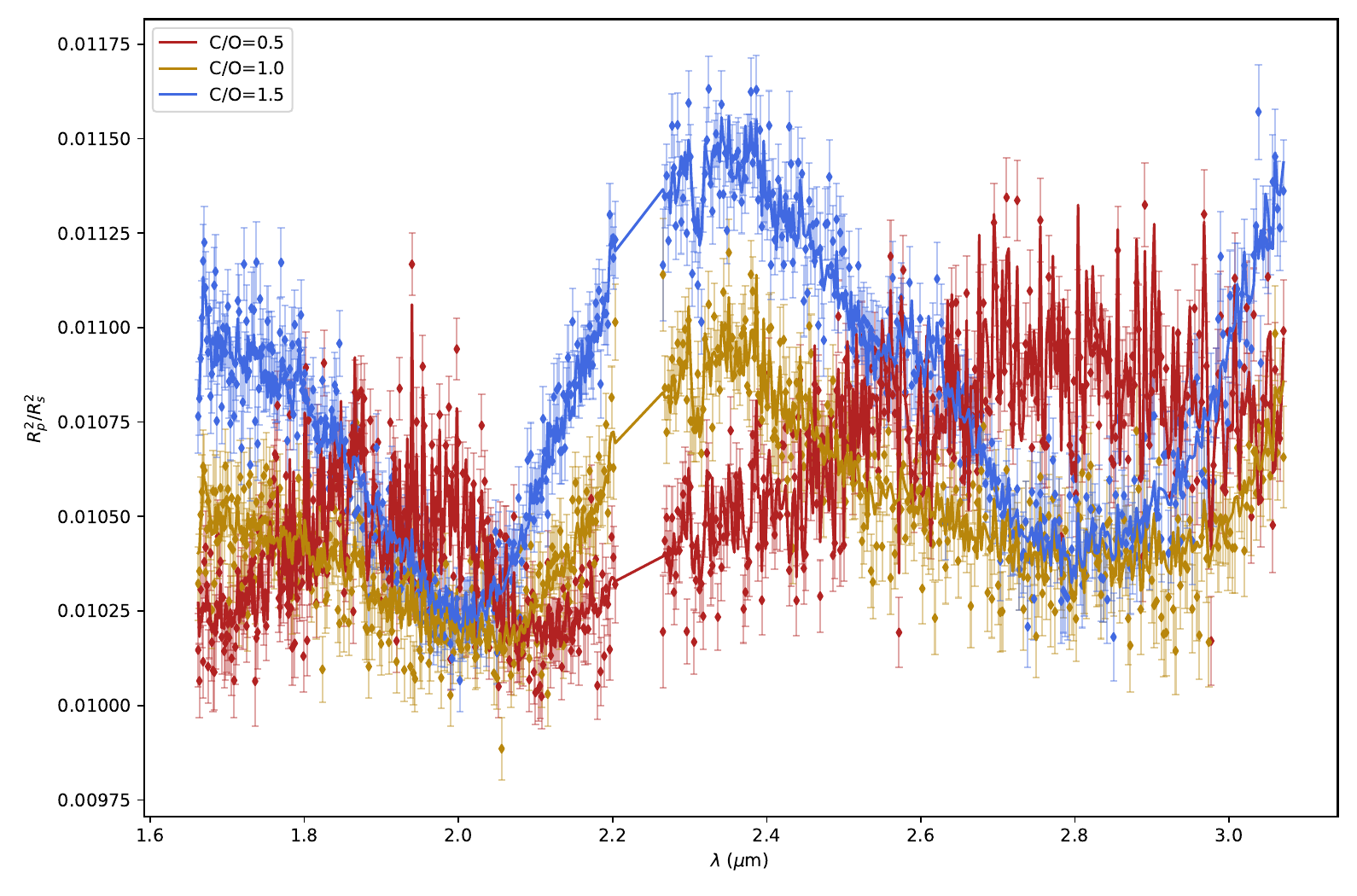}
    \caption{NIRSpec observation simulation using the g235h disperser with f170lp filter (scatter points) and best-fit models from TauREx (lines). The three colours indicate three scenarios: C/O = 0.5 in red, C/O = 1.0 in yellow, and C/O = 1.5 in blue.}
    \label{fig:retrieval-spectra}
\end{figure*}

After generating the transmission spectra using TauREX+FastChem, we simulated a JWST observation using Pandexo \citep{batalha2017}, a software tool specifically developed for the JWST mission. 
The software allows users to model and simulate various atmospheric scenarios, incorporating factors such as atmospheric composition, temperature profiles, and molecular opacities. We simulated a NIRSpec observation in bots mode, using the s1600a1 aperture with g235h disperser, sub2048 subarray, nrsrapid read mode and f170lp filter. We simulated one single transit and an observation 1.75 $T_{14}$ long to ensure a robust baseline coverage. We fixed this instrumental configuration for all three scenarios. In Fig \ref{fig:retrieval-spectra}, we show the resulting spectra for the different C/O ratios and their best-fit models.

We performed three atmospheric retrievals on the NIRSpec/JWST simulations using a Nested Sampling algorithm with the \texttt{nestle}\footnote{\url{https://github.com/kbarbary/nestle}} library with 1000 live points. We used the \texttt{cuda} transmission model with the \texttt{taurex-cuda}\footnote{\url{https://github.com/ucl-exoplanets/taurex_cuda/}} TauREx plugin. We fitted three parameters: the radius of the planet $R_p$, the equilibrium temperature of the atmosphere $T_{\rm eq}$, and the C/O ratios.

Using NIRspec with the g235h disperser wavelength range (1.66\,$\mu$m - 3.07\,$\mu$m), we can assess the C/O ratio under the three assumptions (see Tab \ref{tab:retrieval-results}). In particular, when assuming C/O ratios of 0.5 and 1.0, we can retrieve the correct value within 1$\sigma$ error bar, while under the C/O = 1.5 assumption, we retrieved a value of $1.87 \pm 0.15$, within 2.5$\sigma$ error bar.

\begin{figure}
    \centering
    \includegraphics[width=\hsize]{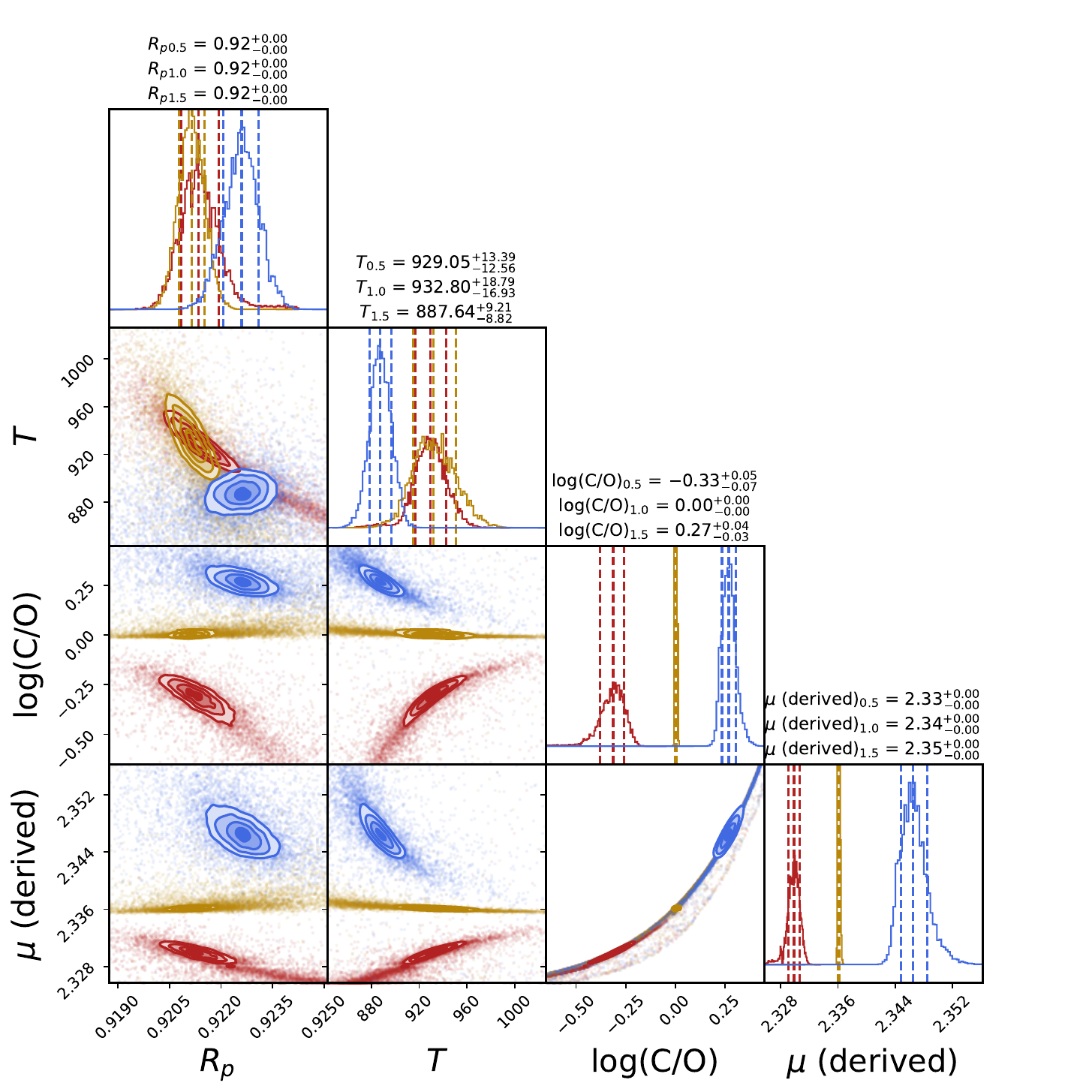}
    \caption{Posterior distributions for the three different scenarios. We show C/O = 0.5 in red, C/O = 1.0 in yellow, and C/O = 1.5 in blue.}
    \label{fig:retrieval-posteriors}
\end{figure}

The three atmospheres can be explained with three distinct sets of parameters (Fig \ref{fig:retrieval-posteriors}). The results of atmospheric retrievals confirm and quantify the feasibility of atmospheric characterisation using NIRSpec@JWST. Furthermore, they demonstrate that TOI-5398 b is an excellent candidate for comprehensive atmospheric analysis, to measure the C/O ratio and, therefore, to constrain planet formation theories for this system.

\begin{table}[h]
\caption{Retrieval results for the three different scenarios.}
\centering
\begin{tabular}{p{1.53cm}| p{1.8cm}| p{1.8cm} |p{1.92cm}}
\hline \hline
Parameter & C/O = 0.5 & C/O = 1.0 & C/O = 1.5 \rule{0pt}{2.3ex} \rule[-1ex]{0pt}{0pt} \\
\hline 
& & &\\
$R_p$ $(R_J)$ &  $0.9214_{-0.0005}^{+0.0006}$ & $0.9211\pm0.0004$ & $0.9226\pm0.0005$ \rule{0pt}{2.3ex} \rule[-1.25ex]{0pt}{0pt} \\

$T_{\rm eq}$ (K) & $929\pm13$ & $932_{-17}^{+19}$ & $887 \pm 9$ \rule{0pt}{2.3ex} \rule[-2.2ex]{0pt}{0pt} \\

C/O & $0.47_{-0.07}^{+0.06}$ & $1.005_{-0.008}^{+0.009}$ & $1.87 \pm 0.15$ \rule{0pt}{2.3ex} \rule[-2.7ex]{0pt}{0pt} \\

$\mu$ (derived) & $2.3298\pm0.0008$ &  $2.3361\pm0.0001$ & $2.3467\pm0.0019$ \rule{0pt}{2.3ex} \rule[-1.5ex]{0pt}{0pt} \\
\hline
\end{tabular}
\label{tab:retrieval-results}
\end{table}

\section{Conclusions}
\label{sec:conclusions}  

In this study, we presented the discovery of the youngest transiting planetary system containing a sub-Neptune  planet orbiting interior to a Saturn-mass planet with $P <$ 15 days. Using HARPS-N radial velocity measurements of the host star TOI-5398 and multidimensional Gaussian processes, we modelled the stellar activity and confirmed the planetary nature of both candidates identified in the \textit{TESS} light curve measuring their masses. Furthermore, our methodology allowed us to accurately determine the stellar parameters. 

With a Transmission Spectroscopy Metric value of around 300, the warm Saturn TOI-5398 b is the most suitable warm giant planet for atmospheric characterisation using JWST. Such investigations are crucial to validate or disprove existing formation and evolution theories. By measuring atmospheric chemistry, we can gain information on metal enrichment and effectively break the degeneracy in determining the planetary bulk composition.

The presence of two planets in this relatively young system offers the opportunity to examine distinct evolutionary paths over the initial hundreds of millions of years following system formation, under the influence of the same host star. In this study, we provided a characterisation of the system, with a special focus on the future evolution of the planetary atmospheres. Future works will focus on investigating the past evolution of the system. We explored the evolution of the atmospheres of both planets, considering the decay of stellar activity and XUV irradiation with time. We estimated that given reasonable assumptions regarding core radius and mass, planet b probably retains a substantial atmosphere, with a mass fraction $\sim 0.5 M_{\rm p}$, making it amenable to be probed with transmission spectroscopy. Conversely, planet c is expected to possess a tiny atmospheric envelope. At the current age, the mass loss rate of planet c exceeds that of planet b by a factor of 7, implying that planet c will completely lose its residual atmosphere within a few hundred million years, while planet b will retain a thick atmosphere even at the solar age.

Notably, \textit{TESS} observed TOI-5398 during sector 48 and no further observations are planned in the current Extended Mission. Consequently, our ground-based light curves play a pivotal role in refining the ephemeris of both planets. The improved ephemeris we calculated are vital for future follow-up observations and surveys, including those conducted by CHEOPS, JWST, and upcoming missions such as PLATO and Ariel, along with telescopes like ELTs. 

\begin{acknowledgements}
We are extremely grateful to the anonymous referee for their meticulous feedback, which undoubtedly improved the quality of this manuscript. Based on observations collected at Copernico 1.82m telescope (Asiago Mount Ekar, Italy) INAF - Osservatorio Astronomico di Padova. This work has made use of data from the European Space Agency (ESA) mission {\it Gaia} (\url{https://www.cosmos.esa.int/gaia}), processed by the {\it Gaia} Data Processing and Analysis Consortium (DPAC, \url{https://www.cosmos.esa.int/web/gaia/dpac/consortium}). Funding for the DPAC has been provided by national institutions, in particular, the institutions participating in the {\it Gaia} Multilateral Agreement. This article is based on observations made with the MuSCAT2 instrument, developed by ABC, at Telescopio Carlos S\'{a}nchez operated on the island of Tenerife by the IAC in the Spanish Observatorio del Teide. This work has been supported by the PRIN-INAF 2019 ``Planetary systems at young ages (PLATEA)''. This work is partly financed by the Spanish Ministry of Economics and Competitiveness through grants PGC2018-098153-B-C31. This work is partly supported by JSPS KAKENHI Grant Numbers JP17H04574, JP18H01265, and JP18H05439, Grant-in-Aid for JSPS Fellows Grant Number JP20J21872, JST PRESTO Grant Number JPMJPR1775, and a University Research Support Grant from the National Astronomical Observatory of Japan (NAOJ). This work makes use of observations from the LCOGT network. Part of the LCOGT telescope time was granted by NOIRLab through the Mid-Scale Innovations Program (MSIP). MSIP is funded by NSF. This research has made use of the Exoplanet Follow-up Observation Program (ExoFOP; DOI: 10.26134/ExoFOP5) website, which is operated by the California Institute of Technology, under contract with the National Aeronautics and Space Administration under the Exoplanet Exploration Program. Funding for the \textit{TESS} mission is provided by NASA's Science Mission Directorate. KAC and CNW acknowledge support from the \textit{TESS} mission via subaward s3449 from MIT. TZi acknowledges NVIDIA Academic Hardware Grant Program for the use of the Titan V GPU card. This work is partly supported by JST CREST Grant Number JPMJCR1761. The postdoctoral fellowship of KB is funded by F.R.S.-FNRS grant T.0109.20 and by the Francqui Foundation. D.V.C. and I.A.S. acknowledge the support of M.V. Lomonosov Moscow State University Program of Development. R.L. acknowledges funding from University of La Laguna through the Margarita Salas Fellowship from the Spanish Ministry of Universities ref. UNI/551/2021-May 26, and under the EU Next Generation funds. This work is partly supported by the CHEOPS ASI-INAF agreement n. 2019-29-HH.0. G.N. thanks for the research funding from the Ministry of Education and Science programme the ``Excellence Initiative - Research University'' conducted at the Centre of Excellence in Astrophysics and Astrochemistry of the Nicolaus Copernicus University in Toru\'n, Poland. We acknowledge the use of public TESS data from pipelines at the TESS Science Office and at the TESS Science Processing Operations Center. Resources supporting this work were provided by the NASA High-End Computing (HEC) Program through the NASA Advanced Supercomputing (NAS) Division at Ames Research Center for the production of the SPOC data products. AM, DL, DT and DP acknowledge partial financial support from the ARIEL ASI-INAF agreement n.2021-5-HH.0. AM also acknowledges ``The HOT-ATMOS Project'' (PRIN-INAF 2019). DP acknowledges the support from the Istituto Nazionale di Oceanografia e Geofisica Sperimentale (OGS) and CINECA through the program ``HPC-TRES (High Performance Computing Training and Research for Earth Sciences)'' award number 2022-05. P.K.
acknowledges the funding from a grant LTT-20015.

\end{acknowledgements}



\bibliographystyle{aa}
\bibliography{references} 

\appendix
\section{Statistical validation}
\label{app:stat}
\begin{figure}
   \centering
   \includegraphics[width=\hsize]{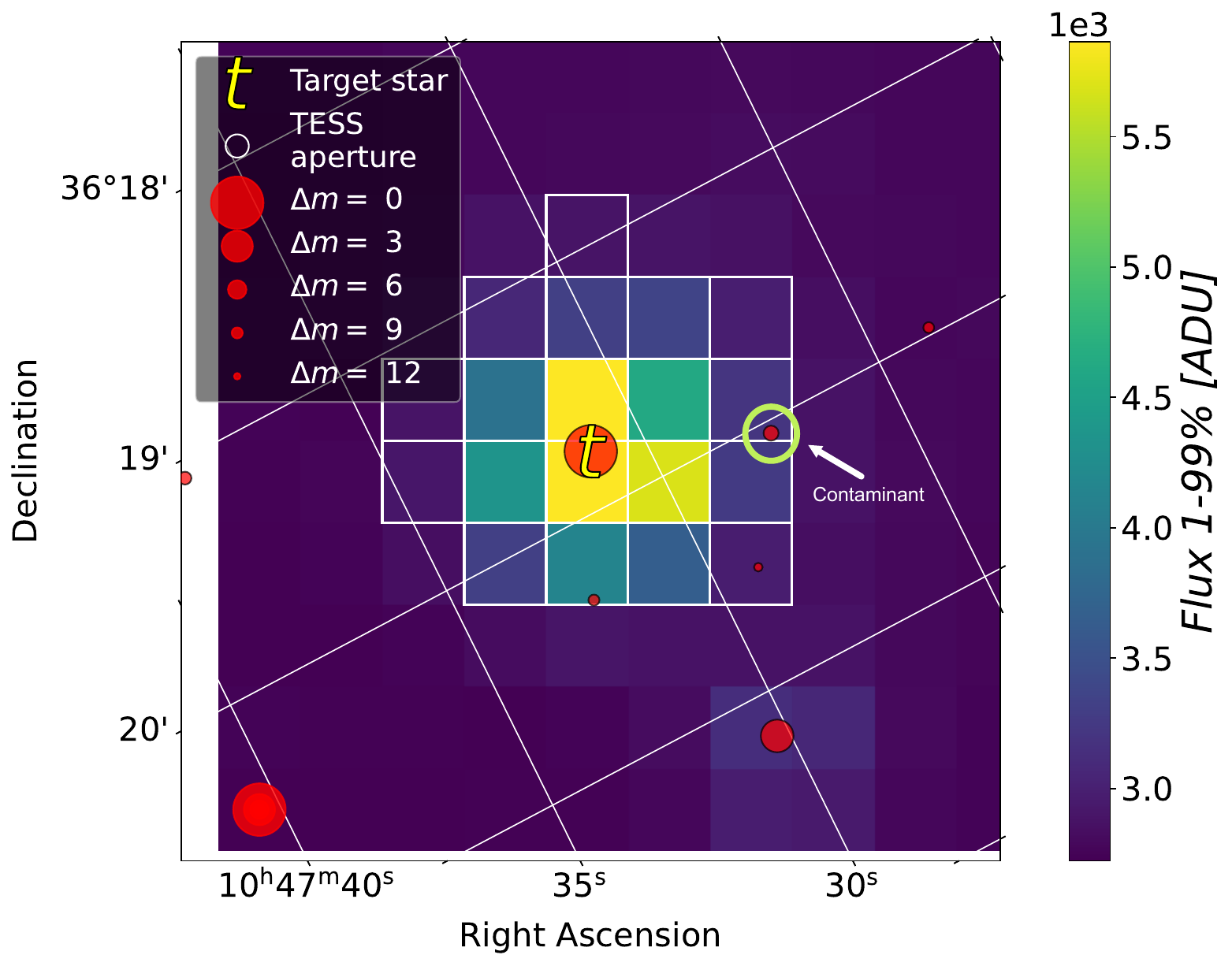}
   \caption{\textit{Gaia} stars identified in the \textit{TESS} field. The letter ``t'' denotes the position of TOI-5398, while the position of the potential contaminant star is enclosed in a green circle. }
   \label{fig:gaia_phot}
\end{figure}

\begin{figure}
   \centering
   \includegraphics[width=\hsize,trim=4 4 4 4,clip]{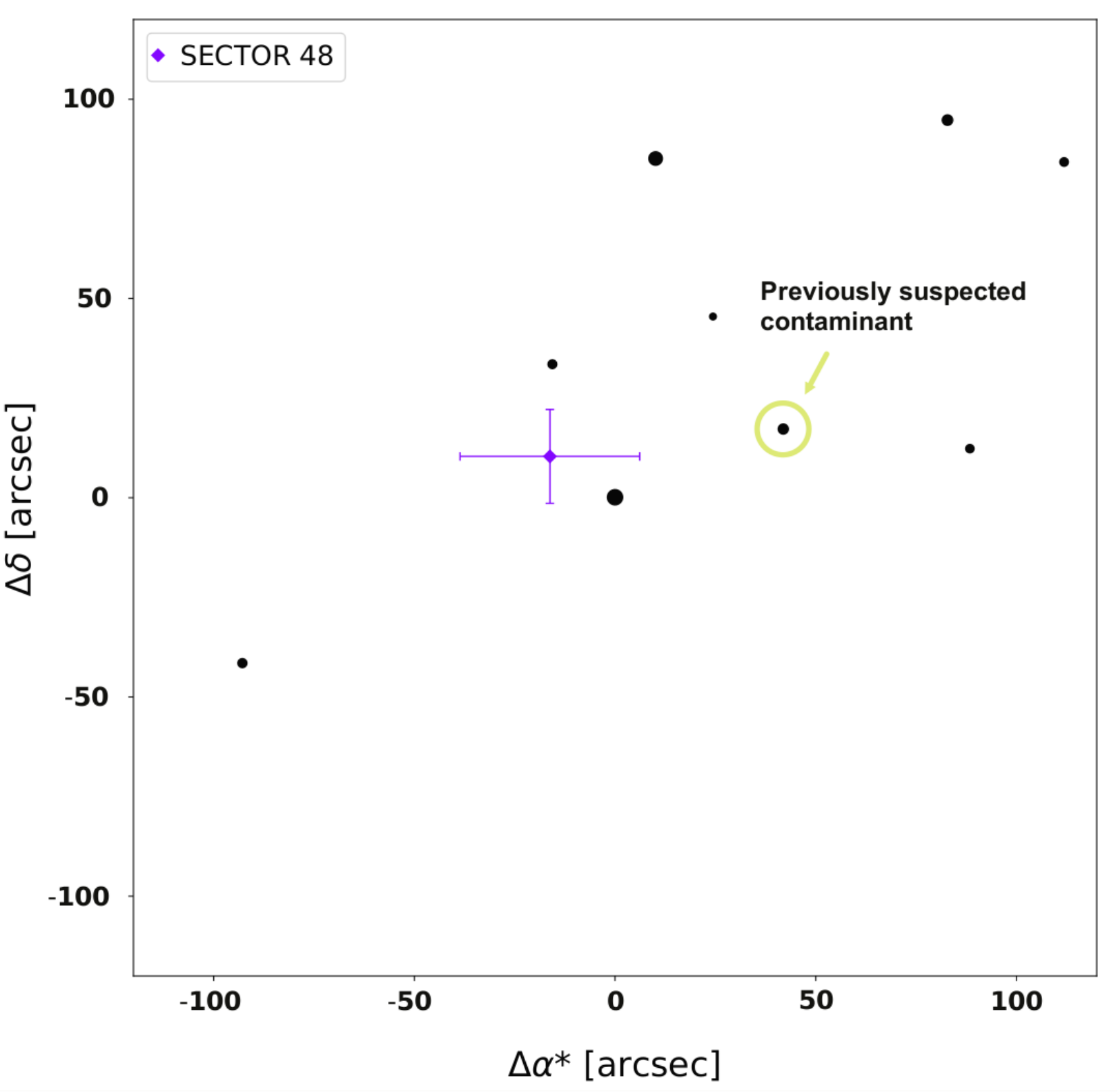}
   \caption{Computation of the in- / out-of-transit difference centroid test for Sector 48 (violet dot and error bars). The potential contaminant star's location is instead surrounded by a green circle. }
   \label{fig:centroid}
\end{figure}

In the first step of our vetting and validation study, we used \textit{Gaia} EDR3 data to identify nearby contaminating stars that might be blended eclipsing binaries (BEBs). This analysis shows that besides TOI-5398, there is a star (\textit{Gaia} DR2 750888586899153536) at 45.3$\arcsec$ separation from TOI-5398 that might reproduce the transit signal of TOI-5398.02, see Fig. \ref{fig:gaia_phot}. To ensure that the transits of TOI-5398.02 are genuine and not caused by contaminant neighbours, we carried out the in- / out-of-transit difference centroid check outlined by \cite{2020MNRAS.495.4924N} and \cite{2020MNRAS.498.5972N}. We show the outcomes in Fig. \ref{fig:centroid}. The in- / out-of-transit mean difference centroids are consistent with TOI-5398's position and far from any potential contaminants, further proving the planetary nature of TOI-5398.02.

To corroborate that TOI-5398.02 is not an FP, we used the VESPA\footnote{\url{https://github.com/timothydmorton/VESPA}} software \citep{Morton_2012,2015ascl.soft03010M} as a final check. We have followed the procedure adopted in \cite{2022MNRAS.516.4432M}, which takes into account the major issues reported in \cite{2023RNAAS...7..107M} and allows us to get reliable results while using VESPA. We used our detrended light curve (see Sect. 2.1), which we normalised using \textsc{wotan}, and we phase-folded after removing TOI-5398 b's signal identified through the TLS analysis (see Sect. 4.1.1). We find a 100\% probability of having a keplerian transiting companion around TOI-5398, while the probability of an FP is of the order of $\sim$ 9 x 10$^{-4}$. It is important to note that candidates in multi-planet systems have a higher probability of being genuine planets \citep{2011ApJ...732L..24L,2012ApJ...750..112L}. Therefore, our FPP (False Positive Probability) should be even lower. All these analyses show that TOI-5398.02 is completely statistically vetted and only needs to go through reconnaissance spectroscopy to be promoted to the status of a \textit{statistically validated} planetary companion \citep{2022MNRAS.516.4432M}. As a result of the extensive HARPS-N spectroscopy examination of the planetary system TOI-5398, the sub-Neptune exoplanet is statistically validated and will henceforth be labelled as TOI-5398 c.

\begin{figure}
   \centering
   \includegraphics[width=\hsize]{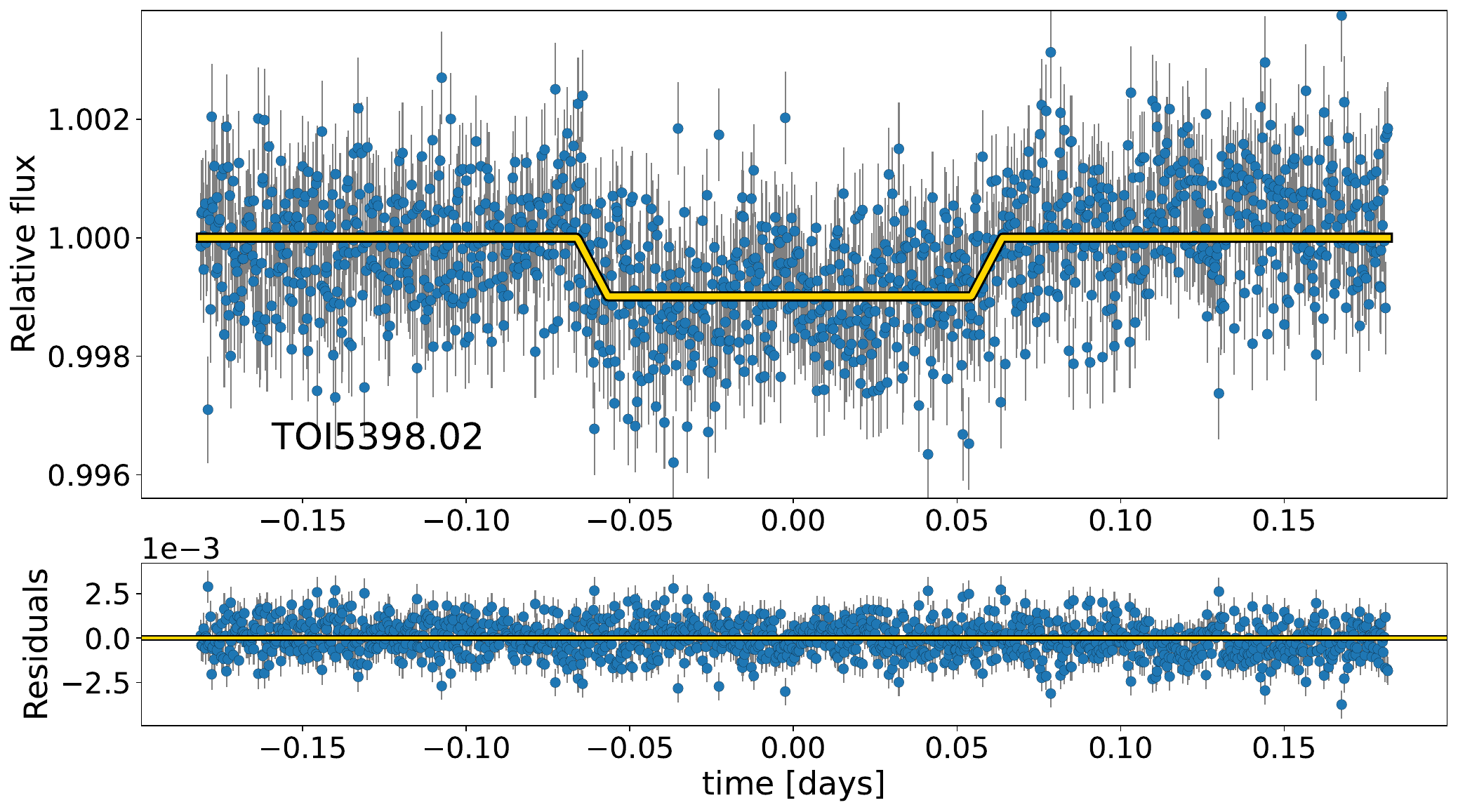}
   \caption{SAP-corrected light curve of TOI-5398 c that has been detrended, normalised, and phase-folded. The gold line is a trapezoidal fitting model produced by VESPA
   .}
   \label{fig:model}
\end{figure}


High angular resolution observations described in Sect.~\ref{sec:highres_obs} rule out -- down to the sensitivity of the data -- the presence of stellar companion which could have been missed by {\it Gaia} photometry. The detection limits for companions ($\Delta I_c$ mag vs separation) were translated into physical properties of companions of the star, using the expected V-I color from the $T_{\rm eff}$ of the target and the stellar models by \cite{2015A&A...577A..42B} for the suitable age of the system. The mass detection limits span from 0.65 $M_\odot$ at 13 au (projected separation) to 0.28 $M_\odot$ at 50 au, and 0.17 $M_\odot$ at 100 au.

\section{Combined analysis of HARPS-N, TRES, and McDonald datasets}
\subsection{Observations and data reduction of TRES data}
Spectra of TOI-5398 were obtained between UT March 28, 2022 - April 19, 2022 using the Tillinghast Reflector Echelle Spectrograph (TRES; \citealt{gaborthesis}) mounted on the 1.5m Tillinghast Reflector telescope at the Fred Lawrence Whipple Observatory (FLWO) atop Mount Hopkins, Arizona. TRES is an optical, fibre-fed echelle spectrograph with a wavelength range of 390-910nm and a resolving power of R~44,000. The TRES spectra were extracted as described in \cite{2010ApJ...720.1118B}. The spectra were visually inspected to check that there were no additional stellar companions contaminating the radial velocities. Then, a multi-order relative velocity analysis was performed by cross-correlating the strongest signal-to-noise observed spectrum as a template, order by order, against the remaining spectra to produce an orbit.
\subsection{Modelling of HARPS-N, TRES, and McDonald RVs}
\begin{figure}
   \centering
   \includegraphics[width=\hsize]{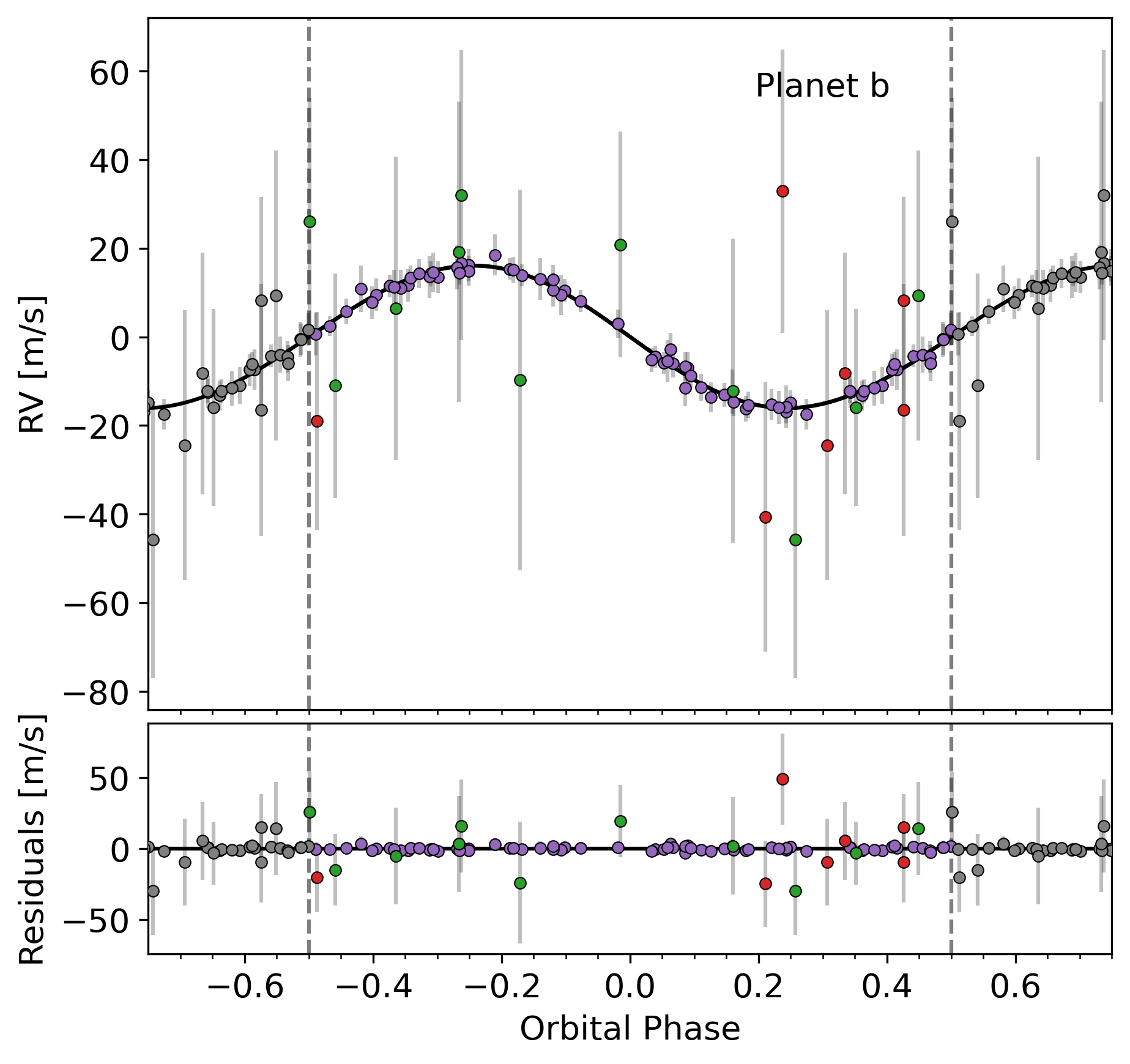}
   \caption{Phase-folded RV fit of TOI-5398 b planetary signal. The three different colours represent HARPS-N (violet), TRES (green), and McDonald 2.7m (red) data. The reported error bars include the jitter term, added in quadrature. The bottom panel displays the residuals of the fit.
   }
   \label{fig:rvb_best_tfop}
\end{figure}

\begin{figure}
   \centering
   \includegraphics[width=\hsize]{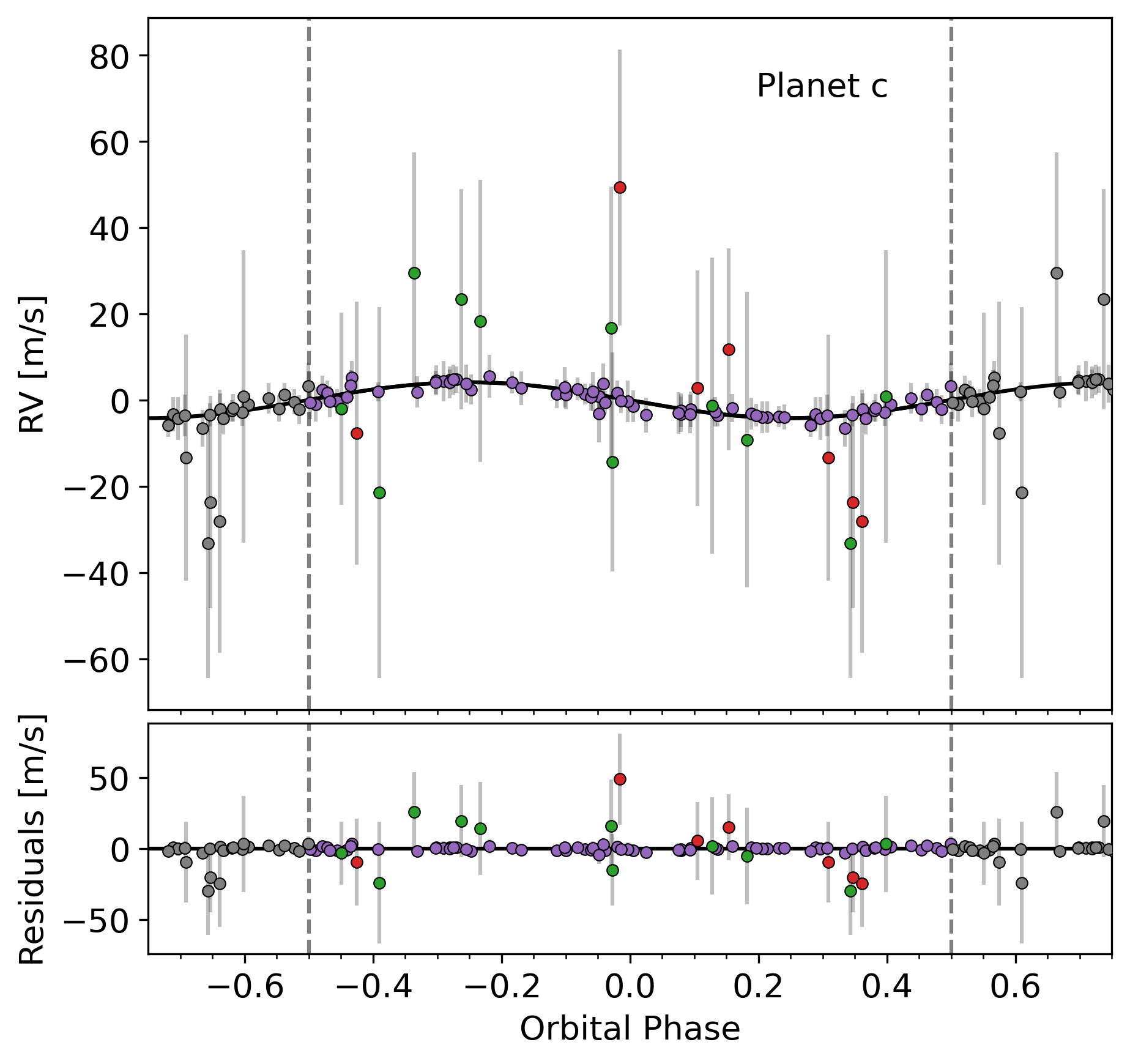}
   \caption{As in Fig. \ref{fig:rvb_best_tfop}, but for planet c.}
   \label{fig:rvc_best_tfop}
\end{figure}

We conducted a modelling analysis using \texttt{PyORBIT} on the HARPS-N, TRES, and McDonald data, by using the same configuration and priors as described in Sect. \ref{sec:rv_analysis} (Case 3). The results of this modelling are reported in Tables \ref{table:model-rv-best_tfop} and \ref{table:model-rv-best_tfop2}, in Fig. \ref{fig:rvb_best_tfop} and \ref{fig:rvc_best_tfop}. Due to the considerable errors associated (see their associated jitters) with these measurements and less precise activity indexes, the TRES and McDonald RVs were not used in determining the planet parameters, as we cannot reliably determined their GP parameters (see their $V_c$, $V_r$ coefficients and error bars). Consequently, we decided to exclude them from the analysis conducted in our study.

\begin{table*}
\caption{ Priors and outcomes of the model of planet b and c from the analysis of combined HARPS-N, TRES, and McDonald spectroscopic series.}
\label{table:model-rv-best_tfop}      
\centering
\begin{tabular}{l c c c}     
\hline\hline     

 \multicolumn{4}{c}{GP framework parameters} \rule{0pt}{2ex} \rule[-1ex]{0pt}{0pt} \\ 
\hline    
Parameter & Unit & Prior & Value \rule{0pt}{2.3ex} \rule[-1ex]{0pt}{0pt}\\ 
\hline    
   Uncorrelated RV jitter HARPS-N ($\sigma^{\rm RV}_{\rm jitter,0}$) & m s$^{-1}$ & ... & 1.28$^{+1.4}_{-0.89}$ \rule{0pt}{2.3ex} \rule[-0.8ex]{0pt}{0pt}\\  
   RV offset HARPS-N ($\gamma^{\rm RV}_{0}$) & m s$^{-1}$ & ... & -9435.1$^{+3.7}_{-3.9}$ \rule{0pt}{2.3ex} \rule[-0.8ex]{0pt}{0pt}\\
   Uncorrelated RV jitter McDonald ($\sigma^{\rm RV}_{\rm jitter,0}$) & m s$^{-1}$ & ... & 26$^{+29}_{-18}$ \rule{0pt}{2.3ex} \rule[-0.8ex]{0pt}{0pt}\\  
   RV offset McDonald ($\gamma^{\rm RV}_{0}$) & m s$^{-1}$ & ... & 27778 $\pm$ 20 \rule{0pt}{2.3ex} \rule[-0.8ex]{0pt}{0pt}\\
   Uncorrelated RV jitter TRES ($\sigma^{\rm RV}_{\rm jitter,0}$) & m s$^{-1}$ & ... & 11.3$^{+13}_{-7.8}$ \rule{0pt}{2.3ex} \rule[-0.8ex]{0pt}{0pt}\\  
   RV offset TRES ($\gamma^{\rm RV}_{0}$) & m s$^{-1}$ & ... & -6$^{+23}_{-20}$ \rule{0pt}{2.3ex} \rule[-0.8ex]{0pt}{0pt}\\
   Uncorrelated BIS jitter ($\sigma^{\rm BIS}_{\rm jitter,0}$) & m s$^{-1}$ & ... & 17.3$^{+1.8}_{-1.6}$ \rule{0pt}{2.3ex} \rule[-0.8ex]{0pt}{0pt}\\
   BIS offset ($\gamma^{\rm BIS}_{0}$) & m s$^{-1}$ & ... & 10.9$^{+3.1}_{-3.0}$ \rule{0pt}{2.3ex} \rule[-0.8ex]{0pt}{0pt}\\
   Uncorrelated log$R^{'}_{HK}$ jitter ($\sigma^{\rm logR^{'}_{HK}}_{\rm jitter,0}$) &  & ... & 0.0132$^{+0.0015}_{-0.0014}$ \rule{0pt}{2.3ex} \rule[-0.8ex]{0pt}{0pt}\\
   log$R^{'}_{HK}$ offset ($\gamma^{\rm logR^{'}_{HK}}_{0}$) &  & ... & -4.4165$\pm $0.0035 \rule{0pt}{2.3ex} \rule[-0.8ex]{0pt}{0pt}\\
   Uncorrelated H${\alpha}$ jitter ($\sigma^{\rm H{\alpha}}_{\rm jitter,0}$) &  & ... & 0.0018$\pm $0.0002 \rule{0pt}{2.3ex} \rule[-0.8ex]{0pt}{0pt}\\
   H${\alpha}$ offset ($\gamma^{\rm H\alpha}_{0}$) &  & ... & 0.1513$\pm $0.0006 \rule{0pt}{2.3ex} \rule[-0.8ex]{0pt}{0pt}\\
   Uncorrelated FWHM jitter ($\sigma^{\rm FWHM}_{\rm jitter,0}$) & km s$^{-1}$ & ... & 0.043$\pm $0.004 \rule{0pt}{2.3ex} \rule[-0.8ex]{0pt}{0pt}\\
   FWHM offset ($\gamma^{\rm FWHM}_{0}$) & km s$^{-1}$ & ... & 11.36$\pm $0.01 \rule{0pt}{2.3ex} \rule[-0.8ex]{0pt}{0pt}\\
   Uncorrelated $W_{\rm CCF}$ jitter ($\sigma^{W_{\rm CCF}}_{\rm jitter,0}$) & km s$^{-1}$ & ... & 0.0014$^{+0.0014}_{-0.0009}$ \rule{0pt}{2.3ex} \rule[-0.8ex]{0pt}{0pt}\\
   $W_{\rm CCF}$ offset ($\gamma^{W_{\rm CCF}}_{0}$) & km s$^{-1}$ & ... & 3.503$\pm $0.004 \rule{0pt}{2.3ex} \rule[-0.8ex]{0pt}{0pt}\\
   $V_c$ HARPS-N & m s$^{-1}$ & $\mathcal{U}$(-100.0, 100.0) & -13.4$^{+3.1}_{-3.7}$ \rule{0pt}{2.3ex} \rule[-0.8ex]{0pt}{0pt}\\ 
   $V_r$ HARPS-N & m s$^{-1}$ & $\mathcal{U}$(0.0, 100.0) & 31.3$^{+6.7}_{-4.6}$ \rule{0pt}{2.3ex} \rule[-0.8ex]{0pt}{0pt}\\ 
   $V_c$ McDonald & m s$^{-1}$ & $\mathcal{U}$(-100.0, 100.0) & 2$^{+30}_{-31}$ \rule{0pt}{2.3ex} \rule[-0.8ex]{0pt}{0pt}\\ 
   $V_r$ McDonald & m s$^{-1}$ & $\mathcal{U}$(0.0, 100.0) & 56$^{+27}_{-28}$ \rule{0pt}{2.3ex} \rule[-0.8ex]{0pt}{0pt}\\ 
   $V_c$ TRES & m s$^{-1}$ & $\mathcal{U}$(-100.0, 100.0) & -13$^{+22}_{-24}$ \rule{0pt}{2.3ex} \rule[-0.8ex]{0pt}{0pt}\\ 
   $V_r$ TRES & m s$^{-1}$ & $\mathcal{U}$(0.0, 100.0) & 18$^{+21}_{-12}$ \rule{0pt}{2.3ex} \rule[-0.8ex]{0pt}{0pt}\\ 
   $B_c$ & m s$^{-1}$ & $\mathcal{U}$(-100.0, 100.0) & 6.1$^{+4.2}_{-3.8}$ \rule{0pt}{2.3ex} \rule[-0.8ex]{0pt}{0pt}\\ 
   $B_r$ & m s$^{-1}$ & $\mathcal{U}$(-100.0, 100.0) & -40.7$^{+6.2}_{-8.0}$ \rule{0pt}{2.3ex} \rule[-0.8ex]{0pt}{0pt}\\ 
   $L_c$ (log$R^{'}_{HK}$) &  & $\mathcal{U}$(-0.1, 0.1) &  -0.012$\pm$0.003 \rule{0pt}{2.3ex} \rule[-0.8ex]{0pt}{0pt}\\ 
   $L2_c$ (H${\alpha}$) &  & $\mathcal{U}$(-0.1, 0.1) & -0.0021$^{+0.0004}_{-0.0005}$ \rule{0pt}{2.3ex} \rule[-1ex]{0pt}{0pt}\\ 
   $L3_c$ (FWHM) & km s$^{-1}$ & $\mathcal{U}$(-0.5, 0.5) & -0.046$^{+0.009}_{-0.011}$ \rule{0pt}{2.3ex} \rule[-1ex]{0pt}{0pt}\\ 
   $L4_c$ ($W_{\rm CCF}$) & km s$^{-1}$ & $\mathcal{U}$(-0.02, 0.02) & -0.015$^{+0.002}_{-0.003}$ \rule{0pt}{2.3ex} \rule[-1ex]{0pt}{0pt}\\ 
\hline         
\multicolumn{4}{c}{Stellar activity} \rule{0pt}{2.3ex} \rule[-1ex]{0pt}{0pt}\\ 
\hline    
Parameter & Unit & Prior & Value \rule{0pt}{2.3ex} \rule[-1ex]{0pt}{0pt}\\ 
\hline 
   Rotational period ($P_{\rm rot}$) & days & $\mathcal{N}$(7.18, 0.21) & 7.42$\pm$0.04 \rule{0pt}{2.3ex} \rule[-1ex]{0pt}{0pt}\\
   Decay Timescale of activity ($P_{\rm dec}$) & days & $\mathcal{U}$(10.0, 2000.0) & 26.8$^{+3.3}_{-3.1}$ \rule{0pt}{2.3ex} \rule[-1ex]{0pt}{0pt}\\
   Coherence scale ($w$) &  & $\mathcal{U}$(0.01, 0.60) & 0.40$^{+0.04}_{-0.03}$ \rule{0pt}{2.3ex} \rule[-1ex]{0pt}{0pt}\\
\hline
\end{tabular}
\end{table*}

\begin{table*}
\caption{Continuation of Table~\ref{table:model-rv-best_tfop}.}
\label{table:model-rv-best_tfop2}      
\centering
\begin{tabular}{l c c c}     
\hline\hline     
\multicolumn{4}{c}{Planet b} \rule{0pt}{2.3ex} \rule[-1ex]{0pt}{0pt}\\ 
\hline    
Parameter & Unit & Prior & Value \rule{0pt}{2.3ex} \rule[-1ex]{0pt}{0pt}\\ 
\hline 
   Orbital period ($P_{\rm b}$) & days & $\mathcal{N}$(10.59049, 0.00002) & 10.59049$\pm$0.00002 \rule{0pt}{2.3ex} \rule[-1ex]{0pt}{0pt}\\
   Central time of the first transit ($T_{\rm 0,b}$) & BTJD & $\mathcal{N}$(2616.4921, 0.0003) & 2616.4921$\pm$0.0003 \rule{0pt}{2.3ex} \rule[-1ex]{0pt}{0pt}\\
   Orbital eccentricity ($e_{\rm b}$) &  & $\mathcal{N}$(0.00, 0.098) & 0.05$^{+0.06}_{-0.04}$ \rule{0pt}{2.3ex} \rule[-1ex]{0pt}{0pt}\\
   Argument of periastron ($\omega_{\rm b}$) & deg & ... & 79$^{+64}_{-105}$ \rule{0pt}{2.2ex} \rule[-0.9ex]{0pt}{0pt}\\
   Semi-major axis to stellar radius ratio ($a_{\rm b}/R_{\star}$) &  & ... & 19.9$\pm$1.2 \rule{0pt}{2.3ex} \rule[-1ex]{0pt}{0pt}\\
   Orbital semi-major axis ($a_{\rm b}$) & au & ... & 0.098$^{+0.004}_{-0.004}$ \rule{0pt}{2.3ex} \rule[-1ex]{0pt}{0pt}\\
   RV semi-amplitude ($K_{\rm b}$) & m s$^{-1}$ & $\mathcal{U}$(0.01, 100.0) & 16.1$^{+1.5}_{-1.5}$ \rule{0pt}{2.3ex} \rule[-1ex]{0pt}{0pt}\\
   Planetary mass ($M_{\rm p,b}$) & $M_{\oplus}$ & ... & 58.8$^{+7.7}_{-7.3}$ \rule{0pt}{2.3ex} \rule[-1.5ex]{0pt}{0pt}\\
\hline
\multicolumn{4}{c}{Planet c} \rule{0pt}{2.3ex} \rule[-1ex]{0pt}{0pt}\\ 
\hline    
Parameter & Unit & Prior & Value \rule{0pt}{2.3ex} \rule[-1ex]{0pt}{0pt}\\ 
\hline 
   Orbital period ($P_{\rm c}$) & days & $\mathcal{N}$(4.7734, 0.0004) & 4.7734$\pm$0.0004 \rule{0pt}{2.3ex} \rule[-1ex]{0pt}{0pt}\\
   Central time of the first transit ($T_{\rm 0,c}$) & BTJD & $\mathcal{N}$(2628.6188, 0.0010) & 2628.6188$\pm$0.0010 \rule{0pt}{2.3ex} \rule[-1ex]{0pt}{0pt}\\
   Orbital eccentricity ($e_{\rm c}$) &  & $\mathcal{N}$(0.00, 0.098) & 0.07$^{+0.07}_{-0.05}$ \rule{0pt}{2.3ex} \rule[-1ex]{0pt}{0pt}\\
   Argument of periastron ($\omega_{\rm c}$) & deg & ... & 132$^{+93}_{-142}$ \rule{0pt}{2.2ex} \rule[-0.9ex]{0pt}{0pt}\\
   Semi-major axis to stellar radius ratio ($a_{\rm c}/R_{\star}$) &  & ... & 11.7$\pm$0.7 \rule{0pt}{2.3ex} \rule[-1ex]{0pt}{0pt}\\
   Orbital semi-major axis ($a_{\rm c}$) & au & ... & 0.058$^{+0.002}_{-0.003}$ \rule{0pt}{2.3ex} \rule[-1ex]{0pt}{0pt}\\
   RV semi-amplitude ($K_{\rm c}$) & m s$^{-1}$ & $\mathcal{U}$(0.01, 100.0) & 4.2$\pm 1.6$ \rule{0pt}{2.3ex} \rule[-1ex]{0pt}{0pt}\\
   Planetary mass ($M_{\rm p,c}$) & $M_{\oplus}$ & ... & 11.6$^{+4.7}_{-4.6}$ \rule{0pt}{2.3ex} \rule[-1.5ex]{0pt}{0pt}\\
\hline
\end{tabular}
\end{table*}

\section{Stellar parameters using TRES spectra}
The TRES spectra were also used to derive stellar parameters using the Stellar Parameter Classification tool (SPC; \citealt{Buchhave2012}). SPC cross correlates an observed spectrum against a grid of synthetic spectra based on the Kurucz atmospheric model \citep{1992IAUS..149..225K} which derives the effective temperature $T_{\rm eff} = 5933 \pm 50$ K, surface gravity $\log g = 4.42 \pm 0.10$, metallicity $[m/H] = 0.03 \pm 0.08$, and rotational velocity of the star $v \sin{i_{\star}} = 8.4 \pm 0.5$ km s$^{-1}$. The stellar parameters align with the estimates that we presented in Sect. \ref{sec:host-star} and Table \ref{tab:star_param}.

\section{Independent RV analysis using {\sc tweaks}
}
The HARPS-N CCFs were independently analysed for planetary signals using {\sc tweaks} (Time and Wavelength domain stEllar Activity mitigation using {\sc \textbf{k}ima} and {\sc \textbf{s}calpels}). This pipeline was designed to achieve a sub-m/s detection threshold at extended orbital periods (Anna John et al. 2023 submitted). This is achieved by obtaining a set of time-domain stellar activity-decorrelation vectors using {\sc scalpels} \citep{2021MNRAS.505.1699C}, by doing principal-component analysis on the autocorrelation function of the CCF. These basis vectors were then used for the spectral line-shape decorrelation \citep{2021MNRAS.505.1699C} in {\sc kima} \citep{2018JOSS....3..487F}, as \citep{2022MNRAS.515.3975J} reported that using the {\sc scalpels} basis vectors to de-trend the RVs for line shape variations results in a model that is significantly better than a model that does not take these stellar activity signatures into account. 

A model with up to five unidentified Keplerian signals was used in our first blind search of the RVs using the {\sc kima} nested-sampling package \citep{2018JOSS....3..487F}. As \citep{2022MNRAS.515.3975J} have found, some planet-like signals elude {\sc scalpels} analysis, so any remaining rotationally-modulated signals were modelled with GP regression applied to the RVs, using a quasi-periodic GP kernel. The joint posteriors showed clear detection of planet b, at orbital period 10.5905 $\pm$ 0.0002 days with an RV semi-amplitude of 14.51 $\pm$ 3.74 ms$^{-1}$. Additionally, the GP strongly constrained the stellar rotation to $P_{\rm rot} = 7.36_{-0.08}^{+0.11}$ days. This independent analysis aligns with the RV reference model solution and the stellar parameters discussed in Sect. \ref{sec:rotation}.

\section{Extra Tables}
\label{app:table_extra}
\begin{table*}
\caption{Second part of Table \ref{table:model_pl}.}             
\label{table:model_pl_2}      
\centering          
\begin{tabular}{l c c c}     
\hline\hline     

 \multicolumn{4}{c}{Photometric time-series fit} \rule{0pt}{2ex} \rule[-0.9ex]{0pt}{0pt} \\ 
\hline    
                          
Parameter & Unit & Prior & Value \rule{0pt}{2.2ex} \rule[-0.9ex]{0pt}{0pt}\\ 
\hline    
   \textit{TESS} dilution factor &  & $\mathcal{N}$(0.00735, 0.00005) & 0.007348$\pm$0.000050 \rule{0pt}{2.2ex} \rule[-0.8ex]{0pt}{0pt}\\
   \textit{TESS} jitter ($\sigma^{\rm TESS}_{\rm jitter}$) & e$^-$ s$^{-1}$ & ... & 5.0$^{+1.1}_{-1.4}$ \rule{0pt}{2.2ex} \rule[-0.8ex]{0pt}{0pt}\\  
   Asiago jitter ($\sigma^{\rm Asiago}_{\rm jitter}$) &  & ... & 0.00074$^{+0.00006}_{-0.00007}$ \rule{0pt}{2.2ex} \rule[-0.8ex]{0pt}{0pt}\\  
   LCO\_McD\_c jitter ($\sigma^{\rm McD\_c}_{\rm jitter}$) &  & ... & 0.00145$^{+0.00011}_{-0.00010}$ \rule{0pt}{2.2ex} \rule[-0.8ex]{0pt}{0pt}\\ 
   LCO\_McD\_b jitter ($\sigma^{\rm McD\_b}_{\rm jitter}$) &  & ... & 0.00073$^{+0.00009}_{-0.00008}$ \rule{0pt}{2.2ex} \rule[-0.8ex]{0pt}{0pt}\\
   MuSCAT2 jitter ($\sigma^{\rm MuSCAT2}_{\rm jitter}$) &  & ... & 0.00014$^{+0.00012}_{-0.00008}$ \rule{0pt}{2.2ex} \rule[-0.8ex]{0pt}{0pt}\\
   KeplerCam jitter ($\sigma^{\rm KeplerCam}_{\rm jitter}$) &  & ... & 0.00279$^{+0.00011}_{-0.00010}$ \rule{0pt}{2.2ex} \rule[-0.8ex]{0pt}{0pt}\\
   Acton jitter ($\sigma^{\rm Acton}_{\rm jitter}$) &  & ... & 0.00349$\pm$0.00010 \rule{0pt}{2.2ex} \rule[-0.8ex]{0pt}{0pt}\\
   LCO\_McD\_bc jitter ($\sigma^{\rm McD\_bc}_{\rm jitter}$) &  & ... & 0.00095$^{+0.00008}_{-0.00008}$ \rule{0pt}{2.2ex} \rule[-0.8ex]{0pt}{0pt}\\
   LCO\_Teid jitter ($\sigma^{\rm Teid}_{\rm jitter}$) &  & ... & 0.00067$^{+0.00012}_{-0.00013}$ \rule{0pt}{2.2ex} \rule[-0.8ex]{0pt}{0pt}\\
   \textit{TESS} quadratic LD coefficient ($u^{TESS}_1$) &  & $\mathcal{U}$(0, 1) & 0.089$^{+0.089}_{-0.061}$ \rule{0pt}{2.2ex} \rule[-0.8ex]{0pt}{0pt}\\
   \textit{TESS} quadratic LD coefficient ($u^{TESS}_2$) &  & $\mathcal{U}$(0, 1) & 0.52$^{+0.11}_{-0.14}$ \rule{0pt}{2.2ex} \rule[-0.8ex]{0pt}{0pt}\\
   Asiago quadratic LD coefficient ($u^{Asiago}_1$) &  & $\mathcal{N}$(0.489, 0.014) & 0.455$\pm$0.013 \rule{0pt}{2.2ex} \rule[-0.8ex]{0pt}{0pt}\\
   Asiago quadratic LD coefficient ($u^{Asiago}_2$) &  & $\mathcal{N}$(0.151, 0.037) & 0.007$^{+0.033}_{-0.032}$ \rule{0pt}{2.2ex} \rule[-0.8ex]{0pt}{0pt}\\
   McD\_c quadratic LD coefficient ($u^{McD\_c}_1$) &  & $\mathcal{N}$(0.34, 0.01) & 0.340$\pm$0.010 \rule{0pt}{2.2ex} \rule[-0.8ex]{0pt}{0pt}\\
   McD\_c quadratic LD coefficient ($u^{McD\_c}_2$) &  & $\mathcal{N}$(0.14, 0.03) & 0.14$\pm$0.03 \rule{0pt}{2.2ex} \rule[-0.8ex]{0pt}{0pt}\\
   McD\_b quadratic LD coefficient ($u^{McD\_b}_1$) &  & $\mathcal{N}$(0.34, 0.01) & 0.343$\pm$0.010 \rule{0pt}{2.2ex} \rule[-0.8ex]{0pt}{0pt}\\
   McD\_b quadratic LD coefficient ($u^{McD\_b}_2$) &  & $\mathcal{N}$(0.14, 0.03) & 0.17$\pm$0.03 \rule{0pt}{2.2ex} \rule[-0.8ex]{0pt}{0pt}\\
   MuSCAT2 quadratic LD coefficient ($u^{MuSCAT2}_1$) &  & $\mathcal{N}$(0.34, 0.01) & 0.337$\pm$0.010 \rule{0pt}{2.2ex} \rule[-0.8ex]{0pt}{0pt}\\
   MuSCAT2 quadratic LD coefficient ($u^{MuSCAT2}_2$) &  & $\mathcal{N}$(0.14, 0.03) & 0.13$\pm$0.03 \rule{0pt}{2.2ex} \rule[-0.8ex]{0pt}{0pt}\\
   KeplerCam quadratic LD coefficient ($u^{KeplerCam}_1$) &  & $\mathcal{N}$(0.34, 0.01) & 0.334$\pm$0.010 \rule{0pt}{2.2ex} \rule[-0.8ex]{0pt}{0pt}\\
   KeplerCam quadratic LD coefficient ($u^{KeplerCam}_2$) &  & $\mathcal{N}$(0.14, 0.03) & 0.10$\pm$0.03 \rule{0pt}{2.2ex} \rule[-0.8ex]{0pt}{0pt}\\
   Acton quadratic LD coefficient ($u^{Acton}_1$) &  & $\mathcal{N}$(0.40, 0.01) & 0.4013$\pm$0.0099 \rule{0pt}{2.2ex} \rule[-0.8ex]{0pt}{0pt}\\
   Acton quadratic LD coefficient ($u^{Acton}_2$) &  & $\mathcal{N}$(0.14, 0.03) & 0.15$\pm$0.03 \rule{0pt}{2.2ex} \rule[-0.8ex]{0pt}{0pt}\\
   McD\_bc quadratic LD coefficient ($u^{McD\_bc}_1$) &  & $\mathcal{N}$(0.34, 0.01) & 0.340$\pm$0.010 \rule{0pt}{2.2ex} \rule[-0.8ex]{0pt}{0pt}\\
   McD\_bc quadratic LD coefficient ($u^{McD\_bc}_2$) &  & $\mathcal{N}$(0.14, 0.03) & 0.14$\pm$0.03 \rule{0pt}{2.2ex} \rule[-0.8ex]{0pt}{0pt}\\
   Teid quadratic LD coefficient ($u^{Teid}_1$) &  & $\mathcal{N}$(0.34, 0.01) & 0.339$\pm$0.010 \rule{0pt}{2.2ex} \rule[-0.8ex]{0pt}{0pt}\\
   Teid quadratic LD coefficient ($u^{Teid}_2$) &  & $\mathcal{N}$(0.14, 0.03) & 0.14$\pm$0.03 \rule{0pt}{2.2ex} \rule[-0.8ex]{0pt}{0pt}\\
   
\hline
\end{tabular}
\end{table*}

\end{document}